\documentclass[twocolumn]{aastex62}

\usepackage{amsmath}
\usepackage{booktabs}

\shortauthors{Hatta}
\graphicspath{{./}{figures/}}
\begin{document}

%\title{Effects of Gradient in Brunt-$\rm{V}\ddot{a}is\ddot{a}l\ddot{a}$ Frequency on G-Mode Period Spacing} 
\title{Semi-Analytical Expression of G-Mode Period Spacing: \\
The Case of Brunt-$\rm{V}\ddot{a}is\ddot{a}l\ddot{a}$ Frequency with Not a Jump But a Ramp}

\correspondingauthor{Yoshiki Hatta}
%\email{yoshiki.hatta@nao.ac.jp}
\email{yoshiki.hatta@isee.nagoya-u.ac.jp}

%\correspondingauthor{August Muench}
%\email{greg.schwarz@aas.org, gus.muench@aas.org}

\author[0000-0003-0747-8835]{Yoshiki Hatta}
\affiliation{Department of Astronomical Science, School of Physical Sciences, SOKENDAI\\
2-21-1 Osawa, Mitaka, Tokyo 181-8588, Japan}
\affiliation{National Astronomical Observatory of Japan \\
2-21-1 Osawa, Mitaka, Tokyo 181-8588, Japan}
\affiliation{Institute for Space-Earth Environmental Research, Nagoya University \\
Furo-cho, Chikusa-ku, Nagoya, Aichi 464-8601, Japan}



\begin{abstract}
%The BV frequency $N$ sensitively depends on ... evolution ... mixing ...
%Deciphering period spacings of high-order g-mode pulsators such as $\Gamma$ Dor stars and SPB stars 
%is quite essential to infer Brunt-$\rm{V}\ddot{a}is\ddot{a}l\ddot{a}$ Frequency ($N$) profiles inside them, 
To decipher complex patterns of gravity-mode period spacings 
observed for intermediate-mass main-sequence stars 
is an important step toward the better understanding of the 
structure and dynamics in the deep radiative region of the stars. 
%In this study, 
%
%The Brunt-$\rm{V}\ddot{a}is\ddot{a}l\ddot{a}$ Frequency ($N$) profile of a star is a quite important parameter 
%on which mixing processes inside the star have significant impacts. 
%A number of main-sequence intermediate-mass stars are exhibiting 
%g-mode period spacings whose patterns are closely related to 
%the Brunt-$\rm{V}\ddot{a}is\ddot{a}l\ddot{a}$ Frequency ($N$) profile of a star
%g-mode period spacing patterns, 
%based on which the deep radiative region of the stars can be probed. 
%
%The period spacing of high-order g modes %($\Delta P_{g}$) 
%which have been often observed for $\gamma$ Dor stars and SPB stars, 
%generally considered to be closely related to $N$ profile of the star, 
%is a fairly informative observable based on which the deep radiative region of stars can be probed. 
%
%are fairly important observables to infer 
%as $\Delta P_{g}$ are generally considered to be closely related to $N$ profile of the star.
%In the context that ... knowing N is important to understand the mixing processes inside stars, 
%period spacings of high-order g-mode pulsators such as ... are quite important observables  
%since they are thought to be closely related to N profile of the star. 
%To better understand the observed ... %$\Delta P_{g}$, 
%deepen the understanding of the relation between the $N$ and $\Delta P_{g}$, %we analyze the effects of the gradient in N on the g-mode period spacing 
%and 
In this study, we apply JWKB approximation to derive 
a semi-analytical expression of the g-mode period spacing pattern, 
for which the gradient in the Brunt-$\rm{V}\ddot{a}is\ddot{a}l\ddot{a}$ frequency is taken into account. 
The formulation includes a %g-mode-period-dependent 
term $P^{-1} B_{\star}$, % $A^{\star}$, 
where $P$ and $B_{\star}$ represent the g-mode period and 
degree of the structural variation, %%sensed by the g modes, % in the deep radiative region of a star, 
the latter of which especially is related to %describes 
the steepness of the gradient of the Brunt-$\rm{V}\ddot{a}is\ddot{a}l\ddot{a}$ frequency. %, 
Tests with 1-dimensional stellar models show that 
the semi-analytical expression derived in this study is useful for %extracting glitch information %inferring the degree of the glitch 
inferring the degree of the structural variation $B_{\star}$ 
with accuracy of $\sim 10\, \%$ in the case of 
relatively massive intermediate-mass models with the mass $M$ 
larger than $3 \,M_{\odot}$. 
%which is in particular the case for relatively massive intermediate-mass main-sequence stars ($M < 3 \,M_{\odot}$). 
%in the deep radiative region of main-sequence stars, 
%especially 
%%
%%is then conducted to evaluate the potential of the semi-analytical expression 
%%for observational studies of the deep radiative region of intermediate-mass stars. 
%%%%there is just at most 10 $\%$ discrepancy ... (systematic errors). 
%%As a case study, the g-mode period spacing pattern observed for one of the Kepler targets, KIC 11145123, 
%%is fitted with the expression, illustrates that ..., 
%%which is consistent with the conclusion of Hatta et al. 2021 that ... .
%Though there remains the need to
%%The derived semi-analytical expression can be used, in particular, 
%%to infer the gradient of the Brunt-$\rm{V}\ddot{a}is\ddot{a}l\ddot{a}$ frequency, 
%Since the gradient of the Brunt-$\rm{V}\ddot{a}is\ddot{a}l\ddot{a}$ frequency is closely related to 
%%%%which corresponds to 
%where the chemical composition gradient is steep, %as constraints on N profile..., 
The newly formulated expression will possibly allow us to put further constraints on, e.g., mixing processes 
inside intermediate-mass main-sequence g-mode pulsators such as $\beta$ Cep, SPB, and $\gamma$ Dor stars 
that have been principal targets in asteroseismology. 

\end{abstract}



\section{Introduction} \label{sec:intro}
Modern space-borne missions such as %CoRoT (ref), 
Kepler \citep{Kepler} and TESS \citep{TESS} 
have enabled us to detect non-radial oscillations of tens of thousands of stars 
with unprecedentedly high precisions ($\sim 10^{-5} \, \mathrm{mag}$), 
bringing about the firm establishment of asteroseismology \citep[e.g.][]{Aerts_text}. 
%Internal gravity waves 
%%Gravity (g) modes are of particular importance 
%since they can support gravity (g) modes, %or mixed modes, 
%based on measurements of which 
Main-sequence gravity-mode pulsators (e.g., $\gamma$ Dor stars, SPB stars, and $\beta$ Cep stars) 
are of particular importance 
since, based on measurements of the g modes, 
we can probe the deep radiative region of these stars \citep[e.g.][]{Unno_text}, 
%such as $\gamma$ Dor stars, SPB stars, and white dwarfsred giants, WD... . 
%(NO?? ANOTHER WAY IS TO ARTICULATE THAT 
%WE ARE NOT GOING TO TALK ABOUT MIXED-MODES IN THIS PAPER 
%(SHOULD BE AROUND THE END OF THE INTRO.??))
% the internal properties of stars %structure and dynamics of the stars 
%%%%especially for the deep radiative region 
%G-mode asteroseismology has thus 
thus far leading to numerous significant constraints on theories of, for instance, 
the angular momentum transfer or the chemical element transport deep inside stars \citep[e.g.][]{Aerts2019}

One remarkable feature often seen in observed spectra 
of g-mode pulsators is that 
g-mode peaks are almost equidistantly spaced in periods 
\citep[e.g.][]{Degroote2010a,Papics2014,Papics2015,VanReeth2015}. 
This observational characteristic can be explained 
by the asymptotic theory of stellar non-radial oscillation \citep[e.g.][]{Tassoul1980} 
%%that predicts, 
where local wavelengths of the waves are assumed to be 
much shorter than the scale height of the background (c.f. geometrical optics). 
In such a high-order limit, 
the g-mode period spacing, %($\Delta P_{g}$ hereafter), 
which is defined as the difference between two g-mode periods 
with the neighboring radial order $n$ and the same spherical degree $l$, 
is constant: 
%The fact that most of the g-mode pulsators observed so far are exhibiting high-order g modes 
%is such a fortune that we can apply the asymptotic theory of stellar non-radial oscillation \citep[e.g.][]{Tassoul1980}. 
%
%to interpret the observation. 
%%%%In particular, the g-mode period spacing $\Delta P_{g}$, which is defined as the difference between two g-mode periods 
%%%%with the neighboring radial order $n$ and the same spherical degree $l$, 
%The g-mode period spacing (delta Pg) is %playing a central role in g-mode asteroseismology. 
%%%%is often used (ref...). % in g-mode asteroseismology. 
%%%%This is because the asymptotic theory tells us that, in a simple case, 
%%%%the g-mode period spacing is constant:  %especially, it is 
%inversely proportional to the following integration 
\begin{equation}
\Delta P_{g} = \frac{2 \pi^2}{\sqrt{l(l+1)}} \biggl [ \int_{r_{0}}^{r_{1}} \frac{N}{r} \, \mathrm{d}r \biggr ]^{-1}, \label{eq01}
\end{equation}
where $N$ and $r$ represent the Brunt-$\rm{V}\ddot{a}is\ddot{a}l\ddot{a}$ 
(BV hereafter) frequency 
and the distance from the center. 
The bottom and top of the g-mode cavity are denoted by $r_0$ and $r_1$, respectively. 
 %of the Brunt-$\rm{V}\ddot{a}is\ddot{a}l\ddot{a}$ frequency. 
%with a certain spherical degree is proportional to  . constant delta Pg. 
%%%%As a star evolves, the nuclear burning core becomes smaller, 
%%%%the central region becomes more dense, and %and so N, 
%%%%the integration accordingly becomes larger, finally causing the g-mode period spacing smaller. 
%%which is also confirmed by 
%%stellar evolutionary codes show that delta Pg monotonically decrease as the star evolves. 
%This is the reason why the g-mode period spacing can be used as an indicator of the age or stellar evolutionary stage. 
%%%%In other words, we can estimate the evolutionary stage of a g-mode pulsator 
%%%%by comparing the mean value of observed g-mode period spacings and 
%%%%those computed based on 1-dimensional stellar evolutionary models. 
Since the integration in the right-hand side of Equation (\ref{eq01}) %value $\int N \mathrm{d} \, \mathrm{ln} \, r$ 
increases almost monotonically 
along with stellar evolution \citep{Miglio2008}, %%%% during the main-sequence (), 
the g-mode period spacing can be used for inferring 
the evolutionary stage of stars \citep[e.g.][]{Bedding2011, Kurtz2014, Saio2015}. %(Bedding+2011; Kurtz+2014; ...). 
%%Since the Brunt-$\rm{V}\ddot{a}is\ddot{a}l\ddot{a}$ frequency is 
%%sensitive to the evolutionary stage, 
%%observed g-mode period spacings are quite useful for inferring ... .

%The g-mode period spacing thus can be used 
%as an indicator of the age or stellar evolutionary stage (Bedding for red giants, ... for MS stars). 

%%(Actually, it is observationally confirmed that ) 
%It is frequently the case that a mean value of observed g-mode period spacings 
%is used for ... modeling ... (ref) 
%%%%More can be done if ... .
Although the almost constant g-mode period spacings indicate 
that the observed g modes are 
indeed high-order enough to closely follow the simple asymptotic relation (\ref{eq01}), 
it is nevertheless usually the case that 
%since 
observed g-mode period spacings exhibit statistically significant deviation from constancy, 
which is often characterized by oscillatory patterns around the constant value 
given by Equation (\ref{eq01}) \citep[e.g.][]{Degroote2010b,VanReeth2015,Pedersen2018,Garcia2022a}. 
%We often see some deviations of observed g-mode period spacings from constancy when the 
%((In reality)), not constant… 
%so, the mean value of the delta Pg … used for modeling… age determination…. 
%But the deviation is ... not merely a ...
%However, more can be done by focusing on the deviation from the constancy. 
These oscillatory patterns % in g-mode period spacings 
are 
interpreted as a representation of mode trapping \citep{Brassard1992,Montgomery2003,Miglio2008}; %,  
%%there should be 
%%%%which can be attributed to % is caused mainly by 
a sharp variation in the BV frequency %Brunt-$\rm{V}\ddot{a}is\ddot{a}l\ddot{a}$ frequency 
must be taking place with the 
%whose scale 
scale height %is 
comparable to (or smaller than) a typical local wavelength of gravity waves there, 
rendering some particular g modes to be ``trapped'' in a %narrowly confined 
certain region. %cavities different from mode to mode. 
Such a sharp variation in the BV %Brunt-$\rm{V}\ddot{a}is\ddot{a}l\ddot{a}$ 
frequency (also called ``buoyancy glitch'') 
is, in the case of main-sequence g-mode pulsators, 
considered to be related to a steep chemical composition gradient %, or the so-called glitch, 
%developed 
in the deep radiative region just above the convective core 
that is left by the receding nuclear burning core \citep{Miglio2008}. 
%Miglio et al. 2008… delta Pg can be oscillatory, 
%representing mode trapping, 
%which is caused by the glitch in the BV frequency. 

%In particular, 
Importantly, %Miglio+2008 have shown that 
it has been theoretically shown that 
the location and degree of a sharp variation %the glitch 
in the BV frequency 
determine the period and amplitude, respectively, 
of the oscillatory component in a g-mode period spacing pattern \citep[e.g.][]{Miglio2008}.
In other words, we can extract the information on %the glitch in 
the BV frequency profile %$N$ 
%the Brunt-$\rm{V}\ddot{a}is\ddot{a}l\ddot{a}$ frequency 
%infer fine structures in the deep radiative region 
by analyzing the observed g-mode period spacing patterns 
($\Delta P_{g}$ patterns hereafter). 
%%In particular, mixing processes inside stars have considerable impacts on the BV frequency profile, 
%%which is the reason why 
%%there have been a number of studies in which 
%%the observed g-mode period spacing patterns have been fitted 
%%with those numerically computed with 1-dimensional stellar models 
%%to put constraints on mixing prescriptions used in the 1-d stellar evolutionary computations. 
For example, various mixing prescriptions used in 1-dimensional stellar evolutionary codes 
can be tested by comparing the observed $\Delta P_{g}$ %g-mode period spacing 
patterns 
with those computed based on 1-dimensional stellar models 
since different mixing prescriptions lead to different BV frequency profiles, 
resulting in different $\Delta P_{g}$ %g-mode period spacing 
patterns. 
For more detail, see, e.g., 
\citet{Moravveji2015}, \citet{Moravveji2016}, \citet{Buysschaert2018}, 
\citet{Wu2019}, \citet{Wu2020}, \citet{Michielsen2021}, \citet{Pedersen2018,Pedersen2022a} 
and references therein. 
There have also been several attempts to extend the theoretical work of \citet{Miglio2008}. 
%In particular, 
Significant progress has been made by \citet{Cunha2019} 
who have explicitly derived 
semi-analytical expressions of the $\Delta P_{g}$ 
%g-mode period spacing 
pattern 
for main-sequence g-mode pulsators 
based on the asymptotic approximation 
and the two-zone modeling of the BV frequency to parameterize the sharp variation in it 
(as will be detailed later in Section \ref{sec:3}). 
%one important based on the asymptotic approximation 
%and the two-zone modeling of the BV frequency to parameterize the buoyancy glitch, 
%Cunha+2019 have derived 
%semi-analytical expressions of the $\Delta P_{g}$ 
%%g-mode period spacing 
%pattern (as will be detailed later in Section \ref{sec:3})
%%Another promising approach ... 
%%%What they have done is ... 
%%%We also have analytical studies such as those conducted by Cunha et al+2019. 
%%They have derived semi-analytical expressions for 
%(gamma Dor and SPB) 
%%%as well as for mixed-mode pulsators, namely, subgiants and red giants.  
%%Their formulation is based on two-zone modeling of the BV frequency 
%%between which is a discontinuity which . 
%%A JWKB solution is assumed for each  between which is a discontinuous ... assuming JWKB 
As a case study, they have successfully demonstrated that the explicit expression can 
reproduce the $\Delta P_{g}$ 
%g-mode period spacing 
pattern of a main-sequence $6 \, M_{\odot}$ model, 
highlighting the high potential of the semi-analytical approach to 
study %the glitch in 
the BV frequency profile in a rather model-independent manner. 
%%%%We can infer … 

In this study, we would like to expand upon %suggest that 
%there is still room for improvements in 
the formulation of \citet{Cunha2019} 
with respect to how we model the BV frequency to derive semi-analytical expressions 
of the $\Delta P_{g}$ pattern. 
As will be explained in Section \ref{sec:2}, 
it becomes rather difficult to parameterize the BV frequency of less massive stars ($\sim 1.3-3 \, M_{\odot}$) 
by two-zone modeling with a discontinuity 
%%as the mass of the star is smaller 
because the BV frequency becomes less discontinuous as the mass of the star gets smaller; 
%%((In other words, there is room for us to consider the … 
%%not a discontinuity but a sharp feature (with a finite width), %(This should be relevant since...)
%%%Such a gradient in $N$ becomes more and more significant as the mass of the star is smaller, 
%%%due to the gradual dominance of the pp-chain reaction 
%%%just outside the convective core. 
not a jump but a ramp is required to model the BV frequency. 
One possible alternative in the formulation of \citet{Cunha2019} is thus 
to take the gradient in the BV frequency into account. % in the formulation. 

It should also be instructive to remind the readers of the fact 
that most of the theoretical and observational studies of 
%%oscillatory components in 
$\Delta P_{g}$ %g-mode period spacing 
patterns to infer the internal structures of g-mode pulsators are focusing on 
relatively massive ($\sim 3-8 \, M_{\odot}$) main-sequence stars %such as SPB stars or $\beta$ Cep stars 
in which the BV frequency %profile %buoyancy glitch %a glitch in the BV frequency 
%%Brunt-$\rm{V}\ddot{a}is\ddot{a}l\ddot{a}$ frequency 
can be parameterized reasonably well by two-zone modeling 
with a discontinuity \citep[e.g.][]{Pedersen2018}. 
%%%which have been often seen especially for the low-mass intermediate-mass stellar models, 
%%%leading to the g-mode period spacings exhibiting the decreasing amplitudes 
%%%as was suggested by Miglio2009 ... .
%%%It is better for us to consider where the glitch is and how sharp the glitch is… 
%%%in a little bit more careful manner… 
In this context, our new formulation might be helpful for carrying out g-mode asteroseismology of 
relatively low-mass intermediate-mass stars, namely, $\gamma$ Dor stars, 
for which a large number of $\Delta P_{g}$ %g-mode period spacing 
patterns have been detected so far \citep[e.g.][]{VanReeth2015}. 
(Note that the $\Delta P_{g}$ patterns of $\gamma$ Dor stars have been intensively analyzed 
to infer the internal rotation rate of the stars \citep[e.g.][]{Li2019, Li2020, Saio2021, Tokuno2022,Garcia2022b}, 
though the oscillatory component around a constant value 
has not been used. 
We will discuss it later in Section \ref{sec:5}.) 

There are two goals in this study. 
One is to present a semi-analytical expression of the $\Delta P_{g}$ pattern 
%g-mode period spacing pattern 
taking into account the gradient in the BV frequency. % into account. 
Another goal is to demonstrate how the semi-analytical expression is useful 
for extracting information on the BV frequency profile. 
We start with a brief review on internal structures of 
main-sequence g-mode pulsators (Section \ref{sec:2}). 
Then, we present the formulation and validation of a new semi-analytical expression (Section \ref{sec:3}). 
The derived expression is tested with realistic stellar models 
and then applied to one of the Kepler targets KIC 11145123 \citep{Huber2014,Kurtz2014}, 
which is a $\delta$ Sct-$\gamma$ Dor hybrid pulsator (Section \ref{sec:4}). 
It should be noted that, in the derivation of the semi-analytical expression, 
we neglect the effects of rotation and convective overshooting on $\Delta P_{g}$ patterns. %, 
%g-mode period spacing patterns, 
%%which have been one of the most actively investigated topics these days (Li+2019, 2021). 
We therefore have a few brief discussions about the impact of the negligence (Section \ref{sec:5}). 
%formulation focusing on the effects neglected in this study (Section 5), 
We finally conclude in Section \ref{sec:6}. 

\section{Structure Around Deep Radiative Region Of %Brunt-$\rm{V}\ddot{a}is\ddot{a}l\ddot{a}$ Frequency of 
Intermediate-Mass Stars} \label{sec:2}
One major goal in this section is to highlight 
how internal structures (especially around the deep radiative region) 
of main-sequence g-mode pulsators vary %are 
%as 
depending on the stellar mass and evolutionary stage. 
We also would like to show that the difference in internal structures 
leads to different behaviors of $\Delta P_{g}$ patterns. 
%compare internal structures of 
%various main-sequence intermediate-mass stellar models with different masses 
To this end, we compute several evolutionary models of main-sequence stars with different masses 
and ages. 
%%In this section, we will present Brunt-$\rm{V}\ddot{a}is\ddot{a}l\ddot{a}$ frequency ($N$) profiles 
%%of various main-sequence intermediate-mass stellar models 
%%to highlight the fact that the slope in $N$ is smaller as the model is less massive. 
%there can be a slope in $N$ of a relatively low-mass intermediate-mass star. 
After we explain basic setups to compute the stellar models in Section \ref{sec:2-1}, 
we show internal structures %$N$ profiles 
of the models in Section \ref{sec:2-2}, where 
a possible cause for a milder variation %smaller and wider glitch 
in the BV frequency for a lower-mass stellar model is discussed. 
The $\Delta P_{g}$ patterns of the corresponding models are shown in Section \ref{sec:2-3} 
in order to qualitatively check the effects of the gradient in the BV frequency on the $\Delta P_{g}$ pattern. 

%Because the period spacing of high-order g modes is essentially determined by 
%the Brunt-$\rm{V}\ddot{a}is\ddot{a}l\ddot{a}$ frequency profile of the star 
%we discuss the Brunt-$\rm{V}\ddot{a}is\ddot{a}l\ddot{a}$ frequency profile of 
%typical g-mode pulsators whose masses, roughly speaking, range from $1.6M_{\odot}$ to $3M_{\odot}$. 
%
%based on 1-dimensional stellar models, 
%we discuss how the Brunt-$\rm{V}\ddot{a}is\ddot{a}l\ddot{a}$ frequency profile inside a star 
%depends on the mass of the star (Section \ref{sec:2-1}). 
%as well as a possible cause of the difference. 

\subsection{Models} \label{sec:2-1}
We have used MESA 
\citep[version 15140;][]{Paxton2011,Paxton2013,Paxton2015,Paxton2018,Paxton2019} 
basically with its default settings 
%Modules for Experiments in Stellar Astrophysics (MESA, ver 15140) (Paxton...) 
to compute stellar evolutionary tracks for different masses, namely, $1.6$, $3$, and $6 \, M_{\odot}$, 
which roughly covers the mass range of main-sequence intermediate-mass g-mode pulsators.  
The same initial chemical composition $(X,Y,Z) = (0.70,0.28,0.02)$ is used for the three evolutionary sequences. 
No magnetic field, rotation, nor convective overshooting has been taken into account. 

%%We have adopted the default settings in MESA regarding 
%%the equation of state, %(OPAL, ...), 
%%the opacity, % (OPAL, ...), 
%%and the nuclear network. % (``basic.net'', see Paxton...). 
%%Convection is described by the mixing length theory \citep{Bohm_Vintense1958} 
%%for which the mixing length parameter $\alpha_{\mathrm{MLT}} = 2$ is assigned in this study. 
One specific MESA option we have additionally activated in the computation 
is elemental diffusion \citep[see, e.g.,][]{Paxton2018} 
for the purpose of eliminating apparently unphysical discontinuous features 
in the BV frequency around the deep radiative region. 
%elemental diffusion is activated \citep[see, e.g.,][]{Paxton2018}. % where 
The diffusion velocities in the envelope are artificially suppressed 
by setting the parameter ``diffusion$\_$v$\_$max'' equal to $10^{-7}$ 
to render the helium in the envelope not to be depleted \citep{Morel2002}. 
Note that radiative levitation \citep[e.g.][]{Michaud2015} is not included. %in the computation of elemental diffusion. 

Along with evolution, 
when the central hydrogen mass content $X_{\mathrm{c}}$ reaches $0.5$ and $0.3$, 
the corresponding equilibrium model is preserved 
whose internal structures will be shown in the next section. 

\subsection{Internal structure around the deep radiative region} \label{sec:2-2}
\begin{figure}[t]
\begin{center}
\includegraphics[scale=0.52]{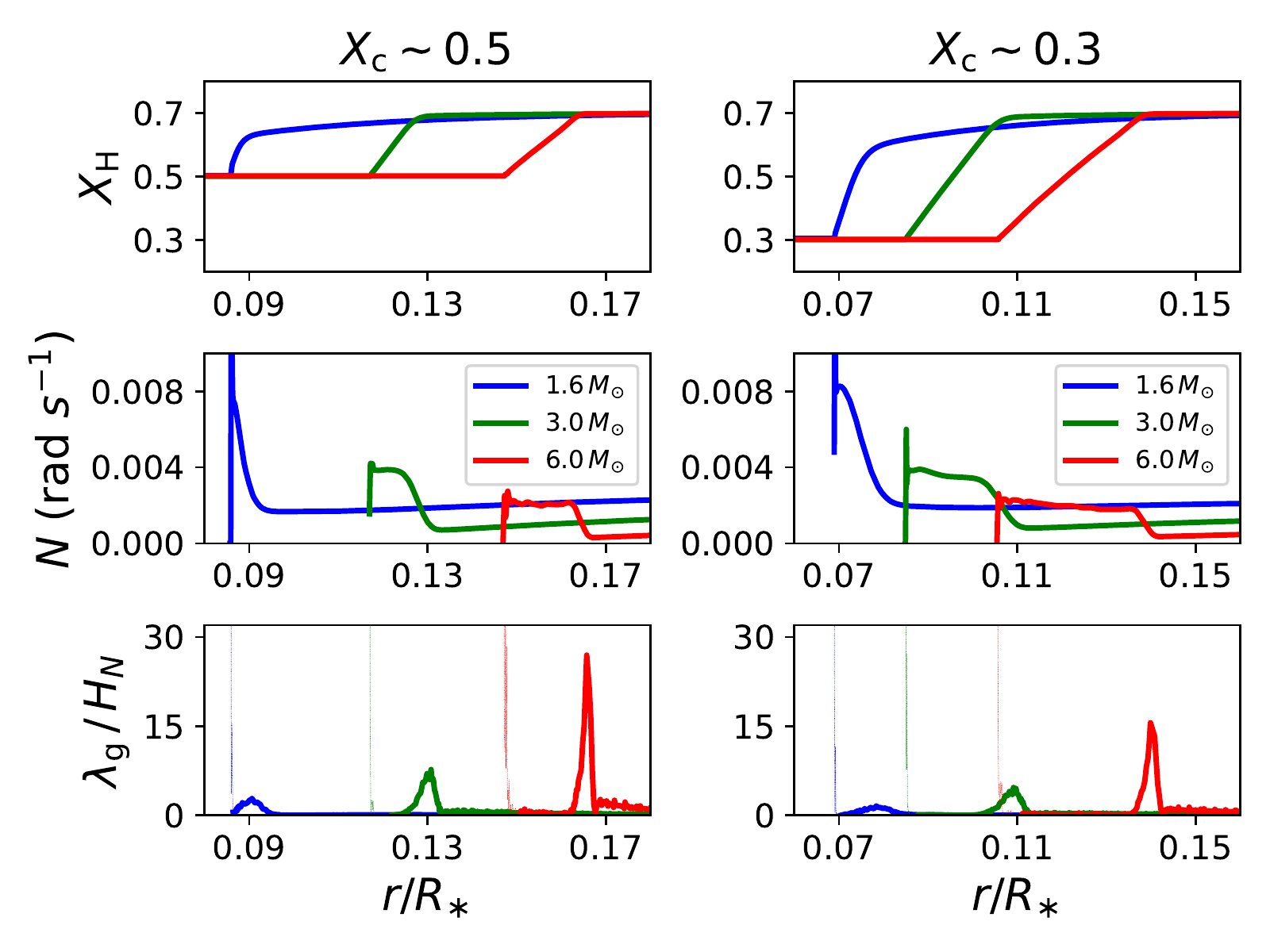}
\caption{\footnotesize Structural parameters in the deep radiative region of stellar models with 
$1.6 \, M_{\odot}$ (blue), $3 \, M_{\odot}$ (green), and $6 \, M_{\odot}$ (red) 
when the central hydrogen mass content $X_{\mathrm{c}}$ is $0.5$ (left column) or $0.3$ (right column). 
The horizontal axis is the fractional radius where $R_{\ast}$ is the radius of the model. 
The hydrogen mass content $X_{H}$, the BV frequency $N$, and the ratio 
(of a typical wavelength of high-order g modes to the scale height of the BV frequency) $\lambda_{g} / H_{N}$ 
are shown in this order from top to bottom. 
Note that $\lambda_{g} / H_{N}$ diverges at the convective-core boundary 
since $H_{N} \sim 0$ at the boundary. 
Because what we will focus on in this article is not the boundary but well inside the g-mode cavity 
(where $N$ is positive), %and thus, 
we represent the irrelevant parts of $\lambda_{g} / H_{N}$ 
by thin dotted curves for visual aid. 
}
\label{fig:2-1}
\end{center}
\end{figure}

As was mentioned in Section \ref{sec:intro}, 
the most important quantity is  
the BV frequency %profile %and the chemical composition profile 
which determines g-mode periods in the high-order limit. %periods. 
We have thus shown BV frequency profiles of the six models 
(see middle row in Figure \ref{fig:2-1}) 
which have been computed based on the settings described in the previous section. 
The corresponding hydrogen abundance profiles are presented as well (see top row in Figure \ref{fig:2-1}) 
since the chemical composition gradient contributes to the BV frequency 
\citep[see, e.g., Equation 2 in][]{Miglio2008}. 
All the models have the convective core where 
the square of the BV frequency is negative and the gravity waves are not propagative. 
In addition, the hydrogen abundance is uniform there. 
%In addition, the square of the BV frequency is negative in the convective core. %, 
%and no value is assigned in the figure. %(see, e.g., the blue curve in middle left panel)
Just above the convective core lies the hydrogen abundance gradient 
which has been developed by the receding nuclear (and convective) core as the star evolves. 
Where the hydrogen abundance gradient is developed is almost identical to 
where the BV frequency is large, 
which plays an important role in mode trapping \citep{Miglio2008}. 
%\begin{figure}[t]
%\plotone{Figure_2_1.pdf}
%\caption{  \label{fig:2-1}}
%\end{figure}

Although all the models have similar internal structures as a whole, 
%BV frequency profiles as a whole, 
%Despite the overall similarities in the BV frequency and hydrogen abundance profiles, 
we notice a trend that a more massive model has a steplike BV frequency profile 
(see, e.g., red curves in middle panels of Figure \ref{fig:2-1}) %gradually changes 
whereas a less massive model has that shaped like a slide 
(see, e.g., blue curves in middle panels of Figure \ref{fig:2-1}). 
%%We would like to stress that %there is one significant difference among them that 
%In other words, 
More precisely, 
the transition of the BV frequency ($N$) between 
an inner region where $N$ is large (large-$N$ region hereafter) %a larger-BV-frequency region %is large %(in the inner region) 
and an outer region where $N$ is small (small-$N$ region hereafter) %a smaller-BV-frequency region 
is sharper for more massive models, %(see the middle panels of Figure \ref{fig:2-1}), 
which affects how strongly g modes are trapped in the large-$N$ region. 
%(where $N$ is large). 
%%The trend implies that 
%Importantly, the degree of the transition and width of the transition region essentially determine 
%whether or not it is reasonable %for us 
%%it is difficult to parameterize the BV frequency profile of less massive stars 
%%by the two-zone modeling 
%%to derive the semi-analytical expression of the $\Delta P_{g}$ pattern 
%%as has been conducted by \citet{Cunha2019}. 
%Importantly, the trend is closely related to ... %can be interpreted as a consequence of 
%the fact that the degree of a sharp variation %the glitch 
%in the BV frequency %differs from model to model; 
%it 
%increases as the mass of the model increases. 
We will explain the point in more detail in the following a few paragraphs. 
%in the internal structures of the stellar models (with different masses), 
%%there is one remarkable feature which is different from model to model; 
%as the mass of the model increases, the degree of the glitch increases. 

Since an important point is how g modes sense the BV frequency transition, 
we firstly would like to quantify %the degree of 
the degree of the BV frequency transition sensed by g modes 
based on %the glitch based on %take a look at 
%which can be confirmed by taking a look at 
the ratio of a typical local wavelength of gravity waves: 
%which shows the ratio of a typical wavelength of high-order g modes: 
\begin{equation}
\lambda_{g} = 2\pi / k_{r} \sim 2\pi \frac{\omega r}{\sqrt{l(l+1)} N} \label{eq02}
\end{equation}
to the scale height of the BV frequency: 
\begin{equation}
H_{N} \equiv \biggl | \frac{\mathrm{d}r}{\mathrm{d} \, \mathrm{ln} \, N} \biggr |.  \label{eq03}
\end{equation}
Note that the approximation $k_{r} \sim \sqrt{l(l+1)} N / \omega r$ is used 
where a typical angular frequency $\omega$ is determined based on high-order g-mode frequencies of each model 
(more details can be found in Section \ref{sec:2-3}), and that 
we only consider dipole modes ($l=1$) here. % from now on. 
%We are also 
Then, %based on the ratio $\lambda_{g} / H_{N}$, we can determine where the glitch is located; 
where the BV frequency transition is sharp (for typical g modes) %glitch 
can be considered 
as a region 
with the value of the ratio $\lambda_{g} / H_{N}$ significantly larger than unity. 
(In other words, the asymptotic limit of g modes corresponds to 
when the ratio approaches zero.) 
%i.e., the asymptotic limit of g modes corresponds to when the ratio $\lambda_{g} / H_{N}$ approaches zero. 
%In other words, the glitch can be interpreted as a region 
%with the value of the ratio $\lambda_{g} / H_{N}$ significantly larger than the unity 
%(see the bottom panels in Figure \ref{fig:2-1}). 
%Note that the approximation $k_{r} \sim \sqrt{l(l+1)} N / \omega r$ is here used 
%where a typical angular frequency $\omega$ is determined based on high-order g-mode frequencies of each model 
%(more details can be found in Section \ref{sec:2-3}). 
%%It should be instructive to mention that the local maximum of the ratio $\lambda_{g} / H_{N}$, % is , 
%%or where the degree of the glitch is largest, corresponds to 
%%the outer edge of the gradient of the BV frequency 
%%(see middle and bottom panels of Figure \ref{fig:2-1}). 
%where the gradient of the BV frequency is fairly steep. 
%as the ratio gets larger, ..., and vice versa. 

%%Broadly speaking, the ratio is close to zero in the deep radiative region of each model 
%%except for regions where chemical composition gradients are developed. 
%
%
The bottom panels of Figure \ref{fig:2-1} show the ratio $\lambda_{g} / H_{N}$ 
around the deep radiative region of the stellar models. 
%Among the models , 
We see that the ratio $\lambda_{g} / H_{N}$ is largest in the cases of the $6 \, M_{\odot}$ models (red curves), 
meaning that the BV frequency transition of the models (sensed by g modes) 
is sharper than that in the other less massive models. 
%glitches the degree of which are stronger than those in the other models. %where 
It should be also noted that, in the $6 \, M_{\odot}$ models, 
the ratio is almost zero everywhere except for in the transition region; %in a narrow region with the glitch, 
g modes behave asymptotically in both the large-$N$ and small-$N$ regions 
that are separated by the sharp transition region. 
This is the reason why we can reasonably approximate the BV frequency of the $6 \, M_{\odot}$ models 
with the two-zone modeling as was done by \citet{Cunha2019}. %Cunha+2019 
%(more details can be found in Section \ref{sec:3}). 

This is nevertheless not the case for, e.g., the $1.6 \, M_{\odot}$ models, 
where the local maxima of the ratio $\lambda_{g} / H_{N}$ are much smaller than those of the $6 \, M_{\odot}$ models. 
Moreover, the width of the sharp transition region (where $\lambda_{g} / H_{N} > 1$) 
in the $1.6 \, M_{\odot}$ models are so broad that 
it overlaps almost the whole large-$N$ region; 
%reflects the fact that the BV frequency profile of a less massive model 
%is shaped like a ramp, 
the large-$N$ and small-$N$ regions are not well separated by the transition region. % of the BV frequency, 
Therefore, it does not seem to be appropriate 
to approximate the BV frequency of the $1.6 \, M_{\odot}$ models 
with the two-zone modeling. 
These properties strongly motivate us to take the gradient in the BV frequency into account 
in derivation of a new semi-analytical expression of the $\Delta P_{g}$ %g-mode period spacing 
pattern. 
%MIGHT BE BETTER TO MENTION THAT 
%WHERE THE RATIO IS LARGE CORRESPONDS TO 
%WHERE THE SLOPE OF BV FREQUENCY IS LOCATED. 
%%derive a new semi-analytical expression of the g-mode period spacing pattern 
%%taking the gradient in the BV frequency into account. 
%In contrast to the case of 6M... where ..., 
%... in the case of 1.6M model... 
%In this paper, we will call ... the asymptoticity ... 
%Therefore, the asymptoticity is high for more massive models. 
%We can see that property by looking at ... (bottom panels), 
%which is the ratio between the scale height ... and the typical wavelength of g modes ... . 
%%Asymptoticity ... 

\begin{figure}[t]
\begin{center}
\includegraphics[scale=0.50]{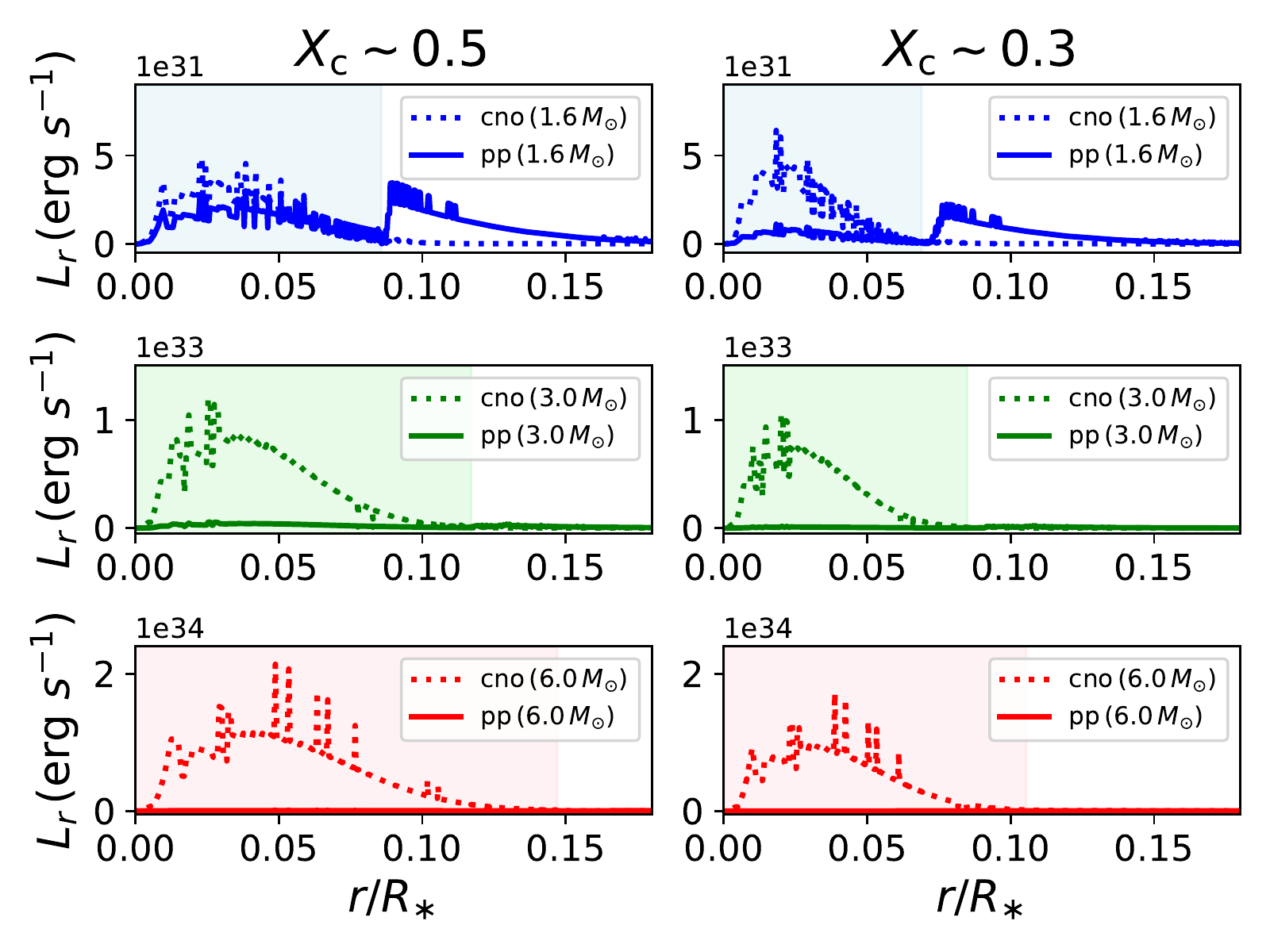}
\caption{\footnotesize Nuclear energy generation rate $L_{r}$ brought about by CNO cycle (dotted) 
or pp-chain reaction (solid) for various models with different masses 
and evolutionary stages (the same colors and columns, respectively, as in Figure \ref{fig:2-1}). 
The horizontal axis shows the fractional radius, and the convective cores are indicated by shaded areas. 
}
\label{fig:2-2}
\end{center} 
\end{figure}

A possible origin for the mass-dependent sharpness of the BV frequency transition is 
the balance between two kinds of hydrogen burning mechanisms, namely, 
CNO cycle and pp-chain reaction \citep[e.g.][]{Kippenhahn_text}.
Figure \ref{fig:2-2} shows the nuclear energy generation rate, %per second, 
which is proportional to the amount of conversion from hydrogen to helium per second, 
brought about by CNO cycle (dotted) 
and pp-chain reaction (solid) for the stellar models. 
We clearly see that pp-chain reaction becomes more and more dominant over CNO cycle 
as the mass of the models decreases. 
Interestingly, in the case of $1.6 \, M_{\odot}$ models, pp-chain cycle is at work 
%%not only in the convective core (shaded area) but also 
across the edge of the convective-core boundary 
(see around the outer edge of the shaded area in the top panels of Figure \ref{fig:2-2}), %in the radiative region 
%just above the convective core, 
which smoothens the chemical composition gradient developed 
just above the convective core %around the edge of the convective boundary 
and renders the BV frequency transition milder 
as seen in the bottom panels of Figure \ref{fig:2-1}.  

%\begin{figure}[t!]
%\plotone{Figure_2_2.pdf}
%\caption{  \label{fig:2-2}}
%\end{figure}

As a final remark in this section, we point out that 
the sharpness of the BV frequency transition depends on the stellar evolutionary stage as well 
(compare left and right at the bottom panels of Figure \ref{fig:2-1}), 
although it is not so evident as the mass dependence. 
Such age dependence is confirmed even for the $6\, M_{\odot}$ models, and thus, 
it should not be solely caused by the balance between the CNO cycle and pp-chain reaction. 
One possibility is that elemental diffusion processes have been smoothening 
the chemical composition gradient inside older stars for longer times, 
leading to a larger scale height of the BV frequency $H_{N}$. 
Combined with the assumption that the typical wavelength of a high-order g mode ($\lambda_{g}$) 
with a certain radial order $n$ is not so changed as the star evolves, 
the smaller $\lambda_{g} / H_{N}$ for the older models 
can be explained by their larger $H_{N}$. 
%because elemental diffusion processes have been smoothening 
%the chemical composition gradient inside older stars for longer times, 
%leading to smaller values of the ratio $\lambda_{g} / H_{N}$. 
%This can be explained by ... more dense deep radiative region ... 
%increasing typical g-mode periods ..., decreasing the ratio. 
%It is therefore, roughly speaking, caused by rather global effect... 
%and thus, we are not going to think about the age-dependence in more detail in this paper. 
%Though 
We will find the age dependence as well later in Section \ref{sec:4}, 
%the mass-dependence has more impacts on the degree of the glitch, 
%and we will not discuss the age-dependence in more detail 
%We will thus concentrate on the mass-dependence more 
%in the rest of the paper. 

\subsection{$\Delta P_{g}$ patterns of the models} \label{sec:2-3}
In this section, we present $\Delta P_{g}$ %g-mode period spacing 
patterns which are 
numerically computed based on the stellar models discussed in the previous section. 
The linear adiabatic oscillation code GYRE \citep{Townsend2013} is used, 
and eigenperiods of high-order ($-40 < n < -13$) and dipole ($l=1$) g modes, 
which are theoretically predicted to be excited 
%correspond to typical g-mode periods observed 
in main-sequence g-mode pulsators 
\citep{Moravveji2016_solo}, 
are calculated for each model. 

\begin{figure}[t]
\begin{center}
\includegraphics[scale=0.52]{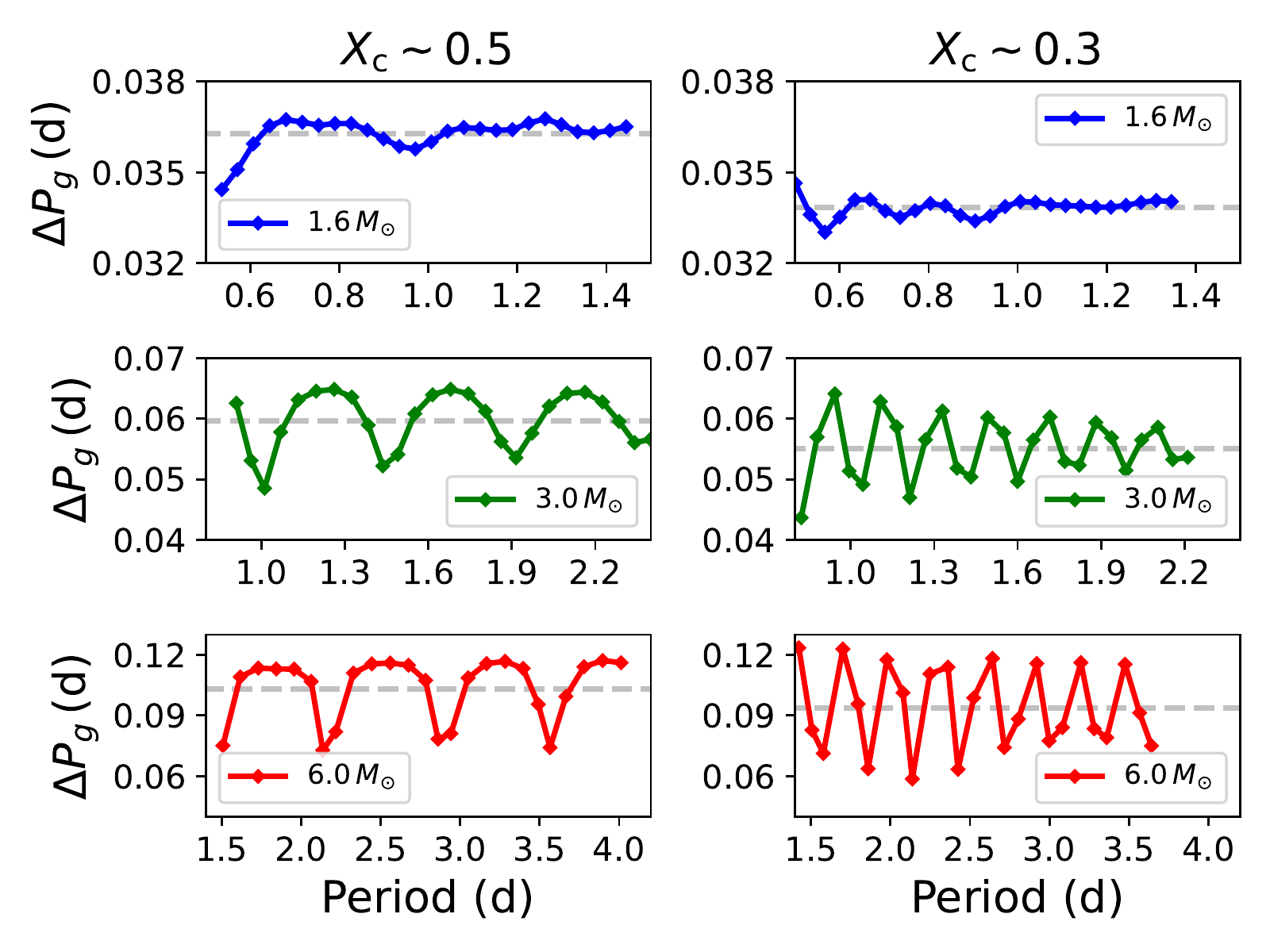}
\caption{\footnotesize G-mode period spacings $\Delta P_{g}$ against the g-mode periods %patterns 
that are numerically computed 
based on various models with different masses and evolutionary stages 
(the same colors and columns, respectively, as in Figure \ref{fig:2-1}). 
%The horizontal axis shows the g-mode period. 
Note that the scales of the vertical axes are different 
depending on the mass of the models; 
the amplitude of an oscillatory component in a $\Delta P_{g}$ pattern is 
larger for massive models.  
}
\label{fig:2-3}
\end{center} 
\end{figure}

Figure \ref{fig:2-3} shows the $\Delta P_{g}$ patterns of the models thus computed. 
In all the cases, we see oscillatory behaviors in the $\Delta P_{g}$ patterns. 
%as have been observationally confirmed and theoretically expected. 
It is also seen that %, for a certain evolutionary sequence, %models with the same mass, 
the mean value of the $\Delta P_{g}$ pattern (dashed grey lines in Figure \ref{fig:2-3}) becomes smaller 
as the model becomes older (compare the left and right panels). 
%\begin{figure}[t!]
%\plotone{Figure_2_3.pdf}
%\caption{  \label{fig:2-3}}
%\end{figure}

One significant difference is %found in the g-mode period spacing patterns is 
the amplitude of the oscillatory component in the $\Delta P_{g}$ pattern; 
the amplitude becomes larger as the model becomes more massive 
(note that the scales of the vertical axes are different in Figure \ref{fig:2-3}). 
This corresponds to the fact that, as discussed in the previous section, 
a more massive model has a sharper BV frequency transition 
(see bottom row in Figure \ref{fig:2-1}), %(for typical g modes), 
resulting in a larger amplitude in the $\Delta P_{g}$ pattern, 
which is consistent with the theoretical prediction \citep[e.g.][]{Miglio2008}%(e.g. Miglio+2008). 

Another important point is that, 
especially in the cases of $1.6 \ M_{\odot}$ and $3 \, M_{\odot}$ models, 
we find that the amplitude becomes smaller as the g-mode period becomes longer. 
This is mainly because 
%the glitches in the models have finite widths 
%(see the bottom panels in Figure \ref{fig:2-1} for the ratio $\lambda_{g} / H_{N}$) 
%and 
the ratio $\lambda_{g} / H_{N}$, %(see the bottom panels in Figure \ref{fig:2-1}), 
which represents the degree of the BV frequency transition sensed by typical g modes, 
%or more specifically, the ratio $\lambda_{g} / H_{N}$ (see the bottom panels in Figure \ref{fig:2-1}), 
is proportional to the g-mode angular frequency $\omega$ 
(see the numerator of Equation \ref{eq02}). 
The ratio $\lambda_{g} / H_{N}$ is therefore smaller for a g mode with a longer period, 
leading to a smaller amplitude of the oscillatory component in the $\Delta P_{g}$ pattern, 
which finally becomes zero in the asymptotic limit. 
%%This can be also explained based on the glitch structures of these models 
%%(see the bottom panels in Figure \ref{fig:2-1} for the ratio $\lambda_{g} / H_{N}$). 
%%Since the ratio $\lambda_{g} / H_{N}$ is proportional to the g-mode angular frequency $\omega$, 
%%the ratio is smaller for a g mode with a longer period, 
%This is also related to the reason why the g-mode period spacing patterns of $6 \, M_{\odot}$ models 
%do not clearly exhibit such a period dependence of the amplitude; 
%the glitches in the $6 \, M_{\odot}$ models are sharp enough that . 
Such a period dependence of the amplitude of the oscillatory $\Delta P_{g}$ pattern 
has been pointed out by numerous authors 
from the theoretical point of view \citep[e.g.][]{Miglio2008, Cunha2019, Wu2020} 
%based on a perturbative approach, 
%but 
although there so far has not been an explicit expression 
that reproduces the period dependence, 
which is another motivation for us to derive a new semi-analytical expression 
of the $\Delta P_{g}$ pattern as we see in the next section. 
%and the explicit expression %has not been given yet 
%which 
%which will be derived in the next section. 
%
%Taking the gradient in the BV frequency is thus important 
%when we ... g-mode period spacing patterns of lower-mass stars... 

\section{Semi-analytical Expression of $\Delta P_{g}$ Pattern} \label{sec:3}
As we see in Section \ref{sec:2}, %the previous section, 
the BV frequency transition (between the large-$N$ and small-$N$ regions) 
becomes milder 
for less massive stars, 
which leads to the period dependence of the amplitude of an oscillatory component in the $\Delta P_{g}$ pattern. 
Such a milder BV frequency transition can also prevent us 
from parameterizing the BV frequency of less massive stars 
with the two-zone modeling as has been conducted by \citet{Cunha2019}. 
In this section, as an alternative way to model the BV frequency profile of less massive stars, 
% first step to ... effects of less? glitch, % as much as we can?, 
we propose to parameterize the BV frequency 
with a ramp function 
%taking the gradient in the BV frequency into account, 
based on which a semi-analytical expression of the $\Delta P_{g}$ pattern is derived. 
%
%the slope in $N$ is a feature common to ..., causing g-mode period spacings with 
%their amplitudes decreasing with the g-mode periods. 
%In this section, the slope in $N$ is taken into account to formulate semi-analytical expression of g-mode period spacing. 
%... is derived taking the slope in $N$ into account. 
We firstly review \citet{Cunha2019} to highlight a basic concept 
in our formulation (Section \ref{sec:3-1}). 
Then, a new semi-analytical expression is given (Section \ref{sec:3-2}). % present the ... in Section ....
Finally, the derived expression will be validated (Section \ref{sec:3-3}). 
%%Note that 

\subsection{Revisiting \citet{Cunha2019}} \label{sec:3-1}
%First of all, we would like to show the fundamental 
%
%a fundamental assumption in the analysis of high-order g modes is 
%the so-called Cowling approximation (ref...) in which we neglect the gravitational potential perturbation. 
%%%Let us start with the second-order differential equation for high-order g modes 
%is the second-order differential equation 
%%%(e.g., ref...) which often plays a central role in the analysis of high-order g modes: 
%%%..., 
%%%where ... . 
%%%In the case that ... no glitch..., the asymptotic approximation allows us to solve the above equation (ref...), 
%Based on the assumptions that ... and that ..., 
%the above equation can be analytically solved (ref...), 
%%%resulting in the constant period spacing (eq...).  
%%%However, in the presence of glitch in the BV frequency, 
%%%it is not straightforward for us to analytically solve the above equation, 
%%%and we need some special treatments as we will see in the subsequent paragraphs. 

%As mentioned in the previous sections, 
%The steep chemical composition gradient and resultant glitch in the BV frequency are 
%common features in the deep radiative region of the main-sequence g-mode pulsators. 
%It is thus of great importance to take effects of the glitch on the g-mode period spacings into account 
%when we analyze high-order g modes based on the asymptotic approximation. 
There are two mainstreams in analytical studies of high-order g modes 
under the existence of the sharp variation in the BV frequency. 
In one approach, the sharp variation is considered as a perturbation to the BV frequency, 
and the $\Delta P_{g}$ pattern is expressed as a result of the perturbed g-mode periods 
\citep[the perturbative approach;][]{Montgomery2003, Miglio2008,Wu2020}. 
In the other approach, the eigenvalue condition is derived 
by linking a few JWKB solutions, 
enabling us to compute the explicit expression of the $\Delta P_{g}$ pattern 
\citep[e.g.][]{Miglio2008, Cunha2015, Cunha2019}. 
%``direct'' approach (in which eigenfunctions are calculated 
%One is based on the perturbative approach (e.g. Montgomery+2003, Miglio+2008), 
%and the other is based on JWKB approximation 
%based on several JWKB solutions, e.g. Miglio+2008, ..., Cunha+2015, 2019). 
%Note that the ``direct'' approach 
Let us call the latter approach as the ``direct'' approach. 
Generally speaking, the perturbative approach is applicable for any variation with arbitrary shapes 
as long as the variation is small enough that it can be considered as a small perturbation. 
In contrast, the ``direct'' approach can only be used for a variation with a rather simple structure, 
but we do not have to assume that it is small. 
One of the reasons why \citet{Cunha2019} utilized the ``direct'' approach may be 
that we often find a fairly sharp BV frequency transition inside stellar models of main-sequence g-mode pulsators; 
it is not reasonable to consider the sharp variation in the BV frequency to be a small perturbation. 
In the same spirit as \citet{Cunha2019}, we would like to focus on the ``direct'' approach, %JWKB approximation, 
and we will briefly present the formulation by \citet{Cunha2019} below. 
\begin{figure}[t]
\begin{center}
\includegraphics[scale=0.5]{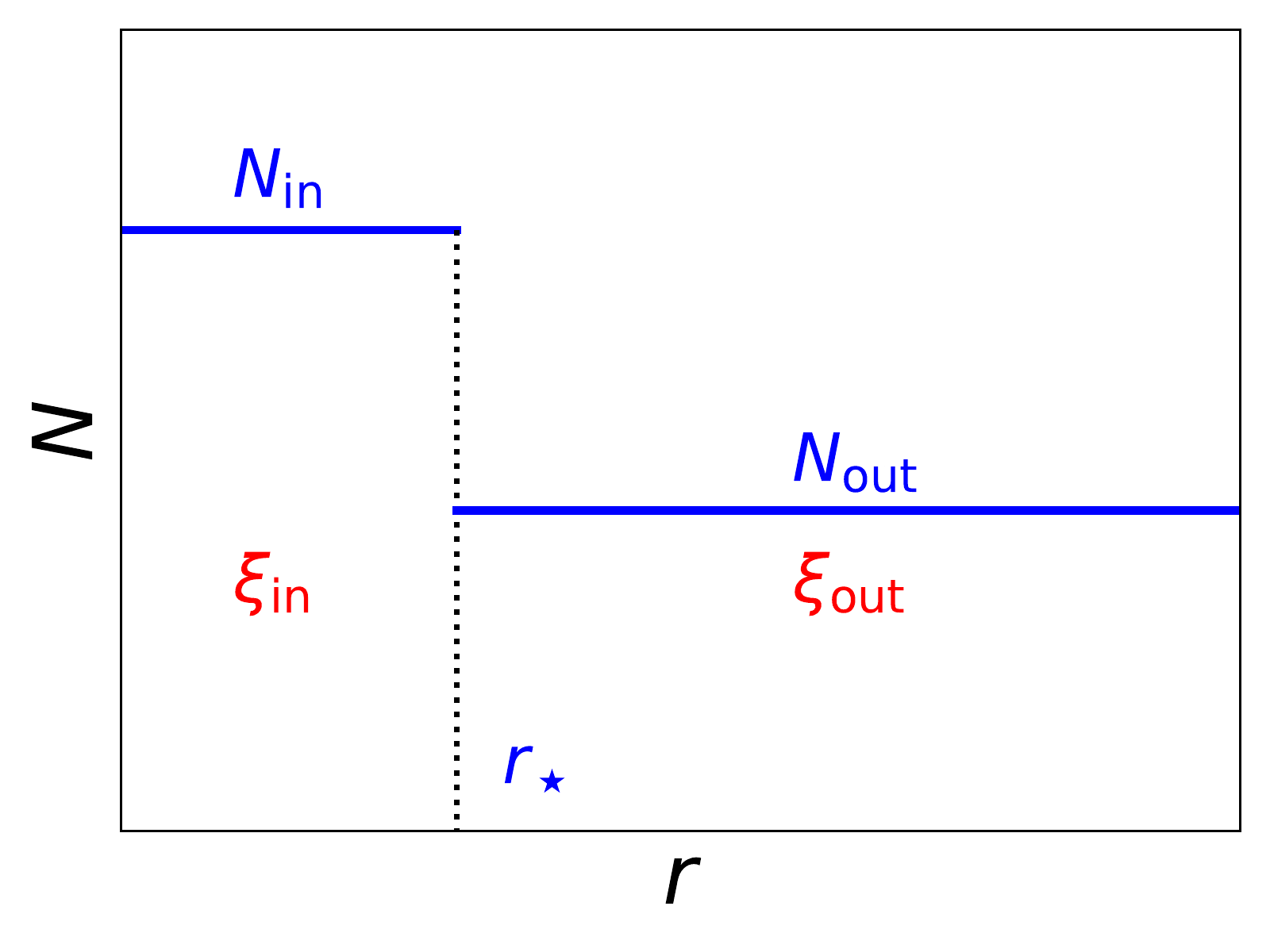}
\caption{\footnotesize Schematic view of the two-zone modeling of 
the BV frequency $N$ in the g-mode cavity, where %radial wavenumber $K_{r} = \sqrt{l(l+1)} N / \omega r$. 
the horizontal axis is the radius $r$. %buoyancy radius at the radius $r$ (see the text for the definition). 
The BV frequency profile %radial wavenumber 
is modeled with the three parameters, namely, 
the value of the BV frequency in the inner region $N_{\mathrm{in}}$, 
that in the outer region $N_{\mathrm{out}}$, and 
the location of the discontinuity $r_{\star}$. 
The eigenfunctions for the inner and outer regions are denoted by $\xi_{\mathrm{in}}$ 
and $\xi_{\mathrm{out}}$, respectively. 
}
\label{fig:3-1-1}
\end{center} 
\end{figure}
The starting point is the two-zone modeling of the BV frequency 
whose inner and outer edges are here assumed to be identical to those of the mode cavity 
of high-order g modes. 
Let us denote the inner and outer edges by $r_{\mathrm{0}}$ and $r_{\mathrm{1}}$, respectively. 
The g-mode cavity is divided 
into two regions by a discontinuity in the BV frequency 
located at $r=r_{\star}$ (Figure \ref{fig:3-1-1}). % ... NOT COMPATIBLE!! MODIFY FIGURES 4 AND 5!!). 

What we would like to solve is the second-order differential equation for the high-order g modes \citep[see, e.g.,][]{Gough1993}: 
\begin{equation}
\frac{\mathrm{d}^2 \xi}{\mathrm{d} r^{2}} + k_{r}^{2} \xi = 0,  \label{Eq_high_ord_g_xi}
\end{equation}
where the eigenfunctions are represented by $\xi$. 
It is assumed that the local wavenumber $k_{r} \sim L N / \omega r$ 
in which $L^{2} = l(l+1)$ following the notation of \citet{Cunha2019}. 
Then, eigenfunctions for the inner and outer regions may be given 
based on the assumption of the Cowling approximation and JWKB approximation \citep[see, e.g.,][]{Gough1993}. 
%(in other words, the wavelengths are much shorter than the scale heights of the backgrounds except for the discontinuity). 
Below are the explicit forms of the eigenfunctions: 
\begin{equation}
\xi_{\mathrm{in}} = \tilde{\xi}_{\mathrm{in}} \frac{1}{\sqrt{k_{\mathrm{in}}}} \mathrm{sin} \biggl (  \int_{r_{0}}^{r} k_{\mathrm{in}} \mathrm{d}r + \frac{\pi}{4}  \biggr ),  \label{Eq_high_ord_g_xi_in}
\end{equation}
and 
\begin{equation}
%\xi_{\mathrm{out}} = \widetilde{\xi_{\mathrm{out}}} \frac{1}{\sqrt{k_{\mathrm{out}}}} \mathrm{sin} \biggl (  \int_{r}^{r_{1}} k_{\mathrm{out}} \mathrm{d}r + \frac{\pi}{4}  \biggr ),  \label{Eq_high_ord_g_xi_out}
\xi_{\mathrm{out}} = \tilde{\xi}_{\mathrm{out}} \frac{1}{\sqrt{k_{\mathrm{out}}}} \mathrm{sin} \biggl (  \int_{r}^{r_{1}} k_{\mathrm{out}} \mathrm{d}r + \frac{\pi}{4}  \biggr ),  \label{Eq_high_ord_g_xi_out}
\end{equation}
where $k_{\mathrm{in}}$ and $k_{\mathrm{out}}$ are the local wavenumbers 
in the inner and outer regions, respectively. 
%defined as (\ref{Eq_kr_N}) for the inner region and that for the outer region. 
There are constants related to the amplitudes of the eigenfunctions, 
$\tilde{\xi}_{\mathrm{in}}$ and $\tilde{\xi}_{\mathrm{out}}$. 

The next thing to do is to link the inner and outer eigenfunctions 
%and the outer one are linked 
continuously and smoothly. 
The conditions can be expressed as the following boundary conditions at the discontinuity $r=r_{\star}$: 
\begin{equation}
\xi_{\mathrm{in}} (r_{\star}) = \xi_{\mathrm{out}} (r_{\star}),  \label{Eq_in_to_out}
\end{equation}
and 
\begin{equation}
\left.\frac{\mathrm{d} \xi_{\mathrm{in}}}{\mathrm{d}r}\right|_{r=r_{\star}} = \left.\frac{\mathrm{d} \xi_{\mathrm{out}}}{\mathrm{d}r}\right|_{r=r_{\star}},   \label{Eq_in_to_out_deriv}
\end{equation}
with which the constants $\tilde{\xi}_{\mathrm{in}}$ and $\tilde{\xi}_{\mathrm{out}}$ can be eliminated. 
After some manipulation (with the assumption that 
the first derivative of the local wavenumber can be neglected), 
we end up with the eigenvalue condition in the presence of a discontinuity 
in the BV frequency \citep[Equation A6 in][]{Cunha2019}: 
\begin{eqnarray}
\lefteqn{\mathrm{sin} \biggl( \int_{r_{0}}^{r_{1}} k_{r} \mathrm{d}r + \frac{\pi}{2}  \biggr ) = } \nonumber \\
 & - & A_{\star} \mathrm{sin} \biggl( \int_{r_{\star}}^{r_{1}} k_{\mathrm{out}} \mathrm{d}r + \frac{\pi}{4}  \biggr ) 
 \mathrm{cos} \biggl( \int_{r_{0}}^{r_{\star}} k_{\mathrm{in}} \mathrm{d}r + \frac{\pi}{4}  \biggr ), \label{Eq_eig_cond_Cunha}
\end{eqnarray} 
where the strength of the discontinuity is represented by $A_{\star}$ defined as below: 
\begin{equation}
A_{\star} \equiv \frac{k_{\mathrm{in}}^{\star}-k_{\mathrm{out}}^{\star}}{k_{\mathrm{out}}^{\star}}, \label{Eq_rtlv_dif}
\end{equation}
in which the local wavelength at the location of the discontinuity for the inner (outer) region is 
denoted by $k^{\star}_{\mathrm{in}}$ ($k^{\star}_{\mathrm{out}}$). 

%%See Appendix of Cunha+2019 for more information on, e.g., explicit forms of the eigenvalue condition. 
%\begin{equation}
%\frac{K_{\mathrm{in}}^{\star}}{K_{\mathrm{out}}^{\star}} \mathrm{sin} \biggl (  \int_{r_{\star}}^{r_{1}} K_{\mathrm{out}} dr +\frac{\pi}{4} \biggr ) 
%\mathrm{cos} \biggl (  \int_{r_{0}}^{r_{\star}} K_{\mathrm{in}} dr +\frac{\pi}{4} \biggr ) 
%+ \mathrm{sin} \biggl ( \int_{r_{0}}^{r_{\star}} K_{\mathrm{in}} dr +\frac{\pi}{4} \biggr ) 
%\mathrm{cos} \biggl (  \int_{r_{\star}}^{r_{1}} K_{\mathrm{out}} dr +\frac{\pi}{4} \biggr ),    \label{Eq_eigenvalue_cond_before1}
%\end{equation}
%where $K_{\mathrm{in}}^{\star}=K_{\mathrm{in}}(r_{\star})$ and $K_{\mathrm{out}}^{\star}=K_{\mathrm{out}}(r_{\star})$. 

With some special techniques developed by \citet{JCD2012} 
\citep[see also][]{Cunha2015}, 
the eigenvalue condition can be further analyzed, %we can further analyze the eigenvalue condition 
finally leading to the semi-analytical expression of the $\Delta P_{g}$ pattern 
in the case of the two-zone modeling of the BV frequency \citep{Cunha2019}: 
\begin{equation}
\frac{\Delta P_{\mathrm{g}}}{\Delta P_{\mathrm{as}}} \sim \biggl [  1-\biggl ( \frac{\Pi_{0}}{\Pi_{\star}} \biggr ) 
\frac{-A_{\star} \mathrm{sin} \tilde{\beta_{1}} + A_{\star}^2 \mathrm{cos}^2 \tilde{\beta_{2}}}{(1+A_{\star} \mathrm{cos}^2 \tilde{\beta_{2}})^2+(0.5 A_{\star} \mathrm{cos} \tilde{\beta_{1}})^2}   \biggr ]^{-1}, \label{Eq_dP_Cunha}
\end{equation}
where the buoyancy radius $\Pi_{r}^{-1}$ is defined as: 
\begin{equation}
\Pi_{r}^{-1} =  \int_{r_{0}}^{r} \frac{N}{r} \, \mathrm{d}r,  \label{Eq_buoy_rad}
\end{equation}
with which $\Pi_{0}^{-1} = \Pi_{r}^{-1}(r_{1})$, $\Pi_{\star}^{-1} = \Pi_{r}^{-1}(r_{\star})$, 
and $\Delta P_{\mathrm{as}}=2\pi^2 \Pi_{0}/L$. 
The phases $\tilde{\beta_{1}}$ and $\tilde{\beta_{2}}$ inside the sinusoidal components in the equation above are defined as
\begin{equation}
\tilde{\beta_{1}} = \frac{2L}{\omega} \Pi_{\star}^{-1} + 2 \delta,  \label{Eq_beta1}
\end{equation}
and 
\begin{equation}
\tilde{\beta_{2}} = \frac{L}{\omega} \Pi_{\star}^{-1} + \frac{\pi}{4} +  \delta,  \label{Eq_beta2}
\end{equation}
respectively. 
The extra phase term represented by $\delta$ is necessary to quantify the phase jump produced around the 
boundaries of the g-mode cavity. 
See \citet{Cunha2015,Cunha2019} for more discussions. 

It should be instructive to mention that Equation (\ref{Eq_dP_Cunha}) contains 
two essential properties expected for the $\Delta P_{g}$ pattern, namely, 
that the amplitude and period of an oscillatory component in the $\Delta P_{g}$ pattern 
are determined by the strength ($A_{\star}$) 
and location ($\Pi_{\star}^{-1}$) of the discontinuity, respectively. 

\begin{figure}[t]
\begin{center}
\includegraphics[scale=0.5]{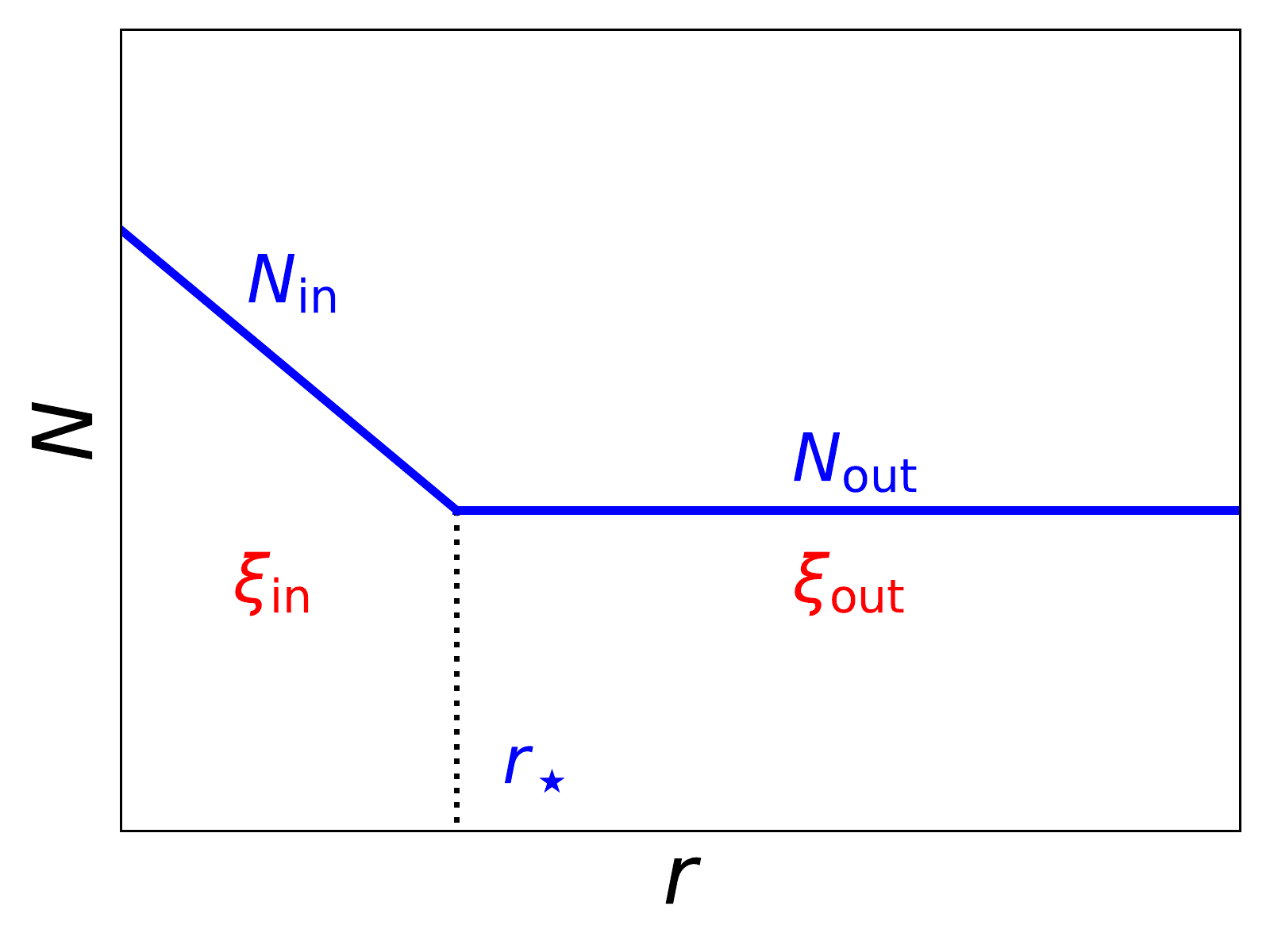}
\caption{\footnotesize Schematic view of 
the BV frequency $N$ %radial wavenumber $K_{r}$ 
whose inner region is modeled with a ramp. 
The variables have the same meaning as in Figure \ref{fig:3-1-1}. 
}
\label{fig:3-2-1}
\end{center} 
\end{figure}

%\subsection{Semi-analytical expression of g-mode period spacing taking ... into account} \label{sec:3-2}
\subsection{A new formulation} \label{sec:3-2}
In this section, a semi-analytical expression of the $\Delta P_{g}$ pattern, 
for which the gradient in the BV frequency is taken into account, 
will be derived based on the ``direct'' approach (Section \ref{sec:3-1}). % JWKB approximation. % as described in the previous section. 

Our starting point is to model %the two-zone modeling of 
the BV frequency 
%that has 
with a ramp (Figure \ref{fig:3-2-1}). 
In this model, the outer edge of the ramp is defined as the location $r_{\star}$ 
that divides the inner and outer regions as we see in Figure \ref{fig:3-2-1}; 
we are thus focusing on a discontinuity in not the BV frequency but the first derivative of the BV frequency. 
In both regions, the corresponding eigenfunctions are assumed to be 
given by the JWKB solutions (Equations \ref{Eq_high_ord_g_xi_in} and \ref{Eq_high_ord_g_xi_out}). 
This assumption seems to be too much of simplification, 
but it can be validated by remembering discussions in Chapter 16 of \citet{Unno_text} %+1989 
where a criterion for whether asymptotic analysis of high-order modes can be used or not has been indicated. 
We will discuss this point in the next two paragraphs. 

One important quantity is $\zeta$ which is related to the buoyancy frequency and is defined as below
\citep[see also Equation 16.22 in][]{Unno_text}: 
\begin{equation}
\zeta = \mathrm{sgn}(k_{r}^2) \biggl ( \biggl | \frac{3}{2} \int_{r_{0}}^{r} |k_{r}| \mathrm{d}r  \biggr |  \biggr)^{2/3}, \label{Eq_zeta}
\end{equation}
where we denote the sign function by $\mathrm{sgn}(\cdot)$. 
The eigenfunctions of high-order g modes, which are solutions for the second-order differential equation 
%of stellar non-radial oscillation 
formulated under the Cowling approximation \citep[see, e.g.,][]{Unno_text}, 
can be expressed by a linear combination of Airy functions when 
the following condition is satisfied: 
\begin{equation}
\zeta \gg f\biggl ( \frac{\mathrm{d}r}{\mathrm{d}\zeta} \biggr ), \label{Eq_to_be_Airy}
\end{equation}
with the function $f$ defined as \citep[see also Equation 16.13 in][]{Unno_text}
\begin{equation}
f(x) \equiv |x|^{1/2} \frac{\mathrm{d}^{2} |x|^{-1/2}}{\mathrm{d} \zeta^{2}}. \label{Eq_f}
\end{equation} 
Note that eigenfunctions represented by the Airy function are asymptotically identical to %the same as 
the JWKB solutions for regions far from the turning point of the Airy function; 
we can thus use JWKB solutions as eigenfunctions for regions 
far from the edges of the g-mode cavity where $\zeta \gg f ( \mathrm{d}r / \mathrm{d}\zeta)$. 

\begin{figure}[t!]
%\plotone{Figure_3_2_2_new2.pdf}
\includegraphics[scale=0.54]{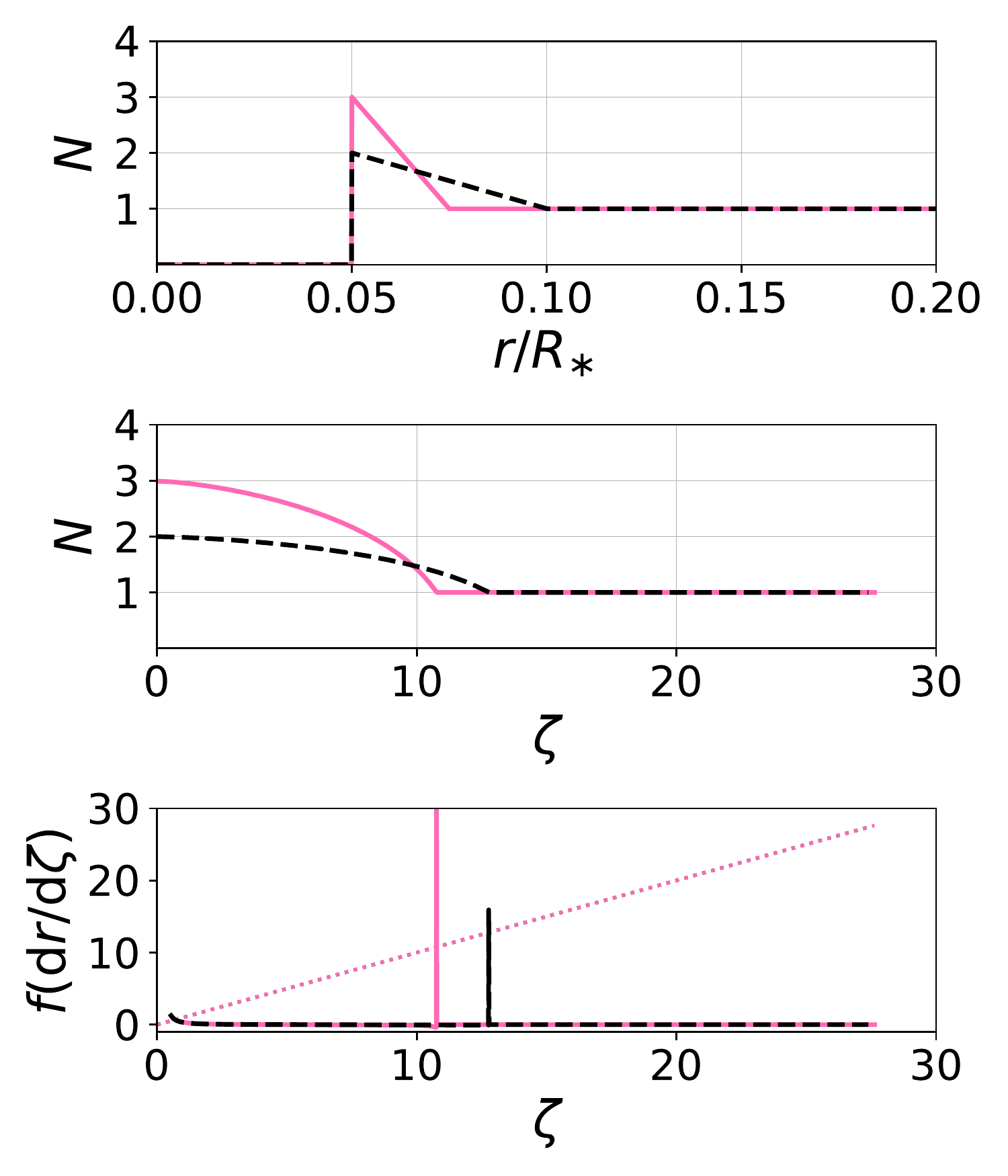}
\caption{\footnotesize Two examples of BV frequency model with a ramp against the fractional radius (top), 
the same BV frequency profiles against $\zeta$ (defined by Equation \ref{Eq_zeta}) (middle), 
and comparison between the corresponding $f(\mathrm{d} r / \mathrm{d} \zeta)$ (solid and dashed curves) 
%$f(\frac{\mathrm{d} r}{\mathrm{d} \zeta})$ (solid and dashed curves) 
and $\zeta$ (dotted lines) (bottom). 
See Equation (\ref{Eq_f}) for the definition of the function $f(\cdot)$. 
%It is seen that, for both cases, the condition $\zeta \gg f(\mathrm{d} r / \mathrm{d} \zeta)$ %(Equation \ref{Eq_to_be_Airy})
%is mostly satisfied in almost the whole regions except for around the outer edges of the ramps 
%(at which the first derivative of the BV frequency is discontinuous). 
Note that the BV frequencies are expressed in arbitrary units. 
 \label{fig:3-2-2}}
\end{figure}

Keeping in mind the discussion above, 
we show comparisons of $\zeta$ and $f (\mathrm{d}r / \mathrm{d}\zeta)$ 
for two examples of the simple BV frequency model with a ramp (the bottom panel of Figure \ref{fig:3-2-2}). 
It is seen that, in both examples, the condition (\ref{Eq_to_be_Airy}) is mostly satisfied 
except for around the location of the discontinuity in the first derivative of the BV frequency 
($r \sim r_{\star}$), 
which guarantees that we can safely use the JWKB solutions 
as the eigenfunctions for the inner and outer regions 
in the case of the BV frequency model with a ramp. 
%It should be noticed that the condition is not satisfied near the inner edge of the g-mode cavity. 
%But 
%as will be discussed later in Section \ref{sec:5}. 
%%Show the schematic view of the assumed $N$ profile. 
%%Formulated basically in the same way except for the assumption 
%%that the wavelength is continuous at the ... in contrast to the assumption of Cunha et al. 2019 
%%that ... .
%\begin{figure}[t!]
%\plotone{Figure_3_2_1}
%\caption{  \label{fig:3-2-1}}
%\end{figure}

%Then, ... two WKBJ solutions. 
%Why is it okay? See $\zeta$. FIGURE!

Once we determine the eigenfunctions for the inner and outer regions, 
we have to then consider the boundary conditions (\ref{Eq_in_to_out}) and (\ref{Eq_in_to_out_deriv}) 
to have the eigenvalue condition. % to be analyzed. 
It should be stressed that, since our BV frequency model is continuous elsewhere, %has a finite ramp, 
%we have an assumption different from that taken in Cunha+2019, 
%namely, 
%we assume that 
the local wavenumbers are the same at the boundary 
($k_{\mathrm{in}}^{\star} = k_{\mathrm{out}}^{\star}$ at $r = r_{\star}$), 
and instead, the first derivative of the local wavenumber is discontinuous there. 
This is different from the two-zone model of \citet{Cunha2019} %(see the previous section). 
where the local wavenumber itself is discontinuous at the boundary. 
%In other words, 

We then have the eigenvalue condition in the case of the BV frequency model with a ramp: 
\begin{eqnarray}
\lefteqn{\mathrm{sin} \biggl( \int_{r_{0}}^{r_{1}} k_{r} \mathrm{d}r + \frac{\pi}{2}  \biggr ) = } \nonumber \\
 & - & A_{\star}' k_{\star}^{-1} \mathrm{sin} \biggl( \int_{r_{\star}}^{r_{1}} k_{\mathrm{out}} \mathrm{d}r + \frac{\pi}{4}  \biggr ) 
 \mathrm{sin} \biggl( \int_{r_{0}}^{r_{\star}} k_{\mathrm{in}} \mathrm{d}r + \frac{\pi}{4}  \biggr ), \nonumber \\ 
 \label{Eq_eig_cond_new}
\end{eqnarray} 
where the strength of the discontinuity (in the first derivative of the local wavenumber) 
$A_{\star}'$ is defined as below: 
\begin{equation}
%%A_{\star}' \equiv \frac{\mathrm{d} \, \mathrm{ln} \, k_{\mathrm{in}}^{-\frac{1}{2}}}{\mathrm{d} r} \biggr |_{\star -} 
%%- \frac{\mathrm{d} \, \mathrm{ln} \, k_{\mathrm{out}}^{-\frac{1}{2}}}{\mathrm{d} r} \biggr |_{\star +}  \label{Eq_rtlv_dif_prime}
A_{\star}' \equiv  k_{\star}^{\frac{1}{2}}\biggl (  \frac{\mathrm{d} k_{\mathrm{in}}^{-\frac{1}{2}}}{\mathrm{d} r} \biggr |_{\star -} 
- \frac{\mathrm{d} k_{\mathrm{out}}^{-\frac{1}{2}}}{\mathrm{d} r} \biggr |_{\star +} \biggr )  \label{Eq_rtlv_dif_prime}
\end{equation}
%We define the strength of the discontinuity (in the first derivative of the local wavenumber) $A_{\mu}^{\star}$ as below: 
%\begin{equation}
%\zeta \gg f\biggl ( \frac{dr}{d\zeta} \biggr ), \label{Eq_to_be_Airy}
%\end{equation}
in which the first (second) term represents the value of the left (right) derivative 
at $r = r_{\star}$. %the outer edge of the ramp ($r = r_{\star}$). 
It might be worth mentioning 
that $A_{\star}'$ is related to the difference in the reciprocal of the local-wavenumber scale height %of the local wavenumber 
and that $k_{\star}^{-1}$ is proportional to the local wavelength ($\lambda_{g}$) 
at $r = r_{\star}$. 
%the term $A_{\star}' k_{\star}^{-1}$ is the 
%ratio of the local wavelength to the scale height of the local wavenumber 
%around the boundary ($r \sim r_{\star}$). 
Thus, the term $A_{\star}' k_{\star}^{-1}$ is similar to the ratio $\lambda_{g} / H_{N}$ 
that is used as an indicator of the degree of the BV frequency transition sensed by g modes 
in Section \ref{sec:2-2}. 
%in terms that 

One remarkable point about the eigenvalue condition (\ref{Eq_eig_cond_new}) 
is that the term $A_{\star}' k_{\star} ^{-1}$ %, which represents the degree of the glitch 
%as mentioned in the last paragraph, 
is inversely proportional to g-mode periods $P$. 
Therefore, for a fixed strength of the discontinuity in the first derivative of the BV frequency ($A_{\star}'$), %become 
the degree of the BV frequency transition sensed by g modes ($A_{\star}' k_{\star} ^{-1}$) 
becomes smaller as the g-mode period is longer. 
This feature is consistent with what we have seen in the $\Delta P_{g}$ patterns 
of the low-mass ($1.3 - 3 \, M_{\odot}$) main-sequence models 
in Section \ref{sec:2-3}. 
%For comparison, we also present the ... (MAYBE IT'S BETTER TO PUT THE EIGENVALUE CONDITION IN CUNHA+2019 IN SEC. 3.1 ... 
%WHICH MAKES IT EASY TO COMPARE... I GUESS) 
%the degree of the jump is dependent on the g-mode period. 
%This is fairly reasonable since a g mode has high (low) aymptoticity 
%if its period is long (short)... . 
A similar discussion on the period dependence of the amplitude of $\Delta P_{g}$ patterns 
can be found in Appendix 1 in \citet{Cunha2019} %Cunha+2019 
where the variation in the BV frequency is assumed to be shaped like a Gaussian distribution. 

By using the techniques of \citet{JCD2012} 
in the same way as in \citet{Cunha2019}, %(see Section \ref{sec:3-1}), 
we can derive a semi-analytical expression of the $\Delta P_{g}$ pattern 
that takes into account the gradient in the BV frequency.  
The explicit expression is actually the same as Cunha et al.'s original one (\ref{Eq_dP_Cunha})
except that the constant $A_{\star}$ is replaced with the period-dependent term $A_{\star}' k_{\star}^{-1}$. 

%%%Before closing this section, 
%%%we would like to clarify what the term ``buoyancy glitch'' would mean in this study. 

%because it might seem to be ambiguous 
%%%Though it often designates the sharp variation in the BV frequency (Section \ref{sec:intro}), 
%%%... 
%%%To avoid any confusion, we are going to refrain from using the term in this paper. 

%which is slightly different from that in Cunha et al.'s BV frequency model. 
%that should be different from the 

\subsection{Validity check of the derived expression} \label{sec:3-3}
In this section, validity check of the semi-analytical expression derived in the last section is conducted. 
To this end, using some simple artificial BV frequency profiles, 
we have compared $\Delta P_{g}$ patterns analytically computed (with the semi-analytical expression) 
and those numerically computed. 
%To validate the expression derived in the last section, 
%we compare g-mode period spacing patterns; 
%ones are numerically computed based on a simple BV frequency 
%by directly solving the second-order differential equation 
%and the other of which are analytically computed using the parameters 
%exploited from the simple $N$. 

We start with somewhat an ordinary case (Section \ref{sec:3-3-1}) in a sense that 
the strength of the discontinuity in the first derivative of the artificial BV frequency ($A_{\star}'$) 
%the degree of the transition assumed for the artificial BV frequency profile 
is comparable to that of the stellar models computed in Section \ref{sec:2-2}. %the of the in %, and then, 
Then, we move on to the case where 
the variation in the BV frequency can be treated as a small perturbation (Section \ref{sec:3-3-2}), 
where g-mode period spacings computed based on the first order perturbation theory 
is compared as well. 
Note that, in both cases, the gradient in the BV frequency is set to be constant 
with respect not to the fractional radius but to the buoyancy radius. 
This enables us to use the analytical expression of the $\Delta P_{g}$ pattern 
derived based on the perturbative approach, 
which is especially relevant in Section \ref{sec:3-3-2}. 
Note also that we are considering only dipole modes ($l=1$) in this section. 
%It should be noted that ... SHOULD MENTION THAT WE ARE CONSIDERING ONLY DIPOLE MODES IN THIS SECTION! 

\subsubsection{When the variation in the BV frequency cannot be treated as a small perturbation} \label{sec:3-3-1}
\begin{figure}[t!]
\includegraphics[scale=0.50]{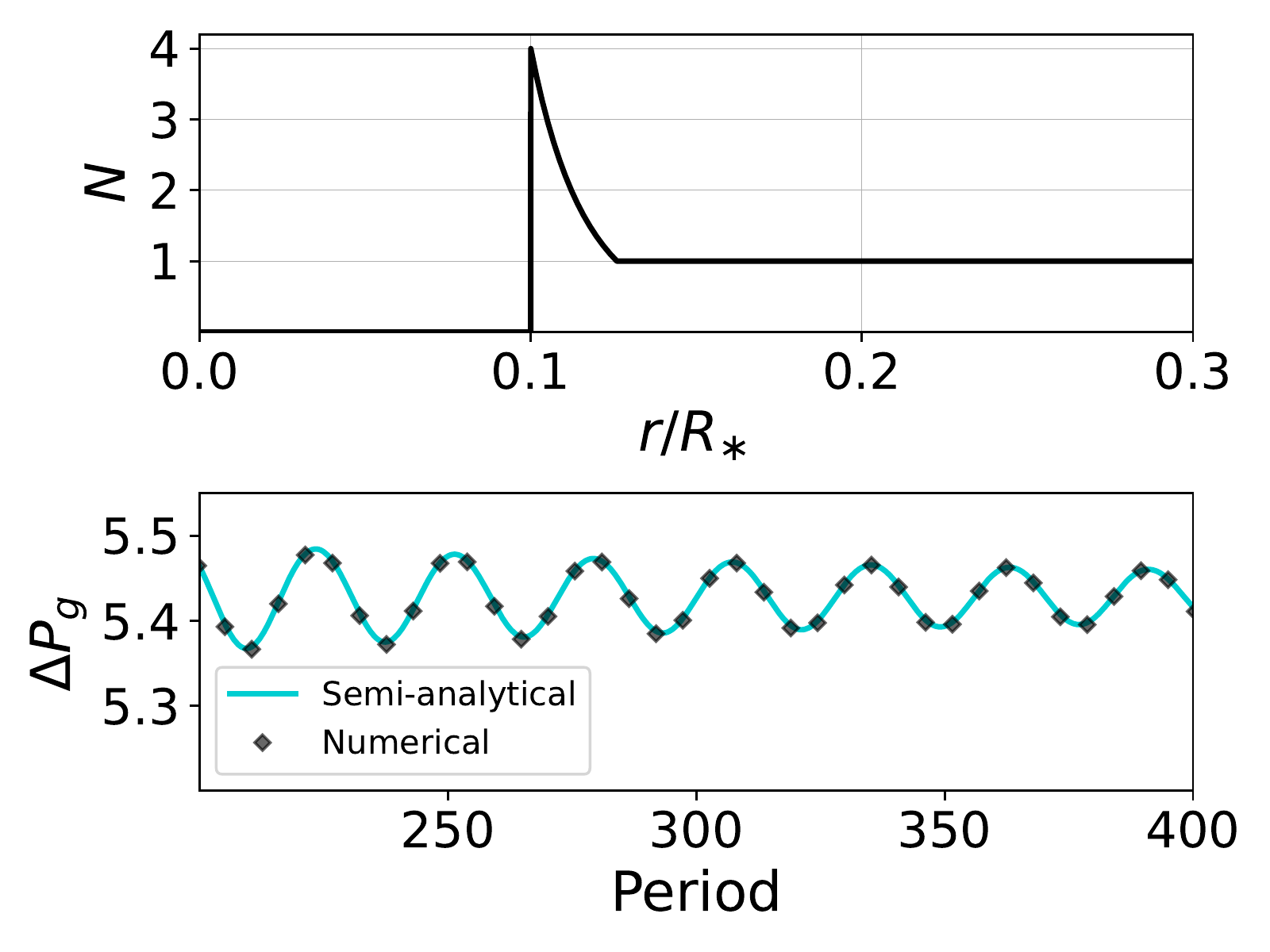}
%\plotone{Figure_3_3_1.pdf}
\caption{\footnotesize Simple artificial model of the BV frequency with a ramp (top), 
based on which the ``numerical'' and ``semi-analytical'' $\Delta P_{g}$ patterns 
have been computed (dark grey diamond and light blue curve, respectively, in the bottom panel). 
%The gradient in the BV frequency is constant 
%with respect to not the fractional radius but the buoyancy radius (see Appendix for more discussions). 
The BV frequency is expressed in arbitrary units, and so are the g-mode periods 
and $\Delta P_{g}$ patterns. 
Note that the gradient in the BV frequency is constant 
with respect to the buoyancy radius. 
\label{fig:3-3-1}}
\end{figure}

The top panel of Figure \ref{fig:3-3-1} shows the BV frequency model 
based on which $\Delta P_{g}$ patterns are to be semi-analytically/numerically computed in this subsection. 
The ratio between the value of the BV frequency at the innermost point ($r/R_{\ast} \sim 0.1$) 
and that of the outer region ($r/R_{\ast} > 0.125$) is determined 
so that it is close to typical values in the case of ordinary stellar models 
(see Figure \ref{fig:2-1}). %the BV frequencies of the stellar models calculated in Section \ref{sec:2-2}). 
%Note that the gradient in the BV frequency is constant 
%with respect to not the fractional radius but the buoyancy radius. 

The ``numerical'' $\Delta P_{g}$ pattern is calculated by 
numerically solving the second-order differential equation (\ref{Eq_high_ord_g_xi}) 
with simple boundary conditions, namely, $\xi = 0$ at the center 
and $\mathrm{d}\xi / \mathrm{d}r = 0 $ at the surface. 
%For simplicity
Then subtraction has been taken between the eigenperiods with the neighboring radial orders ($n+1$ and $n$) 
to obtain 
%The thus obtained eigenperiods $P_{n}$ are then subtracted to have 
the g-mode period spacings $P_{n+1} - P_{n}$. 
%(we have considered only $l=1$ modes.) 
As for the ``semi-analytical'' $\Delta P_{g}$ pattern, 
the parameters $\Pi_{0}$, $\Pi_{\star}$, $A'_{\star}$, and $k_{\star}^{-1}$, %in the semi-analytical expression 
%are directly 
which are directly extracted from the prepared BV frequency profile, 
are substituted for the explicit expression (\ref{Eq_dP_Cunha}) 
(please keep in mind that the constant $A_{\star}$ should be 
replaced with the period-dependent term $A'_{\star} k_{\star}^{-1}$). 
%Note that 
The numerically computed eigenfrequencies $\omega$ are used 
in the phase terms (\ref{Eq_beta1}) and (\ref{Eq_beta2}). 
Note that $\delta$ in these phase terms are treated as a free parameter %to be fitted 
whose value is determined so that 
the phase of the oscillatory component in the 
``semi-analytical'' $\Delta P_{g}$ pattern 
matches that in the ``numerical'' $\Delta P_{g}$ pattern. 
%and then, 

In the bottom panel of Figure \ref{fig:3-3-1}, the ``numerical'' and ``semi-analytical'' $\Delta P_{g}$ patterns 
thus obtained are compared, 
which clearly shows a fairly good agreement between them. 
We especially would like to emphasize that the ``semi-analytical'' $\Delta P_{g}$ patterns 
can successfully reproduce the period dependence of the amplitude of the oscillatory component in the 
``numerical'' $\Delta P_{g}$ patterns; %where 
the amplitude becomes smaller as the g-mode period becomes longer. 

\subsubsection{When the variation in the BV frequency can be treated as a small perturbation} \label{sec:3-3-2}
%In this subsection, we focus on the case where 
%the glitch in the BV frequency is m
Although it is generally considered that 
the strength of the discontinuity in the first derivative of the BV frequency ($A_{\star}'$) 
%the BV frequency transition 
in ordinary main-sequence g-mode pulsators 
is not so small that we cannot treat it as a small perturbation (see discussions in Section \ref{sec:2-2}), 
the perturbative approach \citep[e.g.][]{Miglio2008} 
is still quite useful for validating the derived semi-analytical expression since 
%As we have already see in Section \ref{sec:2}, 
%the glitch in the BV frequency of ordinary main-sequence g-mode pulsators is not a small perturbation..., 
another analytical expression of the $\Delta P_{g}$ pattern can relatively readily obtained 
based on the perturbative approach (see Appendix \ref{ap:sec:a}). 
%PLOT! 

\begin{figure}[t!]
\includegraphics[scale=0.50]{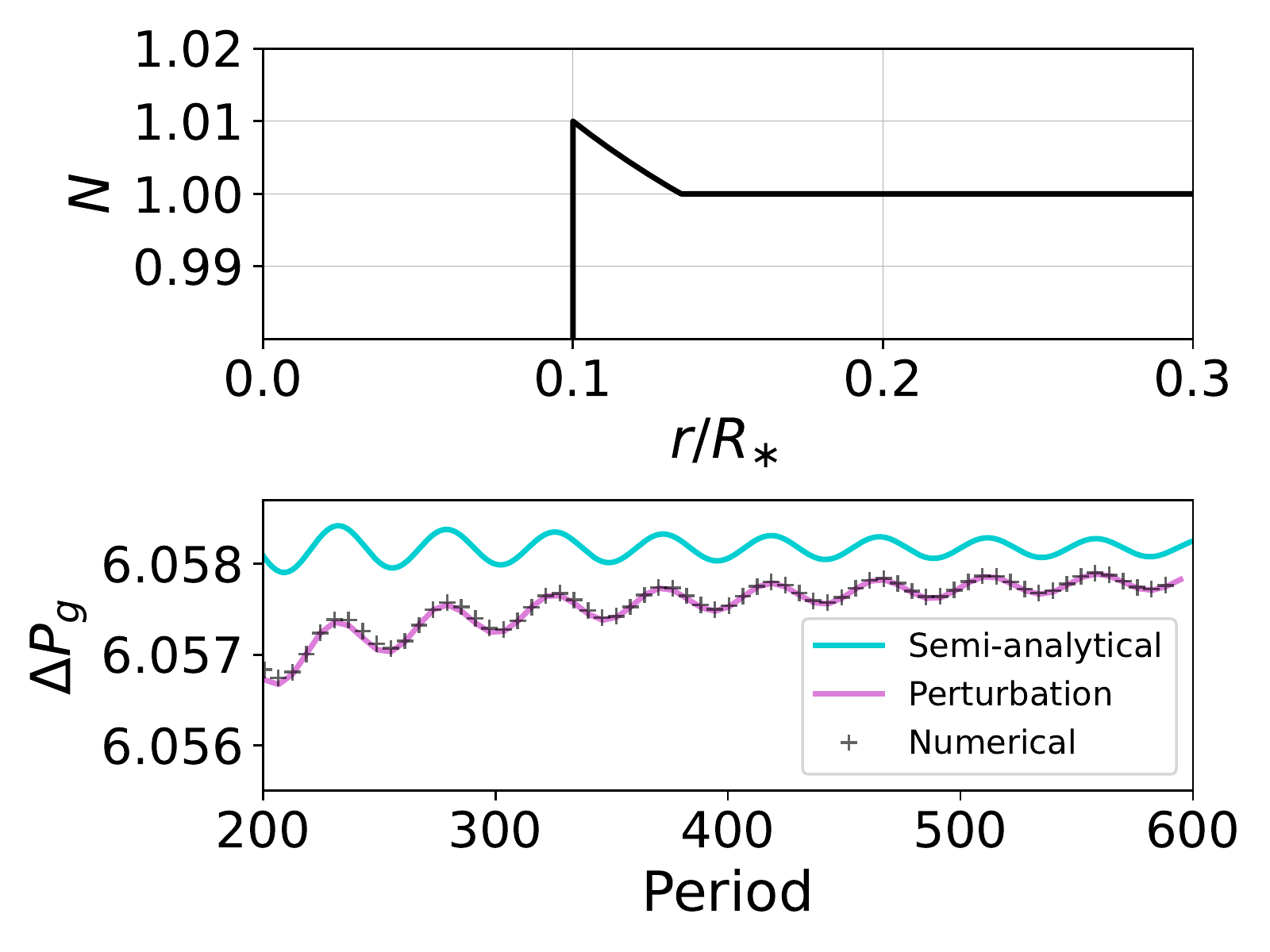}
%\plotone{Figure_3_3_2.pdf}
\caption{\footnotesize Same as Figure \ref{fig:3-3-1} 
except that the variation in the BV is modeled as a small perturbation. 
The $\Delta P_{g}$ pattern calculated based on the perturbative approach (pink curve in the bottom panel) 
are shown in addition to the ``numerical'' (black crosses) and 
``semi-analytical'' (light blue curve) $\Delta P_{g}$ patterns. 
\label{fig:3-3-2}}
\end{figure}

Figure \ref{fig:3-3-2} shows a BV frequency model with a tiny ramp (top) 
and ``numerical'', ``semi-analytical'', and ``perturbative'' $\Delta P_{g}$ patterns 
(black crosses, light blue curve, and pink curve, respectively, in the bottom panel). 
We have followed the same steps as described in the last subsection 
to compute the ``numerical'' and ``semi-analytical'' $\Delta P_{g}$ patterns. 
The ``perturbative'' $\Delta P_{g}$ pattern is calculated based on Equation (\ref{Eq_delP2}), 
in which the g-mode periods numerically computed based on the corresponding unperturbed BV frequency 
are used for $P_{n}$ in the expression.

We have again found a good agreement among the $\Delta P_{g}$ patterns thus computed, 
especially regarding the trend that the amplitude of the oscillatory component 
in the $\Delta P_{g}$ pattern decreases with the g-mode periods. 
(Notice that the scale is significantly different from that 
in the bottom panel of Figure \ref{fig:3-3-1}.) 
It should be mentioned that there exist some offsets 
between the ``semi-analytical'' and the other $\Delta P_{g}$ patterns. 
%the offsets are just 
This may arise from the fact that %is mainly because 
the local wavenumber $k_{r}$ is proportional to $r^{-1}$, 
%which may be caused because 
%that boundary effects are neglected in derivation of the semi-analytical expression. 
which renders $k_{r}$ %The proportionality is relevant 
to be significantly large around the central region ($r \sim 0$), % to %results in a phenomena somewhat similar to mode trapping, 
%which especially affects 
affecting relatively low-order (or shorter-period) modes and 
resulting in a phenomena somewhat similar to mode trapping. 
But such effect has not been taken into account in the derivation of the semi-analytical expression, 
which may cause the offsets between the ``semi-analytical'' and the other $\Delta P_{g}$ patterns. 
We nevertheless would like to emphasize that the offsets approach zero 
as we see the $\Delta P_{g}$ patterns in the higher g-mode period range 
than that shown in the bottom panel of Figure \ref{fig:3-3-2}. 
%%%It is also worth mentioning that, though the offsets do exist in the ordinary case (Section \ref{sec:3-3-1}), 
%%%they are too small to affect the $\Delta P_{g}$ patterns as a whole. %, 
%and thus, we will not discuss this point more in this paper. 

\section{Application} \label{sec:4}
%One of the goals in this study %in the derivation of the semi-analytical expression for the $\Delta P_{g}$ pattern 
%is %to infer glitch properties of main-sequence g-mode pulsators 
%with the semi-analytical expression of the $\Delta P_{g}$ pattern derived in Section \ref{sec:3}. 
In this section, we would like 
to assess how applicable the new semi-analytical expression of the $\Delta P_{g}$ pattern %, 
%which has been derived in Section \ref{sec:3}, 
is for studying the BV frequency profile %glitch properties 
of main-sequence g-mode pulsators. 
%%In this section, we would like to 
%fit the semi-analytical expression of the $\Delta P_{g}$ pattern, 
%which has been derived and validated in Section \ref{sec:3}, 
%to those numerically computed based on stellar models 
%in order to 
%%demonstrate that %check if 
%A primary goal in this section is to demonstrate that 
%%the semi-analytical expression of the $\Delta P_{g}$ pattern, 
%%which has been derived and validated 
%with the simple two-zone modeling of the BV frequency with a ramp (
%%in Section \ref{sec:3}, %), 
%%is in a way useful for inferring glitch properties of main-sequence g-mode pulsators. 
%can be applied to more practical cases. 
%To this end, 
%
We firstly attempt to fit the semi-analytical expression 
% $\Delta P_{g}$ patterns of realistic stellar models %computed in Section \ref{sec:2} 
to $\Delta P_{g}$ patterns of realistic stellar models 
%in order to see if the expression can reproduce the g-mode period spacing patterns of the stellar models 
%or not 
in order to investigate to what extent we can accurately %a parameter range in which we can accurately 
%check if the fitting helps us correctly 
extract the information on the BV frequency transition inside the stellar models 
using the semi-analytical expression 
%or not 
(Section \ref{sec:4-1}). 
%%%A parameter range in which we can accurately 
%%compare estimated parameters obtained via the fitting procedure 
%%with those directly extracted from the stellar models (Section \ref{sec:4-2}). 
%We then present a case study of the fitting procedure for one of the Kepler targets, KIC 11145123 (Section \ref{sec:4-2}). 
%Finally, a brief discussion about degeneracies among the parameters to be estimated %, 
%which we will neglect in Sections \ref{sec:4-1} and \ref{sec:4-2}, 
%be faced with in the fitting procedures in Sections \ref{sec:4-1} and \ref{sec:4-2}, 
Then, as an example to highlight the usefulness of the semi-analytical expression, 
a case study of the fitting procedure for one of the Kepler targets, KIC 11145123, 
is given (Section \ref{sec:4-2}). 

%reinterpret the definition of the parameters 
%in the semi-analytical expression (Section \ref{sec:4-1}) 

%so that we can use the expression to describe g-mode period spacing patterns of realistic stellar models 
%which have BV frequency profiles much more complex 
%compared with the simple two-zone models analyzed in Section \ref{sec:3}. 
%for the case of realistic stellar models (Section \ref{sec:4-1}). 
%Then, based on the semi-analytical expression with the reinterpretation of the parameters in mind, 
%we fit the g-mode period spacing patterns of the stellar models computed in Section \ref{sec:2} 
%in order to see if the expression is useful or not for extracting information on the BV frequencies of the stellar models 
%(Section \ref{sec:4-2}). 

%As was described in Section (\ref{sec:intr}), one of the goals in this study is 

\subsection{Attempts to fit the semi-analytical expression to \\ 
$\Delta P_{g}$ patterns of stellar models} \label{sec:4-1}
The goal of this section is to compare two sets of parameters 
that describe the BV frequency model with a ramp (see Section \ref{sec:3-2} 
for definitions of the parameters). %, namely, 
One is the set of parameters directly extracted from stellar models, 
and the other is that estimated by fitting the semi-analytical expression to 
the $\Delta P_{g}$ patterns numerically computed with the stellar models. 
After presenting stellar models used in this section and 
how to extract the parameters from the models (Subsection \ref{sec:4-1-1}), 
artificial $\Delta P_{g}$ patterns to be fitted are generated in Subsection \ref{sec:4-1-2}. 
A fitting procedure is given in Subsection \ref{sec:4-1-3}, 
based on which parameters are estimated and finally compared with those directly extracted from the stellar models 
(Subsection \ref{sec:4-1-4}). 

\subsubsection{Stellar models and how to extract parameters} \label{sec:4-1-1}
Based on the same settings as described in Section \ref{sec:2-1}, 
we have computed $1.6$, $2$, $3$, and $4 \, M_{\odot}$ models %solar-mass models 
with various evolutionary stages represented by the hydrogen mass content at the center 
$X_{\mathrm{c}} \sim 0.5$, $0.4$, $0.3$, and $0.2$. 

\begin{figure}[t!]
\includegraphics[scale=0.42]{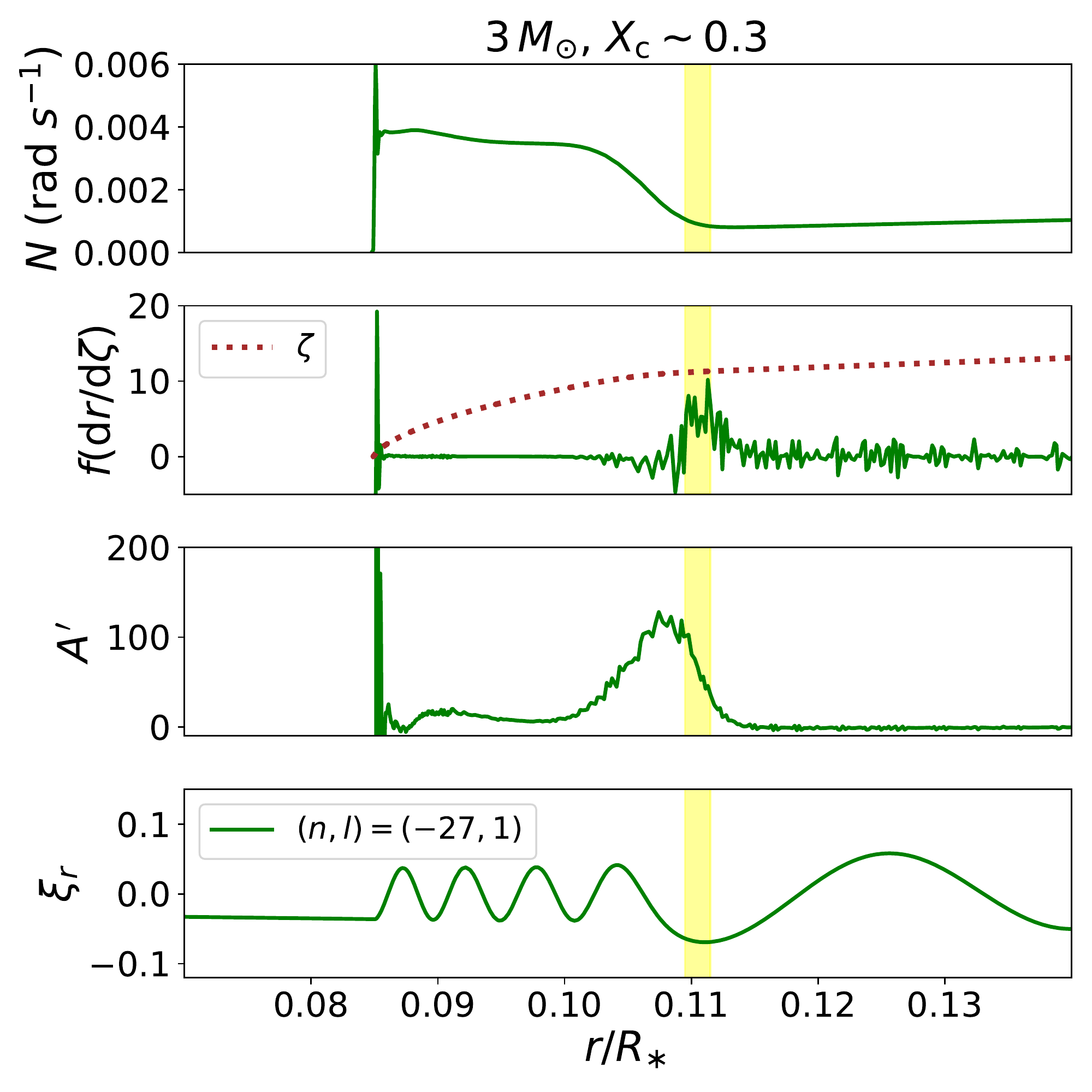}
%\plotone{Figure_4_1_1.pdf}
\caption{\footnotesize Internal properties around the deep radiative region of the $3 \, M_{\odot}$ model 
with the central hydrogen mass content $X_{\mathrm{c}} \sim 0.3$. 
From top to bottom, 
the BV frequency profile, %(top), 
the comparison of $f(\mathrm{d}r/\mathrm{d}\zeta)$ (green curve) and $\zeta$ (red dotted curve), 
the parameter $A'$, 
and the radial component of a g-mode eigenfunction with $(n,l) = (-27,1)$ 
are shown in this order. 
The horizontal axis is the fractional radius $r/R_{\ast}$, 
and the relevant range is common to all the panels. 
Yellow shaded areas indicate where the outer edge of the BV frequency ramp is located. 
Definitions of the parameters and how we decide the location of the outer edge of the ramp can be found in the text. 
\label{fig:4-2-1}}
\end{figure}

For later use in Subsection \ref{sec:4-1-4}, 
we then have to extract the parameters 
%(related to the BV frequency model with a ramp) 
from these models (see Section \ref{sec:3-2} for the definitions), 
which is, however, not a straightforward thing to do. 
This is because the location of the outer edge of the ramp in the BV frequency ($r_{\star}$) 
%In Section \ref{sec:4-1-4}, we will compare 
%One important thing is that it is not straightforward for us 
%to extract the parameters from the stellar models. 
%%For example, 
%%which is implicitly included in $\Pi_{\ast}$ and $B_{\ast}$, %means 
is not obvious at all; %difficult to uniquely determine; 
%For example, 
although the location $r_{\star}$ is defined to be where the first derivative of the BV frequency is discontinuous
in the case of the simple model of the BV frequency with a ramp, 
there is no such discontinuity %in the first derivative of the BV frequency 
in realistic stellar models. 
Then, where should the location $r_{\star}$ be? %in the cases of the stellar models? 
Similarly, how should we compute the strength of the discontinuity (in the first derivative 
of the local wavenumber) $A'_{\star}$? 
%which is implicitly included in $B_{\ast}$?
We would like to articulate these points in the following paragraphs. %%, 
The first parameter to be discussed is the location of the outer edge of the ramp $r_{\star}$. 
%%which was defined as the position where ... in Section \ref{sec:3}. 
%%However, it is not an easy task in the case of realistic stellar models 
%%as it is shown in Figure ....
A hint for where it should be in the case of a stellar model can be obtained 
by remembering the discussions in Section \ref{sec:3-2}. 
The important point is that $r_{\star}$ corresponds to 
where $f(\mathrm{d}r/ \mathrm{d} \zeta)$ is maximum (see Figure \ref{fig:3-2-2}). 
%Since $f(\mathrm{d}r/ \mathrm{d} \zeta)$ is essentially related to the second derivative of the 
%local wavenumber $k_{r}$ (see Equation \ref{Eq_f}), 
%we show in Figure \ref{fig:4-2-1}
As shown in %the middle panels of 
Figure \ref{fig:4-2-1}, %Since, 
in the case of realistic stellar models, % we have computed, 
$f(\mathrm{d}r/ \mathrm{d} \zeta)$ is closely related to $A'$ 
%%the second derivative of $\mathrm{ln} \, k_{r}^{-\frac{1}{2}}$ %$A'$, 
that is defined as
\begin{equation}
A' \equiv  \frac{\mathrm{d}\, \mathrm{ln} \, k_{r}^{-\frac{1}{2}}}{\mathrm{d}r}. \label{Eq_A_prime}
\end{equation}
%(see Equation ...) 
%(that is here defined as the first derivative of $\mathrm{ln} \, k_{r}^{-\frac{1}{2}}$) 
%the second derivative of $k_{r}$, 
%Based on that concept, we 
This is reasonable since $f(\mathrm{d}r/ \mathrm{d} \zeta)$ is essentially related to the second derivative of the 
local wavenumber $k_{r}$ (see Equation \ref{Eq_f}) while $A'$ contains the first derivative of $k_{r}$. 
Accordingly, when we focus on a region 
where the gradient of the BV frequency is steep 
(see around $0.10 < r/R_{\star} < 0.11$ in Figure \ref{fig:4-2-1}), 
we find that the position of the local maximum of $f(\mathrm{d}r/ \mathrm{d} \zeta)$ is almost identical to %correspond to 
that of the local minimum of the first derivative of $A'$ 
(see the yellow shaded areas in Figure \ref{fig:4-2-1}), 
We are thus going to define $r_{\star}$ %(inside stars) 
as the position where the first derivative of $A'$ %$\mathrm{ln} \, k_{r}^{-\frac{1}{2}}$ %$A'$ 
is locally minimum 
in the rest of this paper. 
(For example, in the case of Figure \ref{fig:4-2-1}, $r_{\star} \sim 0.111$.) 
%%The validity of the interpretation can be confirmed when we look at 
%%It should be noted that the inner part ($x \sim ...$) is not chosen as the position of the glitch 
%%because the wavelength of high-order g mode is much longer there ...(Figure...), 
%%meaning that ... . 
%%This is apparent when we see Figure ..., where 
%%the period of an oscillatory component in the period spacing patterns computed with $r_{\ast} = r_{\mathrm{in}}$ 
%%does not match the ones computed numerically. 
%%It is also worth mentioning that ... the position is close to the outer edge of $N_{\mathrm{chem}}$, 
%%where ... ((the same) Figure).

The next parameter to be discussed is the strength of the discontinuity in the first derivative 
of the local wavenumber ($A'_{\star}$). % (defined in Section \ref{sec:3-2}),  
%(THEY SHOULD HAVE A FINITE WIDTH!) 
%In particular, the interval .... is unclear; 
In Section \ref{sec:3-2}, $A'_{\star}$ is defined 
%which is essentially the same 
as the subtraction between left and right derivatives at $r = r_{\star}$ (see Equation \ref{Eq_rtlv_dif_prime}) 
because the first derivative of the local wavenumber $k_{r}$ is discontinuous there. 
This is in contrast not the case for a stellar model 
where $\mathrm{d}k_{r}/\mathrm{d}r$ is continuous 
in the deep radiative region %and the BV frequency transition region accordingly has a finite width 
(see discussions in Section \ref{sec:2-2}). 
We thus have to take subtraction for a certain interval; 
how should we determine the interval then? 
%The buoyancy radius of the glitch is defined as we explained in the last paragraph. 
%It is readily seen that a wider interval is favored to reproduce the numerically computed g-mode period spacings. 
%This can be explained by, again, ... that  
%%We would like to emphasize that 

The point is how high-order g modes sense the sharp variation in the first derivative of $k_{r}$, 
%Since where the 
which should be %qualitatively 
determined by the ratio between a width, with which the sharp variation in the first derivative of $k_{r}$ occurs, 
and a typical local wavelength of the high-order g modes. 
%Based on the assumption that the interval should be related to the wavelength of high-order g modes 
Figure \ref{fig:4-2-1} shows that the width %of the transition region 
(in which $f(\mathrm{dr}/\mathrm{d}\zeta) \sim \zeta$) 
is much smaller than the wavelength of a typical high-order g mode. 
%(Figure ...); 
It is thus expected that the high-order g modes 
should sense the sharp variation (in the first derivative of $k_{r}$) %with a finite width 
as a rather ``abrupt'' change in the structure. 
Together with the fact that $A_{\star}' \sim \bigl [ A' \bigr ]^{\star_{-}}_{\star_{+}}$ 
(see Equations \ref{Eq_rtlv_dif_prime} and \ref{Eq_A_prime}), %can be well approximated as 
%the subtraction between $A'(r_{\ast - })$ and $A'(r_{\ast + })$, 
we choose to define $A_{\star}'$ as the difference 
between the local maximum and local minimum of $A'$ 
that are found in a certain section around the outer edge of the BV frequency ramp ($r \sim r_{\star}$). %region around the glitch. 
The section is defined so that its center and width are 
%the center of it is 
identical to the location $r_{\star}$ and one local wavelength of the typical g mode, 
respectively. 
(For instance, in the case of Figure \ref{fig:4-2-1}, 
the local maximum and local minimum of $A'$ found in a section around $r \sim r_{\star}$ %the region where the sharp variation takes place 
are about $\sim 120$ and $\sim 0$, respectively, 
leading to $A_{\star}' \sim 120$.) 
%()
%%This allows us to explain ...
%%if the wavelength is ... (), the jump should be smaller (larger). 
%In this case, the width of the glitch is wider than one wavelength of the g mode, 
%and thus, ... . 
%In particular, since ..., we interpret the $A_{*}'$ as the difference between the maximum and minimum values of $A'$.
%%\begin{figure}[t!]
%%\includegraphics[scale=0.4]{Figure_4_1_2.pdf}
%\plotone{Figure_4_1_2.pdf}
%%\caption{  \label{fig:4-2-2}}
%%\end{figure}

All the parameters in the semi-analytical expression of the $\Delta P_{g}$ pattern 
can be computed based on the redefinitions demonstrated above. 
These ways of redefining the parameters, %to describe the BV frequency transition in the stellar models, 
which appear to be ad hoc, 
will be validated later in Subsection \ref{sec:4-1-4}. 
Note that the extra phase term $\delta$ is treated as a free parameter to be fitted in this study, 
and we are not going to discuss it in detail. 

\subsubsection{How to generate artificial $\Delta P_{g}$ patterns to be fitted} \label{sec:4-1-2}
For artificial datasets, %to be fitted, 
we have numerically computed g-mode periods and $\Delta P_{g}$ patterns 
with the stellar models that are calculated in the previous subsection. 
The frequency computation has been carried out 
based on the same settings as described in Section \ref{sec:2-2}. 
%%as datasets to be fitted. 
%Using the same models as are shown in Section \ref{sec:2}... .
%%The number of modes used in the fitting is ... . 
%%Only $l=1$. 
To mimic observational uncertainties, random numbers 
are added to the $\Delta P_{g}$ patterns, 
where they are assumed to follow the Gaussian distribution 
with the mean and standard deviation equal to $0$ and $10^{-4}$ (in units of day), respectively, 
and they are statistically independent of each other. 
The value of the standard deviation is determined 
based on typical observational uncertainties of $\Delta P_{g}$ patterns 
\citep[e.g.][]{Papics2014, VanReeth2015}. 

\subsubsection{Fitting procedure} \label{sec:4-1-3}
We conduct the fitting to the artificial datasets %has been conducted 
via the Metropolis method \citep{Metropolis1953}, 
which is one of the standard techniques to carry out %one of the techniques to carry out 
Markov chain Monte Carlo (MCMC) sampling \citep[see, e.g.,][]{Gregory2005}. 
%the setups, namely, 
%the parameters to be estimated and 
%the likelihood and priors used in the MCMC sampling, 
%in the following paragraphs. 
Parameters to be estimated are as follows (see also Sections \ref{sec:3-1} and \ref{sec:3-2}): 
the buoyancy radius at the outer edge of the g-mode cavity ($\Pi_{0}$), 
%that at the outer edge of the ramp $\Pi_{\star}$, 
that at $r=r_{\star}$ %the location of the discontinuity (in the first derivative of the BV frequency) 
($\Pi_{\star}$), 
the extra phase term ($\delta$), 
and the degree of the structural variation ($B_{\star}$) which is defined so that 
$A'_{\star} k_{\star}^{-1} = P^{-1} B_{\star}$. 
The final parameter has been introduced 
by assuming that the term $A'_{\star} k_{\star}^{-1}$ is inversely proportional to the g-mode period ($P$), 
which is correct when we neglect the period dependence of $A'_{\star}$ 
(we will discuss the point later in Subsection \ref{sec:4-1-4}). 
%since g-mode periods ($P$) themselves 
%are not unknowns but inputs to the semi-analytical expression. 
Introducing the parameter is also helpful since it prevents us from suffering from degeneracies 
among parameters such as $A'_{\star}$, $r_{\star}$, and $N_{\star}$. 
%%We will discuss this point later in Section \ref{sec:4-3}. %Based on the assumption of statistical independence, 
From now on, we will denote by $\boldsymbol{\theta}$ the set of the parameters to be estimated. 
%from now on. 

The likelihood of the parameter set $\boldsymbol{\theta}$ 
given a certain g-mode period spacing $\Delta P_{i}^{\mathrm{obs}}$ is here defined in the following way: 
\begin{eqnarray}
p (\Delta P_{i}^{\mathrm{obs}} | \boldsymbol{\theta} ) 
= \frac{1}{\sqrt{2 \pi} e_{i}} 
\mathrm{exp} 
\biggl [ - \frac{1}{2} 
\biggl ( \frac{\Delta P_{i}^{\mathrm{obs}} - \Delta P_{i}^{\mathrm{mod}}(\boldsymbol{\theta})}{e_{i}} \biggr )^2 
\biggr ] , \nonumber \\
\label{Eq_likelihood_i}
\end{eqnarray}
where $\Delta P_{i}^{\mathrm{mod}}(\boldsymbol{\theta})$ is computed by substituting 
the parameter set $\boldsymbol{\theta}$ and corresponding g-mode period $P_{i}^{\mathrm{obs}}$ 
for the derived semi-analytical expression. 
The observational uncertainty $e_{i}$ is assumed to be $10^{-4}$ (in units of day). 
%Based on the assumption that %of statistical independence among the artificial data 
The data $\Delta P_{i}^{\mathrm{obs}}$ is statistically independent of each other 
so that 
the likelihood of the parameter set $\boldsymbol{\theta}$ 
given a certain dataset $\Delta \boldsymbol{P}^{\mathrm{obs}}$ can be expressed as below: 
\begin{eqnarray}
p (\Delta \boldsymbol{P}^{\mathrm{obs}} | \boldsymbol{\theta} ) 
= \prod_{i=1}^{N_{\mathrm{obs}}} p ( \Delta P_{i}^{\mathrm{obs}} | \boldsymbol{\theta}  ). 
\label{Eq_likelihood}
\end{eqnarray}
The number of data used for fitting is denoted by $N_{\mathrm{obs}}$. 

%%$\Delta \boldsymbol{P}^{\mathrm{obs}}$ %is here defined in the following way: 
%%can be expressed as the product of the likelihood 
%%given a certain g-mode period spacing $\Delta P_{i}^{\mathrm{obs}}$: % in the following way: 
%$\boldsymbol{d}$, 
%which corresponds to the $\Delta P_{g}$ pattern of  can be expressed as: 
%g-mode period spacing pattern $\Delta \boldsymbol{P}^{\mathrm{obs}}$ is here given as below: 
%%where $N_{\mathrm{obs}}$ represents 
%%the number of data used for fitting. 
%%The $i$-th component of $\Delta \boldsymbol{P}^{\mathrm{obs}}$ 
%%is represented by $\Delta P_{i}^{\mathrm{obs}}$ 
%%with which we define the likelihood as below: 

With the likelihood (\ref{Eq_likelihood}) as well as prior probabilities of the parameters 
which are assumed to be uninformative to be uniform, 
we have carried out the Metropolis method to sample the posterior probability distribution of the parameters. 
A typical number of iterations for the sampling processes to converge 
and that of the so-called burn-in period 
(during which the obtained samples are not thought to be realizations from the posterior probability distribution) 
are about $5 \times 10^{5}$ and $10^{5}$, respectively. 
Convergence of the sampling has been confirmed by visual inspection. %burn-in period...

\subsubsection{Parameter estimation and comparison} \label{sec:4-1-4}
%Based on these interpretations, we determines the parameters ... which are to be compared with the MAP estimates. 
For each of the artificial datasets which can be labeled by the mass ($M$) and central hydrogen mass content ($X_{\mathrm{c}}$) 
of the corresponding stellar model, %produced with a certain stellar model, % with the mass and central hydrogen mass content, 
we have carried out MCMC fitting based on the procedure described in the previous subsection. 
%to estimate the parameters 
%By taking 
The means of the resultant posterior probability distribution functions 
are chosen as estimates of the parameters, namely, 
$\Pi_{0}$, $\Pi_{\star}$, and $B_{\star}$, for each stellar model. 
\begin{table} [t]
 \begin{center}
  \caption{\footnotesize Relative differences in the parameters, 
  $\Pi_{0}$ (top table), $\Pi_{\star}$ (middle table), and $B_{\star}$ (bottom table), 
  between the estimates obtained via MCMC fitting 
  and those directly extracted from the stellar models. 
  Each element in the tables represents a relative difference in the corresponding parameter 
  obtained in the case of a certain stellar model 
  which is designated by the mass (row) and central hydrogen mass content (column). 
  The relative difference is defined as equation (\ref{Eq_est_vs_prm}), and it is 
  expressed in units of $\%$. 
  %The digit is determined by the estimated uncertainties...  
  \label{tab:1} }
  %\footnotesize 
  $\delta \, \mathrm{ln} \, \Pi_{0}$ ($\%$) 
    \begin{tabular} {c|cccc} \hline\hline
      & $X_{\mathrm{c}} \sim 0.5$ &  $X_{\mathrm{c}} \sim 0.4$ &  $X_{\mathrm{c}} \sim 0.3$ &  $X_{\mathrm{c}} \sim 0.2$  \\ \hline
 %  $\frac{\Pi_{0}^{\mathrm{est}} - \Pi_{0}^{\mathrm{mod}}}{\Pi_{0}^{\mathrm{mod}}}$ & -0.01215  &  -0.01295  &  -0.01663  & -0.02281 \\ 
 %  $\frac{\Pi_{\ast}^{\mathrm{est}} - \Pi_{\ast}^{\mathrm{mod}}}{\Pi_{\ast}^{\mathrm{mod}}}$ & -0.0461  &  -0.0347  &  0.0239  & 0.292   \\ 
 %  $\frac{B_{\ast}^{\mathrm{est}} - B_{\ast}^{\mathrm{mod}}}{B_{\ast}^{\mathrm{mod}}}$ & -0.600  &  -0.552  &  -0.759  &  -0.862  \\  \hline
$1.6 \, M_{\odot}$ & -1.2  &  -1.3  &  -1.7  & -2.3 \\ 
$2 \, M_{\odot}$ & -1.4  &  -1.3  &  -1.1  & -0.83   \\ 
$3 \, M_{\odot}$ & -1.5  &  -1.5  &  -1.6  &  -1.6  \\  
$4 \, M_{\odot}$ & -1.6  &  -1.7  &  -1.8  &  -2.4  \\  \hline
\addlinespace[2.5mm] 
    \end{tabular}
    \\
    
    %\footnotesize 
    $\delta \, \mathrm{ln} \, \Pi_{\star}$ ($\%$) 
    \begin{tabular} {c|cccc} \hline\hline
      & $X_{\mathrm{c}} \sim 0.5$ &  $X_{\mathrm{c}} \sim 0.4$ &  $X_{\mathrm{c}} \sim 0.3$ &  $X_{\mathrm{c}} \sim 0.2$  \\ \hline
 %  $\frac{\Pi_{0}^{\mathrm{est}} - \Pi_{0}^{\mathrm{mod}}}{\Pi_{0}^{\mathrm{mod}}}$ & -0.01215  &  -0.01295  &  -0.01663  & -0.02281 \\ 
 %  $\frac{\Pi_{\ast}^{\mathrm{est}} - \Pi_{\ast}^{\mathrm{mod}}}{\Pi_{\ast}^{\mathrm{mod}}}$ & -0.0461  &  -0.0347  &  0.0239  & 0.292   \\ 
 %  $\frac{B_{\ast}^{\mathrm{est}} - B_{\ast}^{\mathrm{mod}}}{B_{\ast}^{\mathrm{mod}}}$ & -0.600  &  -0.552  &  -0.759  &  -0.862  \\  \hline
$1.6 \, M_{\odot}$ & -4.6  &  -3.5  &  2.4  & 2.9 \\ 
$2 \, M_{\odot}$ & 1.6  &  1.7  &  0.94  & 1.9   \\ 
$3 \, M_{\odot}$ & 1.2  &  0.52  &  -0.020  & 1.5   \\ 
$4 \, M_{\odot}$ & 1.1  &  1.9  &  0.89  &  0.48  \\  \hline
\addlinespace[2.5mm] 
    \end{tabular}
    \\
    
    %\footnotesize 
    $\delta \, \mathrm{ln} \, B_{\star}$ ($\%$) 
    \begin{tabular} {c|cccc} \hline\hline
      & $X_{\mathrm{c}} \sim 0.5$ &  $X_{\mathrm{c}} \sim 0.4$ &  $X_{\mathrm{c}} \sim 0.3$ &  $X_{\mathrm{c}} \sim 0.2$  \\ \hline
 %  $\frac{\Pi_{0}^{\mathrm{est}} - \Pi_{0}^{\mathrm{mod}}}{\Pi_{0}^{\mathrm{mod}}}$ & -0.01215  &  -0.01295  &  -0.01663  & -0.02281 \\ 
 %  $\frac{\Pi_{\ast}^{\mathrm{est}} - \Pi_{\ast}^{\mathrm{mod}}}{\Pi_{\ast}^{\mathrm{mod}}}$ & -0.0461  &  -0.0347  &  0.0239  & 0.292   \\ 
 %  $\frac{B_{\ast}^{\mathrm{est}} - B_{\ast}^{\mathrm{mod}}}{B_{\ast}^{\mathrm{mod}}}$ & -0.600  &  -0.552  &  -0.759  &  -0.862  \\  \hline
$1.6 \, M_{\odot}$ & -60  &  -55  &  -76  & -86 \\ 
$2 \, M_{\odot}$ & -21  &  -36  &  -47  & -57   \\ 
$3 \, M_{\odot}$ & 14  &  -0.30  &  -7.5  & -23   \\ 
$4 \, M_{\odot}$ & 8.3  &  1.2  &  -8.5  &  -18  \\  \hline
\addlinespace[2.5mm] 
    \end{tabular}
    \\
    
\end{center}
%\footnotesize $\mathbf{Note.}$ ... 
%%Decimal logarithm of the global likelihood computed 
%% given a set of the artificially generated rotational shifts (the left column) 
%% based on a model of rotational profile (the top row). 
%% See the text for the meanings of the variables. 
\end{table}

Table \ref{tab:1} shows relative differences between the estimates thus obtained 
and those directly extracted from the models. 
Note that the relative difference is defined as below: %, %result of the MCMC fitting (MAP is used). 
\begin{equation}
\delta \, \mathrm{ln} \, q = \frac{\hat{q} - q_{\mathrm{mod}} }{q_{\mathrm{mod}}}, \label{Eq_est_vs_prm}
\end{equation}
where $\hat{q}$ and $q_{\mathrm{mod}}$ represent an estimate of a particular parameter $q$ 
and that directly extracted from a stellar model, respectively. 

One remarkable point about the result of MCMC fitting is that parameter estimation 
works quite well ($\delta \, \mathrm{ln} \, q \sim$ a few $\%$) 
for the buoyancy radius at the outer edge of the g-mode cavity ($\Pi_{0}$) and 
that at $r = r_{\star}$ ($\Pi_{\star}$), 
%parameters $\Pi_{0}$ and $\Pi_{\star}$, 
which is the case for almost all the stellar models 
considered in this section (see top and middle tables in Table \ref{tab:1}). 
Some systematic underestimation of $\Pi_{0}$ %can be explained by 
%which 
arise probably due to the factor $r^{-1}$ in the local wavenumber $k_{r}$ 
%attributed to the boundary effect of the g-mode cavity 
(see discussions in Section \ref{sec:3-3-2}); 
%fact that 
the asymptotic value (\ref{eq01}) is an upper bound for 
the corresponding $\Delta P_{g}$ pattern numerically computed 
%the $\Delta P_{g}$ patterns numerically computed give 
%are bounded above by the corresponding asymptotic value (\ref{eq01}) 
as we see in Figure \ref{fig:3-3-2}. %, 
%which is caused by the negligence of the boundary effects
%the buoyancy radius of a stellar model is 
%computed 

%The comparison of ... with ... 
%which indicates that, roughly speaking, 
As for the degree of the structural variation ($B_{\star}$), whether parameter estimation works well or not 
depends on the mass and central hydrogen mass content. 
We see that parameter estimation is rather accurate in the case of 
the $3$ and $4 \, M_{\odot}$ models; 
the relative differences are smaller than $25 \, \%$, 
and they are especially small ($< 10 \, \%$) for the younger models 
(see the bottom table in Table \ref{tab:1}), 
%For example, the relative differences between the estimated $B_{\star}$ 
%and those extracted from the models are smaller than $10 \, \%$ for all the evolutionary stages, 
indicating the high potential of the semi-analytical approach 
%expression derived in Section \ref{sec:3} %this study %Section \ref{sec:3} 
to infer the degree of the structural variation inside, e.g., SPB or $\beta$ Cep stars. 

On the other hand, it is evident that estimating $B_{\star}$ is quite difficult 
for the less massive $1.6$ and $2 \, M_{\odot}$ models; 
a typical absolute value of the relative differences is larger than $50 \, \%$. 
%We thus have to be careful 
Therefore, it might be impractical 
to apply the semi-analytical expression to less massive main-sequence g-mode pulsators 
such as $\gamma$ Dor stars for inferring the degree of the structural variation inside them. 

%In Table \ref{tab:1}, we also find a tendency that 
The tendency 
that inaccuracies in parameter estimation of $B_{\star}$ become larger 
as a stellar model becomes less massive or older 
may arise from 
the assumption that we neglect the period dependence of $A'_{\star}$ when we introduce $B_{\star}$. 
However, the interval for which the subtraction is taken (see Equation \ref{Eq_rtlv_dif_prime}) 
should vary as the local wavelength of gravity waves changes; 
$A'_{\star}$ is implicitly period-dependent. 
In particular, it is expected that the milder and wider the BV frequency transition is, 
the more relevant the local wavelength of gravity waves is 
in terms of how to define the width of the interval. 
In other words, the degree of the period dependence of $A'_{\star}$ 
becomes larger for milder and wider BV frequency transitions. 
This is actually in accordance with the result that parameter estimation of $B_{\star}$
does not work well for the less massive or older stellar models 
that exhibit milder and wider BV frequency transitions as discussed in Section \ref{sec:2-2} 
(see also Figure \ref{fig:2-1}).

We could thus refine the semi-analytical expression 
by somehow taking into account the period dependence of $A'_{\star}$. 
Interestingly, we have seen a trend that we are more prone to underestimating $B_{\star}$ 
as the star becomes older even for the massive models 
(though the values of $\delta \, \mathrm{ln} \, B_{\star}$ is not so large), 
which could be resolved as well by the improvement suggested above. 
%But, as it is ... 
We will discuss the improvement in the forthcoming paper. 

%Gradually... , 
%leading to underestimation of ... . 
%there are $\%$, $\%$, and $\%$ systematic errors in the estimation of the parameters. (Figure example?) 
%We can thus claim that ... . 
%(WHAT TO DELIVER SHOULD BE FURTHER CONSIDERED!!)
%%\begin{figure}[t!]
%%\includegraphics[scale=0.52]{Figure_4_1_2_new.pdf}
%\plotone{Figure_4_1_1.pdf}
%%\caption{\footnotesize Comparison of $f(\mathrm{d}r/\mathrm{d}\zeta)$ (solid curves) and $\zeta$ (dotted curves) 
%%around the deep radiative region of $2$ and $4 \, M_{\odot}$ models 
%%(top and bottom panels, respectively) 
%%with the central hydrogen mass contents $X_{\mathrm{c}}\sim 0.4$ or $0.2$ 
%%(left or right column, respectively). 
%%The horizontal axis represents the fractional radius 
%%(note that each panel has a different scale). 
%%A grey shaded area corresponds to where the sharp structural variation is located. 
%%See the text for definitions of the parameters. 
% can be found in the text. 
%Note that the scales of the horizontal axes are different. 
%%\label{fig:4-2-2}}
%%\end{figure}

\begin{figure}[t!]
\includegraphics[scale=0.5]{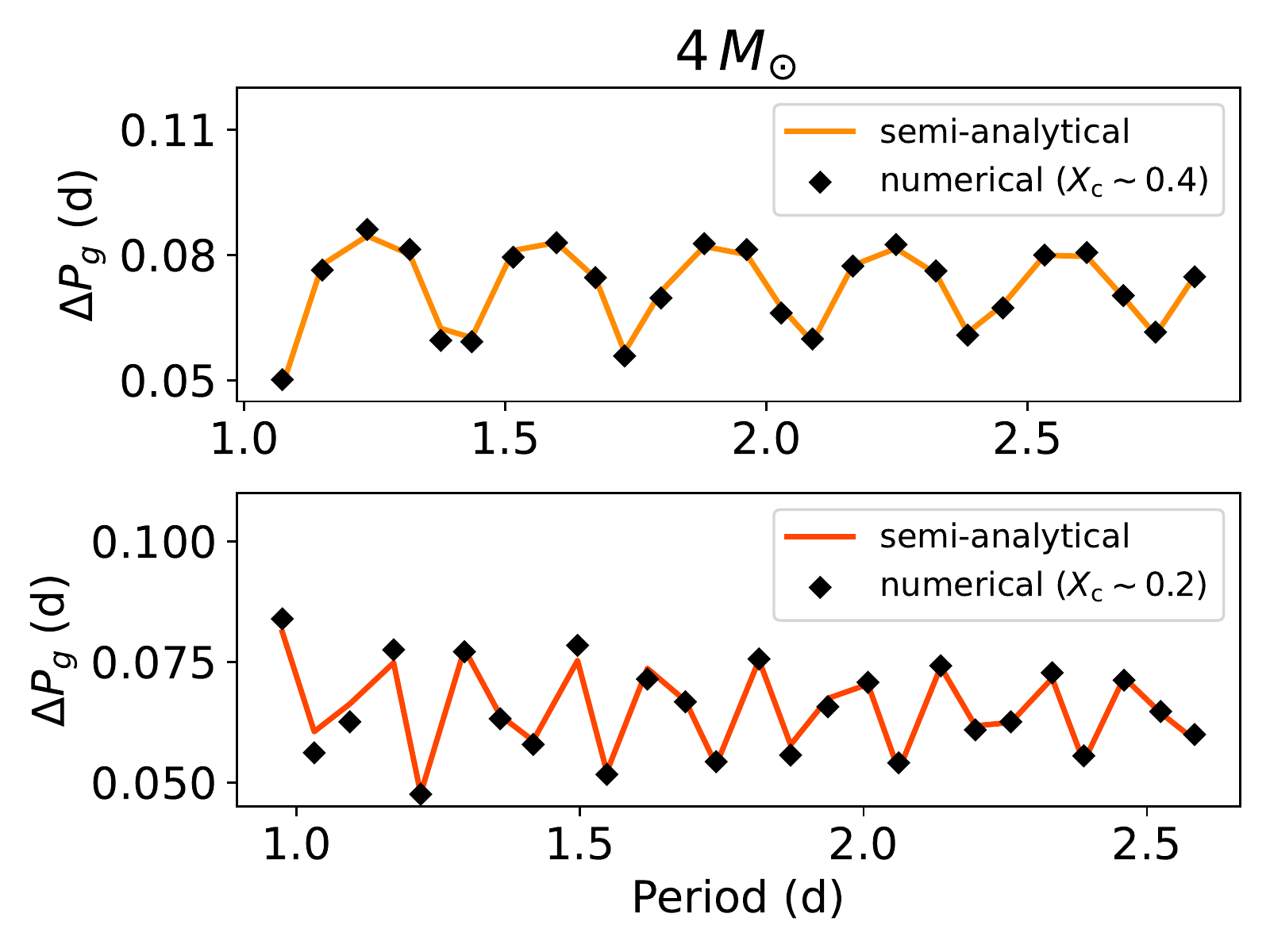}
%\plotone{Figure_4_2_2.pdf}
\caption{\footnotesize Comparison of the fitted $\Delta P_{g}$ patterns 
(warm-color curves) with those numerically computed (dark grey diamonds) 
in the case of the $4 \, M_{\odot}$ models 
%at two different evolutionary stages, namely, when 
with the central hydrogen mass content $X_{\mathrm{c}} \sim 0.4$ or $0.2$ 
(top or bottom, respectively). 
%is $0.505$ (top) and when $X_{\mathrm{c}}$ is 0.305 (bottom). 
The g-mode period (the abscissa) and its period spacing (the ordinate) are expressed in units of day. 
\label{fig:4-1-1}}
\end{figure}

\begin{figure}[t!]
\includegraphics[scale=0.5]{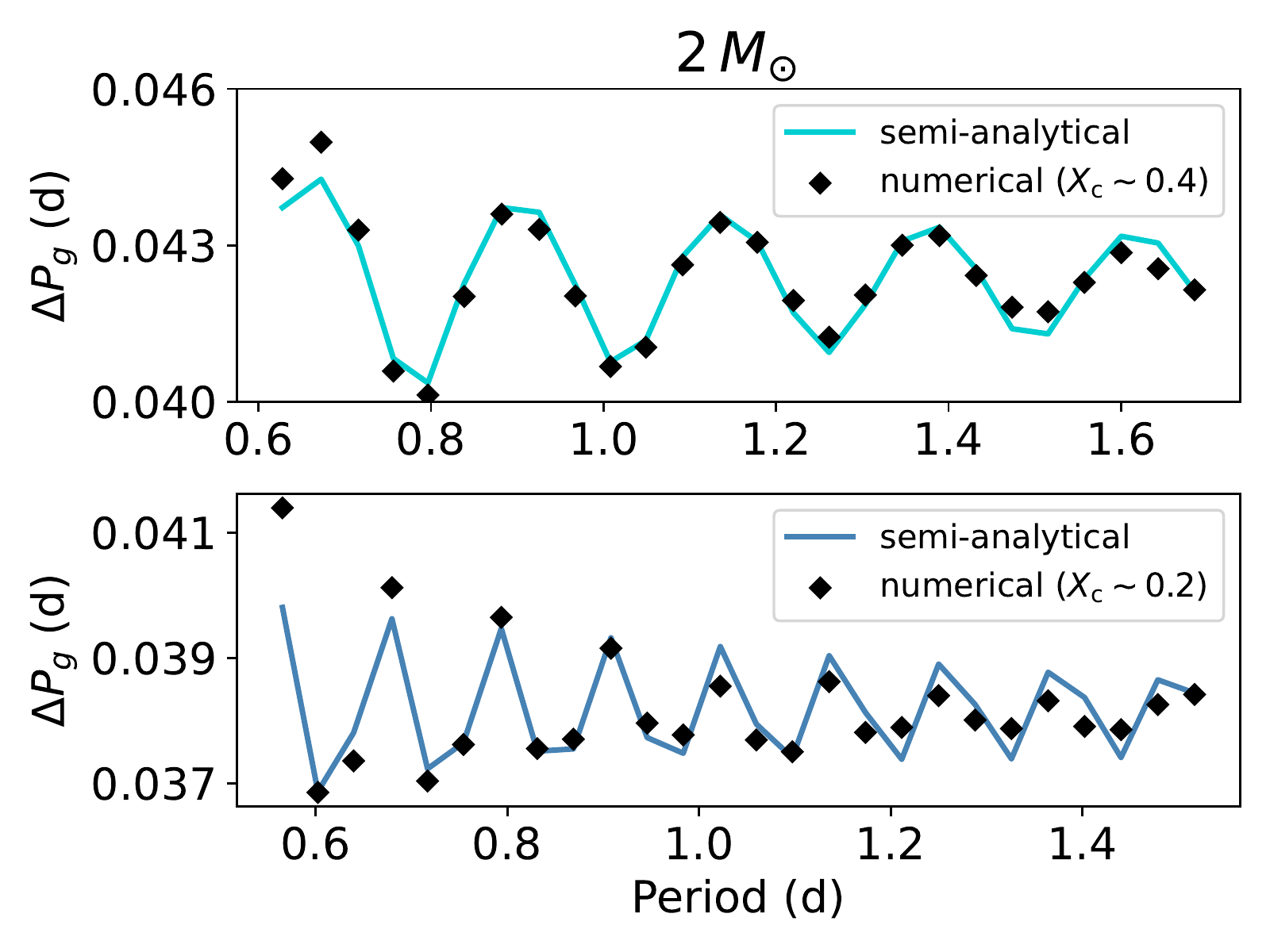}
%\plotone{Figure_4_2_1.pdf}
\caption{\footnotesize 
Same as Figure \ref{fig:4-1-1} 
in the case of the $2 \, M_{\odot}$ models 
except that the fitted $\Delta P_{g}$ patterns are shown in cool colors. 
%(the lime curve in the top panel, and the green curve in the bottom panel). 
\label{fig:4-1-2}}
\end{figure}

Finally, in order to check how well we can reproduce the $\Delta P_{g}$ patterns 
%numerically computed with 
of the stellar models, 
fitted $\Delta P_{g}$ patterns are computed by inserting the so-called 
Maximum A Posteriori (MAP) estimator, which provides us with a parameter set 
that maximizes the posterior probability, 
into the semi-analytical expression. 
Figures \ref{fig:4-1-1} and \ref{fig:4-1-2} show comparisons 
of the fitted $\Delta P_{g}$ patterns thus computed (colored curves) with those  
%using the estimated parameters (MAP is used) which are compared with those 
numerically computed (black diamonds) based on the $4$ and $2 \, M_{\odot}$ models, respectively, 
at two different evolutionary stages $X_{\mathrm{c}} \sim 0.4$ and $0.2$. %, respectively. 
%NOT SO BAD! 

%On one hand, 
In the cases of the $4 \, M_{\odot}$ models (Figure \ref{fig:4-1-1}), 
we see a fairly good agreement between the fitted and numerical $\Delta P_{g}$ patterns, 
which is consistent with the result that parameter estimation works well 
for the relatively massive models. 
On the other hand, it is apparent in the cases of $2 \, M_{\odot}$ models 
that the fitted $\Delta P_{g}$ patterns cannot completely reproduce the numerical results (Figure \ref{fig:4-1-2}), 
%for the both evolutionary stages, 
%%In particular, the period dependence of the amplitude of the $\Delta P_{g}$ patterns 
%%of the $2 \, M_{\odot}$ models %numerically computed 
%%is different from that of the semi-analytical expression; 
%%the amplitude becomes smaller against the g-mode period more quickly than the semi-analytical expression does. 
which may be related to the negligence of the implicit period dependence of $A'_{\star}$ 
as we mentioned a few paragraphs before. 
It should especially be noted that the amplitude of the oscillatory $\Delta P_{g}$ pattern is 
overestimated (underestimated) for the shorter (longer) g-mode period (see Figure \ref{fig:4-1-2}), 
indicating that the amplitude decreases more rapidly with respect to the g-mode period 
compared with what the semi-analytical expression predicts. 
%%This is nevertheless reasonable since, as mentioned in a few paragraphs before, 
%%the degree of the sharp variation in the first derivative of $k_{r}$ 
%%in these less massive models is too small 
%can be attributed to the fact that ... 
%%to maintain the assumption %condition (\ref{Eq_to_be_Airy}) 
%%that is necessary for validating the derivation of the semi-analytical expression. 
Nevertheless, we also would like to stress 
that the semi-analytical expression at least succeeds in reproducing 
the decreasing amplitude of the oscillatory component in the $\Delta P_{g}$ pattern 
against the g-mode period, %becomes larger, 
which certainly is an advantage the semi-analytical expression 
derived in this study has. 

%%\subsection{Can we correctly extract glitch information \\
%%by fitting the semi-analytical expression?
%Reinterpreting the parameters in the semi-analytical expression
%Interpreting the parameters to describe the g-mode period spacing \\
%based on stellar models
%%} \label{sec:4-2}

%%As shown in the last section, the new semi-analytical expression can more or less 
%%reproduce the $\Delta P_{g}$ patterns of the stellar models. 
%%Then, the next question is whether or not 
%%we can correctly estimate the parameters, namely, $\Pi_{0}$, $\Pi_{\ast}$, and $B_{\ast}$ 
%%(note that the extra phase term $\delta$ is treated as a free parameter in this study), 
%%based on the fitting of the semi-analytical expression. % to the $\Delta P_{g}$ patterns. 
%Is it possible for us to ... ? 
%%To check this point, we will compare the MAP estimates obtained in Section \ref{sec:4-1} 
%determined via the MAP estimator) 
%%with those directly extracted from the $1.6 \, M_{\odot}$ and $3.0 \, M_{\odot}$ models. 

\subsection{A case study of KIC 11145123} \label{sec:4-2}
As another example of applications of the derived semi-analytical expression for the $\Delta P_{g}$ pattern, 
we will present a case study where the expression has been fitted to 
the observed $\Delta P_{g}$ pattern of 
one of the Kepler targets, KIC 11145123. 

KIC 11145123 is a $\delta$ Sct-$\gamma$ Dor hybrid pulsator 
observed by the Kepler probe for four years \citep{Huber2014}, 
so far found to be exhibiting 15 high-order ($l=1$) g modes, 2 low-order ($l=1$) p modes, 
and 6 %low-order 
($l=2$) mixed modes \citep{Kurtz2014}. 
%Due to its ..., 
%it is one of the Kepler targets that are most actively studied via asteroseismology \citep[e.g.][]{...}. 
The observed $\Delta P_{g}$ pattern has been used for 
carrying out asteroseismic modeling of the star \citep[][]{Kurtz2014,Takada_Hidai2017,Hatta2021}, 
%determining its evolutionary stage 
%for example revealing 
finding, for instance, that the star has exhausted almost all the hydrogen at the nuclear burning core %is at the 
(namely, that the star is at the terminal-aged main-sequence stage) and 
%and 
%helping us estimate 
that the mass of the star is around $1.4 \, M_{\odot}$. 
Note also that the well-resolved rotational splitting of the star indicates that the star is a slow rotator 
with the rotation period of $\sim 100 \, \mathrm{days}$ \citep[e.g.][]{Kurtz2014}. 

% based on asteroseismic modeling. 

%One interesting argument about the internal structure of the star in \citet{Hatta2021} is that 
%
Interestingly, \citet{Hatta2021} have claimed that, 
in order to reproduce 
the observed $\Delta P_{g}$ pattern of the star, 
%\citet{Hatta2021} have constructed ..., 
%the asteroseismic analysis of the observed $\Delta P_{g}$ pattern of the star 
%has led \citet{Hatta2021} to 
%claiming that 
the chemical composition gradient in the deep radiative region of the star %, 
%%which can be represented by $B_{\star}$, % and can be determined base on a way described in Section \ref{sec:4-1}, 
should be much steeper than those of 1-dimensional evolutionary models 
%calculated with the default settings in ... 
in which mixing processes such as elemental diffusion and convective overshooting 
are treated basically with the default settings in MESA 
\citep[see, e.g.,][]{Paxton2011,Paxton2013,Paxton2015,Paxton2018,Paxton2019}. 
To emphasize the point, they have presented a seismic model of the star that is constructed 
by artificially increasing the chemical composition gradient just above the convective core 
of a $1.4 \, M_{\odot}$ evolutionary model 
%and they have additionally constructed another asteroseismic model 
\citep[see more details for][]{Hatta2021}. 
%For convenience in the following paragraphs, 
%let us denote the evolutionary model (from which \citet{Hatta2021} constructed the seismic model) 
%and the seismic model 
%by the ``unmodified'' model and the ``modified'' model, respectively. 

\begin{figure}[t!]
\includegraphics[scale=0.50]{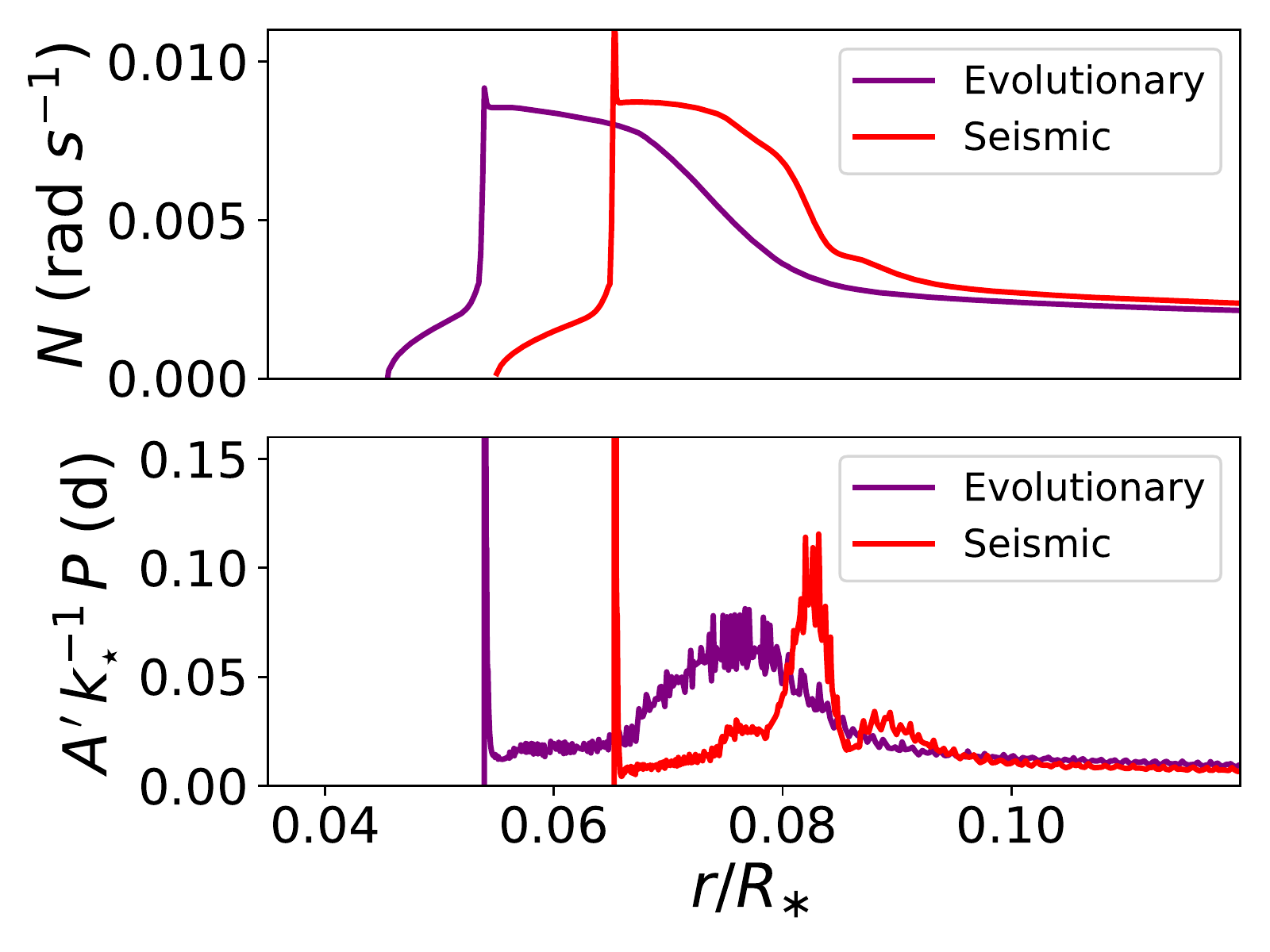}
%\plotone{Figure_4_3_2.pdf}
\caption{\footnotesize 
Structural parameters %BV frequency $N$ (top panel) and the parameter $A' k_{\star}^{-1} P$ (bottom panel) 
around the deep radiative region 
of the %$1.4 \, M_{\odot}$ 
evolutionary (purple curves) and seismic (red curves) models of KIC 11145123 
constructed by \citet{Hatta2021} via asteroseismic modeling. 
The top and bottom panels show BV frequency $N$ and 
a parameter $A' k_{\star}^{-1} P$, % (which is related to the degree of the glitch $B_{\star}$), 
respectively, against the fractional radius $r/R_{\ast}$. 
Note that the degree of the structural variation ($B_{\star}$) can be expressed as 
$B_{\star} = [A' k_{\star}^{-1} P]_{\star -}^{\star +}$, 
and that we thus have to take subtraction between the maximum and minimum values 
of $A' k_{\star}^{-1} P$ to compute $B_{\star}$ (see also discussions in Section \ref{sec:4-1-1}). 
Definitions of the parameters can be found in the text. 
%which is defined so that $...$. 
 \label{fig:4-3-1}}
\end{figure}

Although the inferred mass of the star %inferred via asteroseismic modeling 
($\sim 1.4 \, M_{\odot}$) 
seems to be too small for us to apply the semi-analytical expression 
based on the result of Section \ref{sec:4-1}, 
this is not the case. 
The reason is that 
the peculiarly sharp chemical composition gradient inferred based on the %``modified''
seismic model constructed by \citet{Hatta2021} can %be assessed 
%in a more qualitative manner 
%from a different perspective %may be worth comparing 
%An important point is that they 
%%the large amplitude of the observed $\Delta P_{g}$ pattern of the star 
%%has led \citet{Hatta2021} to claim that the chemical composition gradient of the star, 
%%which can be represented by $B_{\star}$, % and can be determined base on a way described in Section \ref{sec:4-1}, 
%%should be much steeper compared with that of their model, which may be worth testing 
%with the fitting procedure based on 
%%by 
justify applying the derived semi-analytical expression to the star, 
which has motivated us to carry out the case study with the star in this section. 
It should also be noticed that 
the BV frequencies of the models have ramp structures (see top panel in Figure \ref{fig:4-3-1}), 
preventing us from using the semi-analytical expression derived in \citet{Cunha2019}. 
%%... 

We have followed the same procedure as described in Section \ref{sec:4-1} 
to fit the semi-analytical expression to the observed $\Delta P_{g}$ pattern of the star. % via the MH method. 
The values of the observed $\Delta P_{g}$ spacings and the corresponding observational uncertainties can be found 
in \citet{Kurtz2014}. 

The estimate of the degree of the structural variation $B_{\star}$ thus obtained is $0.093 \pm 0.002$ 
(in units of days) that is significantly larger than  
that of Hatta et al.'s $1.4 \, M_{\odot}$ evolutionary model ($B_{\star} \sim 0.06$) 
%which is obtained by taking subtraction between the maximum and minimum values of 
(see the purple curve in the bottom panel of Figure \ref{fig:4-3-1}, 
and take the difference between the ``peak'' and ``baseline'' values %local 
%maximum and local minimum found 
found around the sharp transition region. See also Section \ref{sec:4-1-1}). 
We have thus confirmed that the degree of the structural variation $B_{\star}$ 
in the deep radiative region of the star 
should be significantly larger than that predicted based on ordinary 1-dimensional evolutionary computations, 
which is consistent with the argument of \citet{Hatta2021}. %that 
%the chemical composition gradient in the deep radiative region of the star, 
%which is here represented by the parameter $B_{\star}$, is 
%somehow much steeper than that predicted based on ordinary 1-dimensional evolutionary computations. 
%in a rather qualitative manner
Moreover, the estimate ($B_{\star} = 0.093 \pm 0.002$) is close to 
%to 
that directly extracted from the seismic model of \citet{Hatta2021} ($B_{\star} \sim 0.09$) 
(see the red curve in the bottom panel of Figure \ref{fig:4-3-1}, 
and take the difference between the ``peak'' and ``baseline'' values 
%the local maximum and local minimum %values 
found around the sharp transition region), 
clearly indicating that the seismic model is more favored than the original evolutionary model 
in terms of the observed $\Delta P_{g}$ pattern of the star. 
The semi-analytical expression derived in Section \ref{sec:3} is 
thus shown to be quite useful in the case of the star. 
%for inferring chemical composition gradient in the deep radiative region 
%compared with that of their original $1.4 \, M_{\odot}$ evolutionary model ($B_{\star} \sim 0.050 \, d$) (Figure \ref{fig:4-3-1}). 
%We have thus confirmed the argument of \citet{Hatta2021} that 
%the chemical composition gradient of in the deep radiative region of the star is 
%somehow much steeper than that of an ordinary 1-dimensional evolutionary model. 
%We have also found a hint that the seismic model

The result that the fitting procedure works well for KIC 11145123 
whose mass has been estimated to be around $1.4 \, M_{\odot}$ %via asteroseismic modeling 
seems to contradict with what we find in Section \ref{sec:4-1} %the previous section 
that it is not appropriate to apply the semi-analytical expression of the $\Delta P_{g}$ pattern 
derived in Section \ref{sec:3} to lower-mass main-sequence g-mode pulsators. 
%However, it should be noted that  just simple models... ; 
However, we would like to emphasize that 
the tests in Section \ref{sec:4-1} have been carried out based on the rather simple stellar models 
%(see the setups for Section \ref{sec:2-1}) 
%We also would like to emphasize that 
%an important point is that 
%The reason why the fitting procedure works well for the star 
%in spite of its relatively low mass ($\sim 1.4 \, M_{\odot}$, inferred via asteroseismic modeling) 
%is that the condition ... is somehow satisfied; 
and that the most essential parameter that determines 
whether or not we can describe the $\Delta P_{g}$ pattern with the semi-analytical expression 
is not the stellar mass 
but the degree of the sharp variation (in the first derivative of $k_{r}$) 
%in the deep radiative region 
inside a star 
(see discussions in Subsection \ref{sec:4-1-4}). 
%If the 
%%Therefore, it might be better to check if 
%%we can use the semi-analytical expression of the $\Delta P_{g}$ pattern or not 
%%to check it might be better to ... different ... star to star. 
%In the forthcoming paper. 
Therefore, in a sense, it may be such a coincidence that KIC 11145123 has a sharp structural variation 
in its deep radiative region 
whose degree is large enough that the semi-analytical expression is applicable. %this study. 
%Though it is not conclusive yet, 
%It might be instructive to refer to the suggestion by 
Related to this point, 
\citet{Hatta2021} have suggested that 
elemental diffusion much weaker than 
that predicted by ordinary 1-dimensional evolutionary codes 
can produce a steep chemical composition gradient 
(which leads to the large value of $B_{\star}$) 
even inside the low-mass main-sequence g-mode pulsators. 
%can partly resolve the discrepancy 
%between stellar models and the star. 

\begin{figure}[t!]
\includegraphics[scale=0.50]{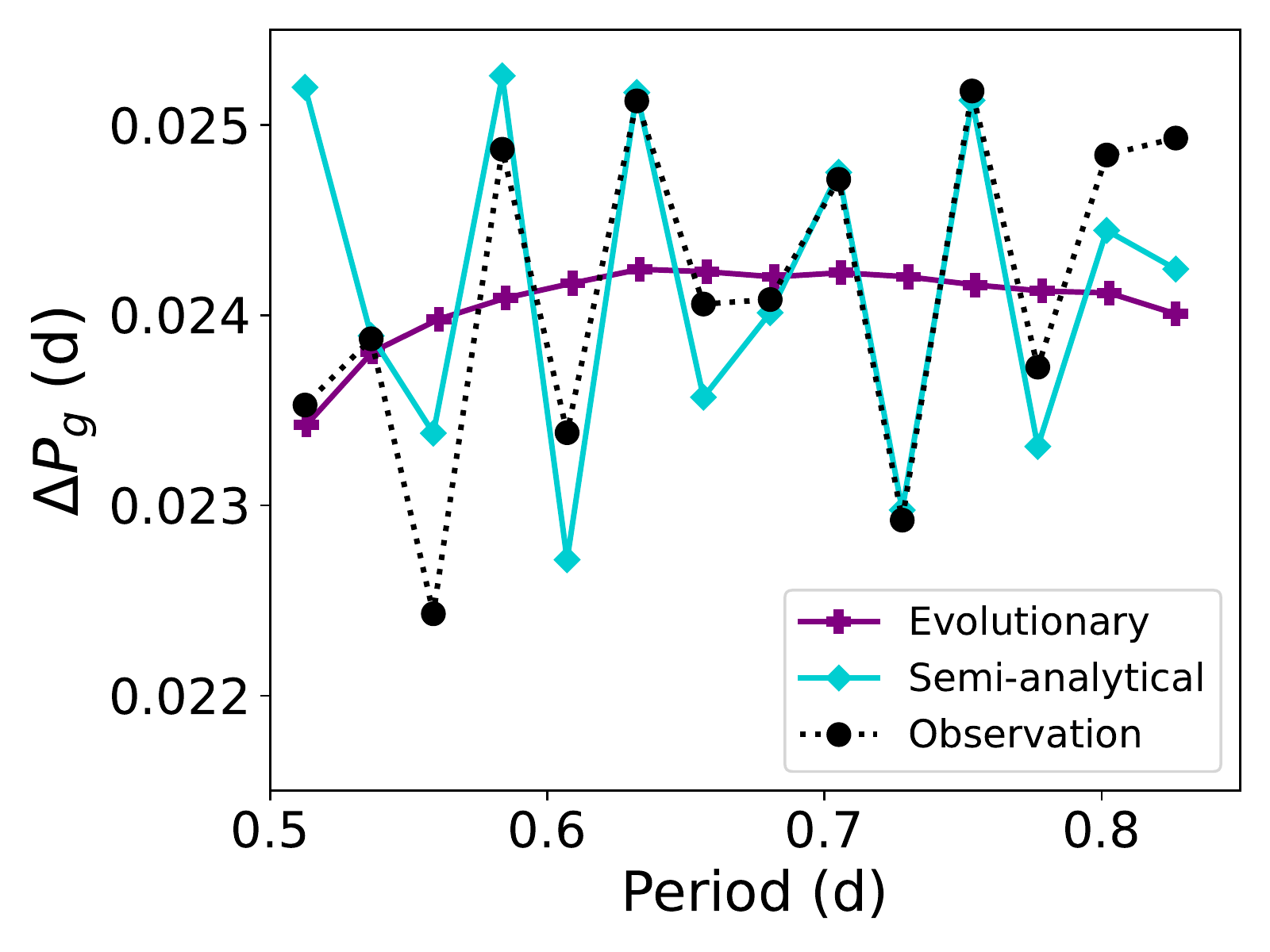}
%\plotone{Figure_4_3_1.pdf}
\caption{\footnotesize Comparison of %three kinds of $\Delta P_{g}$ patterns, 
the observed $\Delta P_{g}$ pattern for KIC 11145123 (black dotted curve with circles), 
the fitted $\Delta P_{g}$ pattern obtained with the semi-analytical expression (light blue curve with diamonds), 
and the $\Delta P_{g}$ pattern numerically computed with the evolutionary model of \citet{Hatta2021} (purple curve with crosses). 
%
%(warm-color curves) with those numerically computed (dark grey diamonds) 
%in the case of the $4 \, M_{\odot}$ models 
%at two different evolutionary stages, namely, when 
%with the central hydrogen mass content $X_{\mathrm{c}} \sim 0.4$ or $0.2$ 
%(top or bottom, respectively). 
%is $0.505$ (top) and when $X_{\mathrm{c}}$ is 0.305 (bottom). 
The g-mode period (the abscissa) and its period spacing (the ordinate) are expressed in units of day. 
Note that the observational uncertainties of the $\Delta P_{g}$ spacings ($\sim 10^{-5}$ in units of days) are 
much smaller than the size of the markers. 
  \label{fig:4-3-2}}
\end{figure}

We finally show three $\Delta P_{g}$ patterns, namely, 
the one observed for KIC 11145123, 
the one obtained by inserting the estimated parameters into the semi-analytical expression, 
and the one numerically computed based on the evolutionary $1.4 \, M_{\odot}$ model of \citet{Hatta2021} 
%and numerically computed g-mode period spacings... 
(Figure \ref{fig:4-3-2}). 
It is clearly seen that 
the fitted semi-analytical expression (turquoise curve) 
reproduces the amplitude of the oscillatory component in the observed $\Delta P_{g}$ pattern (black dotted curve)
much better than %with a shorter period much better 
the $\Delta P_{g}$ pattern of the $1.4 \, M_{\odot}$ evolutionary model (purple curve). 
Note that there still exists a systematic deviation between 
the observed $\Delta P_{g}$ pattern and the fitted semi-analytical one, 
which is thought to be caused by convective overshooting \citep{Hatta2021}. 
We will discuss the point in the next section as well as the effect of rotation. 

\section{Discussion} \label{sec:5}
We have seen that the semi-analytical expression derived in Section \ref{sec:3} is 
%more or less working appropriately 
useful for studying the sharp BV frequency variation in intermediate-mass main-sequence g-mode pulsators 
($1.3 \, M_{\odot} < M < 4 \, M_{\odot}$) 
given that the degree of the variation is 
large enough to validate the assumptions adopted in the derivation of the semi-analytical expression 
(e.g. the condition \ref{Eq_to_be_Airy}). 
%to reproduce the g-mode period spacings of several stellar models and KIC 11145123. 
We then would like to remind the readers of another important assumption 
in the derivation of the semi-analytical expression, i.e., 
%that 
we neglect rotation and convective overshooting, 
both of which play important roles inside the intermediate-mass main-sequence g-mode pulsators \citep{Maeder_text}. 
%are typical agents in the case of the intermediate-mass main-sequence g-mode pulsators (ref...). 
%are negligible. 
In this section, we give brief reviews and discussions about impacts rotation and convective overshooting have 
on $\Delta P_{g}$ patterns. 

%Let us start with rotation. 
It is generally considered that early-type stars are rotating rapidly 
($v \, \mathrm{sin} \, i \sim$ one hundred $\mathrm{km} \, \mathrm{s}^{-1}$) \citep[e.g.][]{Royer2007}, 
%in contrast to slowly rotating late-type stars ($v \, \mathrm{sin} \, i \sim$ a few tens $\mathrm{km}\, \mathrm{s}^{-1}$) (ref...), 
%Rotation is ... relatively well-understood (tilt). 
which realizes the coupling between the gravity mode and inertial mode 
(whose restoring force is the inertial force) 
to establish the gravito-inertial mode inside the stars \citep[e.g.][]{Mathis2014,Ouazzani2017}. 
The asymptotic analysis of the gravito-inertial mode 
has been often performed 
based on the so-called traditional approximation of rotation \citep[TAR;][]{Eckart1960,Lee1997}, 
where uniform rotation is assumed and the horizontal component of the rotation angular velocity is neglected. 
\citet{Bouabid2013} have studied the $\Delta P_{g}$ pattern of the fast-rotating stars based on TAR, 
finding that the $\Delta P_{g}$ pattern is quasi-linearly related to the g-mode period 
and that the gradient of the general trend in the $\Delta P_{g}$ pattern 
(as a function of the g-mode period) is determined by the rotation rate. 
%leading to 
%shows us that the $\Delta P_{g}$ pattern of a fast-rotating intermediate-mass main-sequence star 
%should be inclined against the period of the gravito-inertial mode \citep{Bouabid2013}. 
%%Since the formulation via TAR includes the rotation parameter, 
%The gradient of the ramp is determined by the rotation rate of the star, 
The property has enabled us to infer the internal rotation rates of the g-mode cavity of 
the fast-rotating stars \citep{Li2019, Li2020,Pedersen2022b}. 
%fitting the observed $\Delta P_{g}$ patterns of, for instance, fast-rotating $\gamma$ Dor stars, 
%based on TAR has enabled us to infer the internal rotation rates of the g-mode cavity of 
%the stars \citep{Li2019, Li2020}. 

Another mechanism has been suggested by \citet{Ouazzani2020} %have suggested another mechanism 
that the inertial mode in the convective core can be coupled with the g mode, 
which can produce the dip structure in the $\Delta P_{g}$ pattern of rapidly rotating stars. 
\citet{Saio2021} have analyzed the dips in the $\Delta P_{g}$ patterns of 
16 $\gamma$ Dor stars observed by the Kepler probe 
to infer the convective-core rotation rates of the stars, 
concluding that most of the stars studied are rotating almost rigidly. 
Similar discussions can also be found, e.g., \citet{Lee2020} and \citet{Tokuno2022}. 
%Another effect rotation can have on the $\Delta P_{g}$ pattern, Saio+2021... convective modes causing dip. 

As such, %demonstrated in the previous paragraphs, 
rotation certainly has significant effects on 
$\Delta P_{g}$ patterns of intermediate-mass main-sequence g-mode pulsators 
\citep[for more thorough review, see][]{Reese2022}. 
It is still possible that the effects of rotation on the $\Delta P_{g}$ pattern can be separable 
from those of the sharp BV frequency variation inside stars 
because signatures imprinted by rotation is not oscillatory %periodic ones 
with respect to the g-mode period \citep{Bouabid2013, Saio2021}. 
We thus might be able to, for example, analyze 
a component in the $\Delta P_{g}$ pattern originating from rotation and 
that originating from the sharp BV frequency variation independently 
by assuming a simple linearity among the effects. %, 
Alternatively, we can generalize the newly derived semi-analytical expression %derived in this study 
for rotating stars within the traditional approximation. 
These should be worth investigating in our future work. 

%Then, we move on to convective o
Convective overshooting is, theoretically speaking, considered as a natural consequence of 
uprising mass elements having inertia \citep[e.g.][]{Kippenhahn_text}. 
Multiple observational studies of stellar clusters 
have also favored the existence of convective overshooting 
in early-type main-sequence stars with the convective core 
\citep[e.g.][]{Maeder1981,Meynet1993,VandenBerg2004,Rosenfield2017}. 
It is however quite difficult to evaluate effects of convective overshooting on the internal structure of stars 
mainly because of a lack of the understanding of physics around the convective boundary, 
so far forcing us to rely on phenomenological schemes to describe convective overshooting 
%such as the so-called convective penetration \citep{Viallet2015} and 
%exponential diffusive overshooting \citep{Herwig2000} 
\citep[e.g.][]{Viallet2015}. 
An important point is that different prescriptions of convective overshooting 
lead to different BV frequency profiles, %chemical composition profiles, 
then resulting in different morphologies of $\Delta P_{g}$ patterns. 
%whose parameter ranges have been constrained by asteroseismic studies such as 
This has strongly motivated us to carry out 
asteroseismic modeling to determine which prescription is most favorable 
to reproduce the observed $\Delta P_{g}$ patterns 
\citep[see][and references therein]{Pedersen2022a}. 
\begin{figure}[t!]
\includegraphics[scale=0.54]{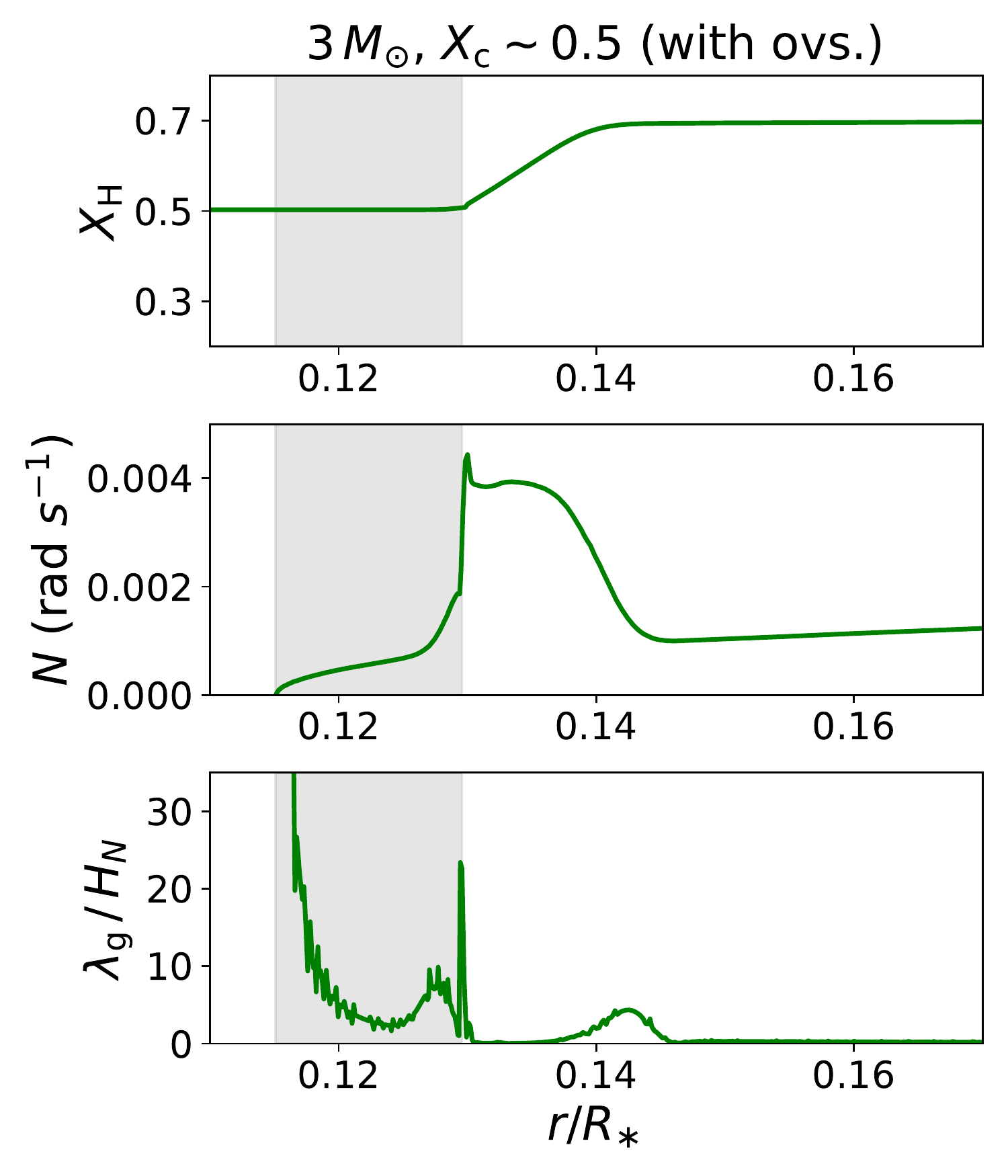}
%\plotone{Figure_5_1.pdf}
\caption{\footnotesize Internal properties around the convective-core boundary of 
the $3 \, M_{\odot}$ model with $X_{\mathrm{c}} \sim 0.5$ 
for which convective overshooting is taken into account. 
Hydrogen mass content ($X_{H}$), BV frequency ($N$), 
and the ratio of the typical wavelength of high-order g modes 
to the scale height of the BV frequency ($\lambda_{g} / H_{N}$) 
are shown from top to bottom. 
The horizontal axis is the fractional radius. 
Convective overshooting is treated as exponential diffusive overshooting \citep{Herwig2000} where 
the overshooting parameter ($f_{\mathrm{ov}} = 0.02$) is used. 
The overshoot zone is indicated by the grey shaded region, 
in which the chemical composition is uniform and the square of the BV frequency is positive. 
%Same as Figure \ref{fig:2-1} 
%in the case of the $2 \, M_{\odot}$ models 
%except that convective overshooting is taken into account 
%based on the exponentially-decaying scheme (ref...). 
%It is seen that the BV frequencies are positive at the overshoot zones 
%even though the chemical composition is uniform there 
%(see, for instance, )
 \label{fig:5-1}}
\end{figure}

\begin{figure}[t!]
\includegraphics[scale=0.5]{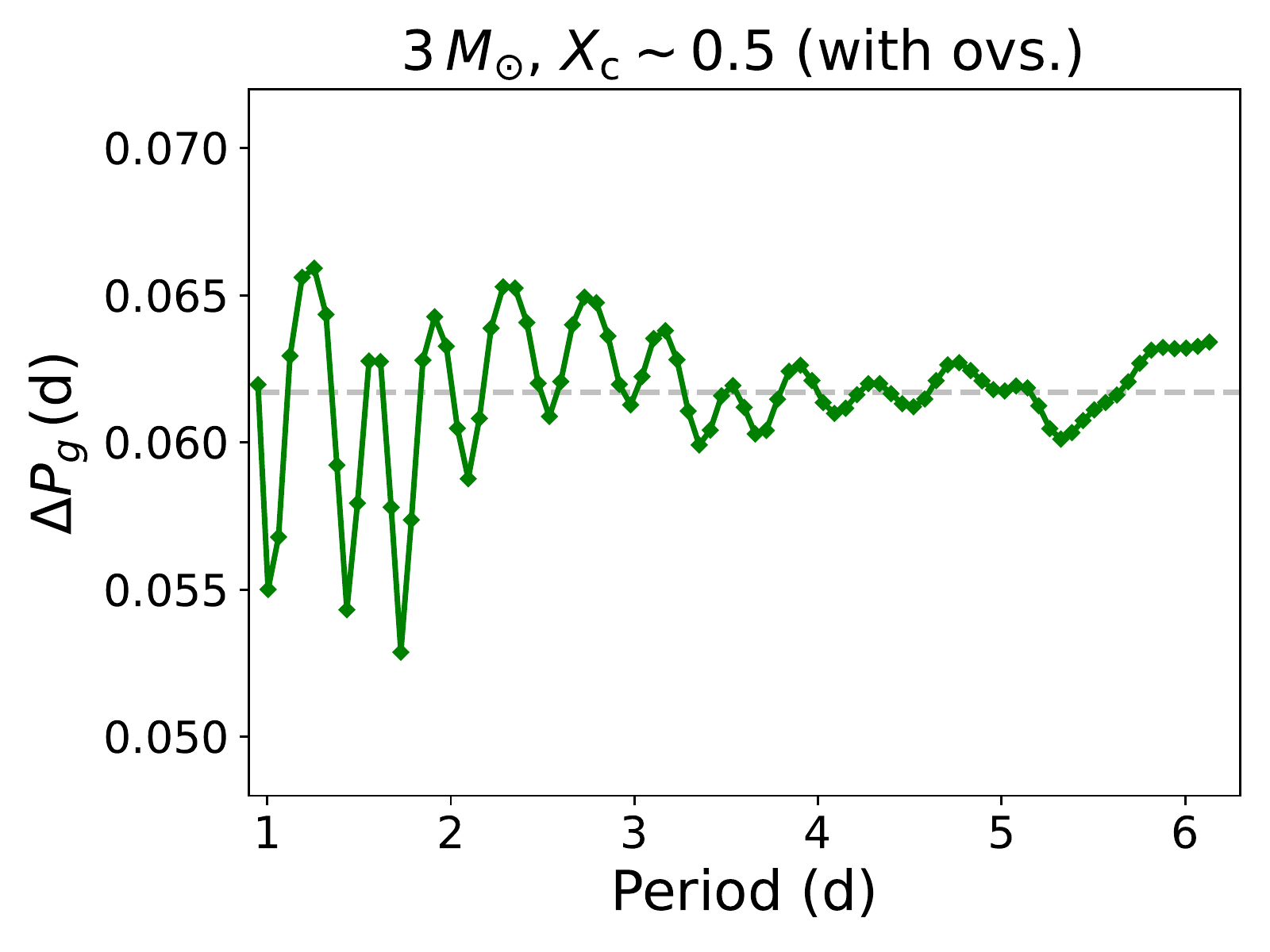}
%\plotone{Figure_5_2.pdf}
\caption{\footnotesize $\Delta P_{g}$ pattern numerically computed 
based on the $3 \, M_{\odot}$ model with $X_{\mathrm{c}} \sim 0.5$ 
for which convective overshooting is taken into account 
(the same model as that shown in Figure \ref{fig:5-1}). 
The horizontal axis shows the g-mode period. 
The mean value of the $\Delta P_{g}$ pattern is indicated by the grey dashed line. 
%Same as Figure \ref{fig:2-3} 
%in the case of the $2 \, M_{\odot}$ models 
%except that the stellar models shown in Figure \ref{fig:5-1}, 
%for which convective overshooting is taken into account, 
%are used. 
%See the caption of Figure \ref{fig:5-2} for details in the prescription 
%used to compute convective overshooting. 
  \label{fig:5-2}}
\end{figure}

%more complex to evaluate, 
One of the effects of convective overshooting 
on the internal structure of a star %the $\Delta P_{g}$ pattern 
is to extend the uniformly mixed region above the convective-core boundary 
where the convective stability is satisfied and the squared BV frequency is positive. 
This could lead to a jump-like structure 
in the BV frequency %that is closer to the inner edge of the g-mode cavity 
\citep[see around the outer edge of the grey shaded area in the middle panel of Figure \ref{fig:5-1},
in which convective overshooting is treated as exponential diffusive overshooting;][]{Herwig2000}. 
It is therefore expected that there is another oscillatory component in the $\Delta P_{g}$ pattern 
in addition to that caused by the sharp BV frequency variation 
located around the outer edge of the chemical composition gradient. 
%g-mode period spacings 
%since the period of the oscillatory component in the $\Delta P_{g}$ pattern 
%is determined by the ratio $\Pi_{0}/\Pi_{\star}$ (see Equations \ref{Eq_delP2} and \ref{Eq_delP3}). 
Such a $\Delta P_{g}$ pattern with double oscillatory components 
can be confirmed by the analytical expression of the $\Delta P_{g}$ pattern 
derived based on the perturbative approach (see Equation \ref{Eq_delP3}). 
Note that the JWKB approximation is broken near the inner edge of the g-mode cavity 
where the typical wavelength of the g mode is much longer than 
the scale height of the BV frequency 
(see around the inner edge of the grey shaded area in the bottom panel of Figure \ref{fig:5-1}). 
Accordingly, $\Delta P_{g}$ patterns numerically computed with stellar models 
(for which convective overshooting is taken into account) exhibit rather complex behaviors 
compared with what the analytical expression predicts 
%We can also see complex behaviors in $\Delta P_{g}$ patterns numerically computed with stellar models 
%for which convective overshooting is taken into account 
(Figure \ref{fig:5-2}). 
%\citet{Hatta2021} used the property to reproduce the observed $\Delta P_{g}$ pattern of the star 
%and to determine the extent of convective overshooting.  

Since the location of the jump-like structure in the BV frequency is 
closer to the inner edge of the g-mode cavity 
(compared with that of the BV frequency transition 
caused by the chemical composition gradient), 
the oscillatory component in the $\Delta P_{g}$ pattern 
%caused by the sharp BV frequency variation 
related to convective overshooting 
should have much longer period than that of the oscillatory component 
related to 
%caused by the glitch located around 
the outer edge of the chemical composition gradient 
(see Equation \ref{Eq_delP3}). 
%(see Figure \ref{fig:5-2}). 
This implies the possibility that we can study these effects independently 
in a similar way as we discussed effects of rotation. 
However, one difficulty is that %the period of the longer-period oscillatory component is too long... 
it should be often the case that the number of detected modes is not enough to observe 
the whole structure of a long-period component in the $\Delta P_{g}$ pattern. 
In addition, the dip structure caused by the fast rotation may complicate the situation 
\citep[see discussions in Appendix A of][]{Saio2021}. 
Further analyses would be thus expected for comprehensive analysis of the observed $\Delta P_{g}$ patterns. 

\section{Conclusion} \label{sec:6}
%\begin{enumerate}
%   \item 
1-dimensional stellar models of intermediate-mass stars 
($1.6 \, M_{\odot} < M < 6 \, M_{\odot}$) show that %, in this study) show that 
not only the size of the convective core but also the BV frequency profile 
around the deep radiative region just above the convective core depends strongly 
on the balance between the pp-chain reaction and CNO cycle. 
In particular, as the mass of the star becomes smaller, 
the pp-chain reaction becomes dominant over the CNO cycle more 
in the core region. 
For the less massive stars, the region where the pp-chain reaction is at work 
extends beyond the convective-core boundary and 
effectively smoothens the chemical composition gradient there, 
leading to the smaller steepness of the BV frequency gradient that is shaped like a ramp rather than a jump. 

Based on the assumption that the BV frequency of less massive stars
($M < 4 \, M_{\odot}$) can be modeled with a ramp function, 
we apply the JWKB approximation to derive a semi-analytical expression 
of the $\Delta P_{g}$ pattern for high-order g modes. 
The formulation includes a period-dependent term 
that represents the degree of the sharp variation in the BV frequency sensed by g modes 
and enables us to reproduce a trend 
that the amplitude of the oscillatory component in the $\Delta P_{g}$ pattern decreases 
for longer g-mode periods. 
The semi-analytical expression has been validated %carried out 
by comparing it with $\Delta P_{g}$ patterns numerically computed based on simple BV frequency models. 
%   \item Generally speaking, as an intermediate-mass main-sequence star evolves, 

Tests with 1-dimensional stellar models show that 
the semi-analytical expression can be used 
for inferring the degree of the structural variation ($B_{\star}$) in the BV frequency, 
achieving accuracy around $\sim 10 \, \%$ % at best, 
given that the BV frequency transition in the deep radiative region is 
sharp enough to validate neglecting the period dependence of 
the strength of the discontinuity (in the first derivative of the local wavenumber) $A'_{\star}$. 
%structural variation %is large enough to validate the JWKB approximation. 
For instance, the condition mentioned above is satisfied 
in the case of ordinary 1-dimensional stellar models with the mass larger than $3 \, M_{\odot}$. 
Thus, we might apply the fitting procedure with the semi-analytical expression 
presented in this study to SPB or $\beta$ Cep stars. 
Parameter estimation of $B_{\star}$ with the semi-analytical expression, however, 
does not work well for the less massive stellar models ($M \le 2 \, M_{\odot}$), 
which could be resolved 
by taking into account the period dependence of $A'_{\star}$ 
in the semi-analytical expression. %might improve the accuracy of parameter estimation 
%for stellar models with the smaller mass $< 2 \, M_{\odot}$. 
%which will be 
We nevertheless would like to emphasize that 
the most important parameter is not the mass 
but the sharpness of the BV frequency transition. 
Some less massive stars ($M \le 2 \, M_{\odot}$) might have 
such a sharp BV frequency transition, as is the case for KIC 11145123, 
that we can possibly justify applying the semi-analytical expression 
to these stars. 

Finally, it should be noticed that dynamics such as rotation and overshooting 
have significant impacts on the $\Delta P_{g}$ pattern as well, 
and therefore, further comprehensive analyses of the $\Delta P_{g}$ pattern would be desirable. 

%\end{enumerate}
%The BV frequency of a star predominantly determines the g-mode period 
%in the high-order limit. 
%In particular, the gradient of the BV frequency around the deep radiative region of a star 
%becomes smaller as the star is less massive or older, 
%The g-mode period is asymptotically determined by 
%the BV frequency of a star in the high-order limit. 
%In particular, in the case of massive stars, we can parameterize the BV frequency profile 
%with the two-zone modeling, leading to the ... . 
%The BV frequency, however, becomes ... as the star is less massive or older 
%The steeper the chemical composition gradient is, 
%the larger the degree of the glitch in the BV frequency is; 
%this is especially the case for relatively massive stars 

%The gradient in the BV frequency is a common feature in intermediate-mass main-sequence g-mode pulsators, 
%which has prevented us from parameterizing the BV frequency profile inside the stars 
%with the two-zone modeling and from 

\begin{acknowledgments}

We would like to express our gratitude to the NASA and \textit{Kepler} team for the precious data. 
Y. H. acknowledges T. Sekii for his insightful comments. 
M. Takata, O. Benomar, and 
%for their constructive comments. 
M. Cunha are also thanked for their constructive advices. 
%Y.H. acknowledges the Research Fellowship from the Japan Society for the Promotion 
%of Science for Young Scientist. 
%\edit1{We also would like to thank the anonymous referee for his or her constructive comments on 
%our future work.} 
This work was supported by 
MEXT as “Program for Promoting Researches on the Supercomputer Fugaku” 
(Toward a unified view of the universe: from large scale structures to planets, JPMXP1020200109), 
JSPS Grant-in-Aid for Scientific Research (A) Grant Number JP21H04492, 
and JSPS Grant-in-Aid for JSPS Research Fellow Grant Number JP20J15226. 
%\edit1{Fugaku grant...}

\end{acknowledgments}

%\bibliography{sample631}{}
%\bibliographystyle{aasjournal}

\appendix
\section{Analytical Expression of G-Mode Period Spacing \\
Derived Based on Perturbation Theory} \label{ap:sec:a}
%MENTION WU ET AL. 2020 AS WELL!! 
%There have thus been several attempts to theoretically analyze the oscillatory $\Delta P_{\mathrm{g}}$ patterns, 
%e.g. \citet{Montgomery2003}, \citet{Miglio2008}, \citet{Cunha2019}. 
%These studies can be divided into two ways based on how they theoretically quantify the $\Delta P_{\mathrm{g}}$ patterns.  
%One is based on the variational principle where 
We here present analytical expressions of the $\Delta P_{g}$ pattern 
derived based on the perturbative approach (see also Section \ref{sec:3-1}), 
%In the perturbative approach, 
in which the oscillatory component in the $\Delta P_{\mathrm{g}}$ pattern is considered 
as a result of perturbed g-mode eigenfrequencies 
that originate from a perturbation in the BV frequency. 
We will show two examples of how we model the perturbation in the BV frequency ($\delta N$ hereafter): 
in one case, $\delta N$ is modeled with a ramp (Appendix \ref{ap:sec:a1}), 
and in the other case, it is modeled with a convective overshoot zone (Appendix \ref{ap:sec:a2}). 
%Another case 
%As mentioned in Section \ref{sec:3-1}, 
%the perturbative approach is 
%This approach has been intensively discussed by \citet{Montgomery2003} and \citet{Miglio2008}, 
%We will start with ... and then, ... . 
%For more discussions, see, e.g., \citet{Montgomery2003}, \citet{Miglio2008}, and \citet{Wu2020}. 
%and we followed the  in this appendix. 
%with a few new expressions derived by myself. 
%
%%Though the analysis based on the variational principle is applicable 
%%for any perturbation in the $\rm{Brunt}$-$\rm{{V\ddot{a}is\ddot{a}l\ddot{a}}}$ frequency, 
%%the perturbation has to be small enough that 
%%terms higher than the second-order can be neglected and that the variational principle is valid; 
%%this assumption is often not acceptable when we deal with, 
%%for instance, chemical composition gradients inside low-mass main-sequence stars, 
%%where even a discontinuity may develop.  
%%.In that case, we can derive solutions for g-mode eigenfrequencies (as we did in the main body of this text). 

Before going into detail, %deriving the explicit analytical expressions of the $\Delta P_{g}$ pattern 
let us present an important equation that relates the perturbation in the BV frequency ($\delta N$) 
to the perturbed g-mode period to first order: 
\begin{equation}
-\frac{\Pi_{0}^{-1}}{2} \frac{\delta P_{n}}{P_{n}} =  \int \biggl (  \frac{\delta N}{N} \biggr ) \mathrm{sin}^2 \biggl (\frac{L P_{n}}{2 \pi} \Pi_{r}^{-1} +\frac{\pi}{4}   \biggr ) \mathrm{d}\Pi_{r}^{-1},  \label{Eq_var_N_to_delP}
\end{equation}
where $P_{n}$ and $\delta P_{n}$ represent the period of a g mode with the radial order $n$ 
and the perturbed period of the g mode, respectively. 
The meanings of the other parameters can be found in the main text. 
This equation can be derived based on the variational principle 
\citep[see, e.g.,][for more information on the derivation]{Montgomery2003,Miglio2008,Wu2020}.

By parameterizing the perturbation in the BV frequency 
$\delta N$ so that we can analytically compute the 
integration in Equation (\ref{Eq_var_N_to_delP}), 
we can then derive explicit expressions of the $\Delta P_{\mathrm{g}}$ pattern as a function of the g-mode periods 
as we will see in Appendices \ref{ap:sec:a1} and \ref{ap:sec:a2}. 
%With Equation (\ref{Eq_var_N_to_delP}), 
%we can derive analytical expressions of $\Delta P_{\mathrm{g}}$ pattern as a function of the g-mode periods 
%once we parameterize %the perturbation in the BV frequency 
%$\delta N$, % as \citet{Miglio2008} has done, 
%and analytically compute the integration in Equation (\ref{Eq_var_N_to_delP}). 
%Three specific ways of parameterization of $\delta N$ are presented in the following small subsections. 

\subsection{The case of $\delta N$ parameterized with a ramp function} \label{ap:sec:a1}
%Here, the results of the analysis by \citet{Miglio2008} are again reproduced with the same purpose in the preceding small subsection. 
%For more detailed derivations and confirmation of the analyses can be found in the 
%original paper \citep{Miglio2008}. 
The parameterization of $\delta N$ is done with a ramp function as below:  
\begin{equation}
\frac{\delta N}{N} = \frac{1- \alpha^2}{\alpha^2} \frac{\Pi_{\star}^{-1} - \Pi_{r}^{-1}}{\Pi_{\star}^{-1} } H(\Pi_{\star}^{-1}-\Pi_{r}^{-1}),  \label{Eq_delN_param2}
\end{equation} 
where $H(\Pi_{\star}^{-1}-\Pi_{r}^{-1})$ represents a step function in terms of the buoyancy radius $\Pi_{r}^{-1}$. 
We follow the notation of \citet{Miglio2008} to express 
the strength of the perturbation with the term $(1 - \alpha^{2}) / \alpha^{2}$. 
%see an example of the profile in Figure \ref{param_delN}. It should be noted that 
Note that 
the buoyancy radius of the inner edge of the g-mode cavity ($r=r_{0}$) is zero. 
An example of $\delta N$ parameterized in this way is shown in Figure \ref{fig:3-3-2} in the main text 
where $\alpha^{2} = 0.99$. 

We can analytically compute the integral (\ref{Eq_var_N_to_delP}) 
by inserting Equation (\ref{Eq_delN_param2}) 
and integrating by parts once. 
The result is 
\begin{equation}
\frac{\delta P_{n}}{P_{n}}=\biggl (  - \frac{1-\alpha^2}{\alpha^2} \Pi_{0} \Pi_{\star} \biggr ) 
 \biggl [  \frac{1}{2} (\Pi_{\star}^{-1})^2  +\frac{\omega_{n}}{2L} \Pi_{\star}^{-1} - \biggl ( \frac{\omega_{n}}{2L}  \biggr )^2 
 \mathrm{sin}\biggl ( 2 n \pi \Pi_{0} \Pi_{\star}^{-1}  \biggr ) \biggr ],   \label{Eq_delP2}
\end{equation} 
in which $\omega_{n} = 2 \pi / P_{n}$.  
%The expression above seems to be much complex compared with the expression (\ref{Eq_delP1}), but 
When we focus on the sinusoidal component in the expression, 
it is readily seen that the amplitude of the oscillatory $\Delta P_{\mathrm{g}}$ pattern 
is proportional to the strength of the perturbation $(1-\alpha^2)/\alpha^2$ %as in the 
%expression (\ref{Eq_delP1}) 
and that the period of the oscillatory component is determined by a ratio between $\Pi_{0}^{-1}$ and $\Pi_{\mu}^{-1}$. 
%The essential points are thus not changed. 
In addition, we see that the amplitude depends on %the buoyancy radius of the discontinuity $\Pi_{\mu}^{-1}$ and on 
the g-mode periods $P_{n}$. 
These properties are consistent with what have been pointed out by, e.g., \citet{Miglio2008}. 
%which is not indicated in the case of the simpler parameterization of $\delta N$ and it is later demonstrated that 
%the parameterization in this small subsection is better-describing (than that in the preceding small subsection) oscillatory $\Delta P_{\mathrm{g}}$ patterns calculated based on 
%realistic stellar models. 

\subsection{The case of $\delta N$ parameterized with an overshoot zone} \label{ap:sec:a2}
%A novel formulation is presented in this small subsection with a parameterization of $\delta N$ by a modified ramp function as shown 
%in the right panel of Figure \ref{param_delN}. 
In Appendix \ref{ap:sec:a1}, it is assumed that $\delta N$ is non-zero %finite 
near the inner edge of the g-mode cavity where $\Pi_{r}^{-1} \sim 0$. 
But it is ordinarily seen based on 1-dimensional stellar evolutionary calculations (see discussions in Section \ref{sec:5} 
of the main text) 
that the BV frequency is rather smooth around regions
just above the convective core. 
This is due to the presence of the overshoot zone, 
in which the chemical compositions are well-mixed 
despite the convective stability there 
(the adiabatic temperature gradient $\nabla_{\mathrm{ad}}$ is larger than 
the radiative temperature gradient $\nabla_{\mathrm{rad}}$). 
%the BV frequency is finite there. 
As a result, we can assume $\delta N \sim 0$ in the overshoot zone. 

Then, the parameterization of $\delta N$ with an overshoot zone is given as below:  
\begin{equation}
\frac{\delta N}{N} = \left \{
\begin{array}{ll}
0 & (0 \leq  \Pi_{r}^{-1} \leq \Pi_{\star_{1}}^{-1}) \\
\frac{1- \alpha^2}{\alpha^2} \frac{\Pi_{\star_{2}}^{-1} - \Pi_{r}^{-1}}{\Pi_{\star_{2}}^{-1} - \Pi_{\star_{1}}^{-1}} H(\Pi_{\star_{2}}^{-1}-\Pi_{r}^{-1}) & (\Pi_{\star_{1}}^{-1} \leq \Pi_{r}^{-1}),
\end{array}
\right. \label{Eq_delN_param3}
\end{equation}
where $\Pi_{\star_{1}}^{-1}$ and $\Pi_{\star_{2}}^{-1}$ represent the buoyancy 
radii with which the interval $0 \leq \Pi_r^{-1} \leq \Pi_{\star_{1}}^{-1}$ 
is the overshoot zone (where $\delta N$ is zero) 
and the interval $\Pi_{\star_{1}}^{-1} \leq \Pi_r^{-1} \leq \Pi_{\star_{2}}^{-1}$ 
is the sharp buoyancy ramp above the overshoot zone. %radius for an innermost position and 
%that for an outermost position, respectively, between which $\delta N$ is non-zero. 

We can thus obtain the analytical expression of the $\Delta P_{\mathrm{g}}$ pattern 
in the case of $\delta N$ with the overshoot zone 
almost in the same way as in Appendix \ref{ap:sec:a1}: 
\begin{eqnarray}
\lefteqn{\frac{\delta P_{n}}{P_{n}} = \biggl (  - \frac{1-\alpha^2}{\alpha^2} \Pi_{0} (\Pi_{\star_{2}}^{-1} - \Pi_{\star_{1}}^{-1})^{-1}\biggr )} \nonumber \\
 & \times &  \biggl [  \frac{1}{2} (\Pi_{\star_{2}}^{-1} - \Pi_{\star_{1}}^{-1})^2  + \frac{\omega_{n}}{2L} (\Pi_{\star_{2}}^{-1} - \Pi_{\star_{1}}^{-1}) \mathrm{cos} \biggl ( 2 n \pi \Pi_{0} \Pi_{\star_{1}}^{-1}  \biggr  )
  - \biggl ( \frac{\omega_{n}}{2L}  \biggr )^2 
\biggl ( \mathrm{sin}\biggl ( 2 n \pi \Pi_{0} \Pi_{\star_{2}}^{-1}  \biggr ) - \mathrm{sin}\biggl ( 2 n \pi \Pi_{0} \Pi_{\star_{1}}^{-1}  \biggr ) \biggr ) \biggr ]. \nonumber \\ \label{Eq_delP3}
\end{eqnarray} 
%Note that we have extra terms which, in the case of $\delta N$ modeled with a ramp function, 
%vanish where $\Pi_{\mu_{1}}^{-1} \to 0$ and $\Pi_{\mu_{2}}^{-1} \to \Pi_{\mu}^{-1}$, 
We obtain the expression (\ref{Eq_delP2}) 
if we take limits $\Pi_{\star_{1}}^{-1} \to 0$ and $\Pi_{\star_{2}}^{-1} \to \Pi_{\star}^{-1}$. 

Equation (\ref{Eq_delP3}) contains two oscillatory components, one of which has a period determined by a ratio 
between $\Pi_{0}^{-1}$ and $\Pi_{\star_{1}}^{-1}$, and the other has a period determined by a ratio 
between $\Pi_{0}^{-1}$ and $\Pi_{\star_{2}}^{-1}$, 
indicating that not only a sharp feature in the BV frequency 
but also the extent of convective overshooting 
have impacts on the $\Delta P_{\mathrm{g}}$ patterns. 



\end{document}